\documentclass[twocolumn, tighten, times]{aastex62}

\PassOptionsToPackage{hyphens}{url}\usepackage{hyperref}

\usepackage{amsmath,amssymb}
\usepackage{aas_macros}
\usepackage{graphicx}
\usepackage{CJKutf8}

\usepackage[all]{hypcap} %Links go to figures;breaks on deluxetables

\newcommand{\mum}{\ifmmode{\rm \mu m}\else{$\mu$m}\fi}

\newcommand{\sevenrm}{\rm\scriptsize}

\newcommand{\cosmology}{$\Omega _m = 0.27$,~$\Omega_{\Lambda}=0.73$~and~$H_0=71$km~s$^{-1}$Mpc$^{-1}$}

\submitjournal{ApJ}

\received{2018 May 15}
\revised{2018 August 29}
\accepted{2018 September 8}
\shorttitle{Reconcilation of Type-1 AGN IR SED}
\shortauthors{Lyu \& Rieke}

\begin{document}

\title{\bf\Large  
    Polar Dust, Nuclear Obscuration and IR SED Diversity in Type-1 AGNs \footnote{
	\bf The AGN template library developed in this work can be obtained on
	\href{https://github.com/karlan/AGN_templates}{github}
	(\url{https://github.com/karlan/AGN_templates})}
  }

\correspondingauthor{Jianwei Lyu}
\email{jianwei@email.arizona.edu}

\author[0000-0002-6221-1829]{Jianwei Lyu (\begin{CJK}{UTF8}{gkai}吕建伟\end{CJK})}
\affiliation{ Steward Observatory, University of Arizona, 
933 North Cherry Avenue, Tucson, AZ 85721, USA}

\author[0000-0003-2303-6519]{George H. Rieke}
\affiliation{ Steward Observatory, University of Arizona, 
933 North Cherry Avenue, Tucson, AZ 85721, USA}

\begin{abstract}
    Despite the hypothesized similar face-on viewing angles, the infrared
    emission of type-1 AGNs has diverse spectral energy distribution (SED)
    shapes that deviate substantially from the well-characterized quasar
    templates. Motivated by the commonly-seen UV-optical obscuration and the
    discovery of parsec-scale mid-IR polar dust emission in some nearby AGNs,
    we develop semi-empirical SED libraries for reddened type-1 AGNs built on
    the quasar intrinsic templates, assuming low-level extinction caused by an
    extended distribution of large dust grains. We demonstrate that this model
    can reproduce the nuclear UV-to-IR SED and the strong mid-IR polar dust
    emission of NGC 3783, the type-1 AGN with the most relevant and robust
    observational constraints.  In addition, we compile 64 low-$z$ Seyfert-1
    nuclei with negligible mid-IR star formation contamination and
    satisfactorily fit the individual IR SEDs as well as the composite UV to
    mid-IR composite SEDs.  Given the success of these fits, we characterize
    the possible infrared SED of AGN polar dust emission and utilize a simple
    but effective strategy to infer its prevalence among type-1 AGNs. The SEDs
    of high-$z$ peculiar AGNs, including the extremely red quasars, mid-IR
    warm-excess AGNs, and hot dust-obscured galaxies, can be also reproduced by
    our model. These results indicate that the IR SEDs of most AGNs, regardless
    of redshift or luminosity, arise from similar circumnuclear torus
    properties but differ mainly due to the optical depths of extended
    obscuring dust components.

\end{abstract}


\section{Introduction}

Accreting supermassive black holes (BHs) are surrounded by copious dust and
generally emit strongly in the infrared (IR). Compared with star-forming or
quiescent galaxies whose emission is dominated by stars and H{\sevenrm\,II}
regions, the active galactic nuclei (AGNs) have distinctive IR spectral energy
distributions (SEDs), offering a unique window to look for obscured AGNs
\citep[e.g.,][]{Lacy2004, Lacy2007, Stern2005, Stern2012, Donley2012}, a
critical means to explore the properties of dusty structures around their
central engines \citep[e.g.,][]{Fritz2006, Nenkova2008a, Nenkova2008b,
Stalevski2012, Stalevski2016}, and a powerful tool to constrain their host
galaxy properties by SED decompositions \citep[e.g.,][]{Bongiorno2007,
Bongiorno2012, Xu2015a, Lyu2016, Collinson2017}.  More importantly, the dusty
structures around the central engine, commonly termed as ``torus'', lay the
foundation of AGN unification and have physical scales that bridge the BH
accretion disk and the host galaxy \citep[e.g.,][]{Antonucci1993a, Urry1995}.
The IR SEDs have become a crucial probe for the AGN phenomenon and the
BH-galaxy coevolution \citep[e.g.,][]{Heckman-Best2014, Caputi2014,
Netzer2015}. Nevertheless, most studies are focused on individual objects or
some particular narrowly defined AGN population. In this work, we propose and
test a model to reconcile the various IR SEDs of type-1 AGNs over a broad range
of luminosity and redshift.

The intrinsic SEDs of the most luminous AGNs, or quasars (bolometric luminosity
$L_{\rm AGN, bol}\gtrsim10^{11}~L_\odot$), are observationally
well-characterized \citep[e.g.,][]{Elvis1994, Richards2006, Shang2011,
Krawczyk2013, Scott2014, Lyu2017, Lyu2017b, Lani2017}.  The average SEDs of
unobscured quasars, whether the parent sample is (mainly) optically-selected
\citep[e.g.,][]{Elvis1994, Krawczyk2013}, or combined with mid-IR selection
\citep[e.g.,][]{Richards2006}, or X-ray-selected \citep[e.g.,][]{Polletta2007,
Elvis2012}, have been found to be strikingly similar. The classical
\citet{Elvis1994}-like template has proven to be a realistic representation of
the infrared emission by hot dust in most quasars \citep{Lyu2017}. The
exceptions, including hot-dust-free or hot-dust-poor objects
\citep[e.g.,][]{Jiang2010, Hao2010, Hao2011}, are part of the quasar intrinsic
IR SED diversity seen at all redshifts. For unobscured quasars at $z\sim$0--6,
the AGN IR SEDs can be generally grouped into three basic types; besides the
normal AGNs described by the Elvis-like template, there are also the
warm-dust-deficient (WDD) AGNs and the hot-dust-deficient (HDD) AGNs
\citep{Lyu2017}. Despite the variations in the near-IR or mid-IR, the shapes of
quasar far-IR intrinsic SEDs have an identical pattern in a statistical sense
\citep{Lyu2017b}, which drops quickly at $\lambda\gtrsim20~\mum$
(\citealt{Xu2015a, Lyu2017b}; see also \citealt{Lani2017}).

By comparison, the behavior of the intrinsic emission from relatively faint
AGNs ($L_{\rm AGN,bol} \sim 10^8$--$10^{11} ~L_\odot$) is elusive because of
the significant galaxy contamination as well as the possible line-of-sight
(LOS) extinction. A frequent hypothesis is that the intrinsic SEDs of these
Seyfert nuclei are identical to those of quasars, thus one single template for
AGNs with a very broad luminosity range is assumed (e.g.,
\citealt{Hopkins2007}; \citealt{Assef2010}; \citealt{Donley2012}). However,
notable differences seem to exist, particularly in the IR. The AGN emission of
many Seyfert-1 galaxies is peaked at $\lambda\gtrsim10~\mum$ and the near- to
mid-IR broad-band SEDs are commonly described by a single power-law
\citep[e.g.,][]{Spinoglio1995,Alonso-Herrero2003, Prieto2010}. On the other
hand, normal quasars exhibit a prominent SED bump peaked in the UV and emit
relatively less strongly in the IR with an obvious SED jump due to emission by
hot dust starting at 1.3~$\mum$ and an almost flat ($\nu F_\nu\propto\nu^0$)
mid-IR SED at $\sim$3--20~$\mum$ \citep[e.g.,][]{Sanders1989, Elvis1994}. As
suggested by e.g., \citet{Prieto2010}, such different behavior might be caused
by obscuration. However, frequently only the reddening of the UV-optical SED is
considered, while the accompanying effect on the IR emission is ignored.

In fact, an equatorial optically-thick torus is only a first-order
approximation of the circumnuclear dust environment around AGN. Evidence for
polar dust at $\sim10^2$ pc scale has been suggested since the early 1990s
\citep[e.g.,][]{Braatz1993, Cameron1993, Bock2000} and the AGN obscuration is
known to happen at a range of different scales \citep[e.g., see review
by][]{Bianchi2012}. In theory, the AGN torus could form during the gas
accretion of the central black hole and have material exchanges with the ambient
environment \citep[e.g.,][]{Hopkins2012}. Although it cannot survive very close
to the central engine, dust is expected to be found in many other AGN
components, e.g., narrow-line regions \citep[e.g.,][]{Groves2006b, Mor2009}
and/or AGN-driven outflows \citep[e.g.,][]{Fabian1999, Murray2005}. As argued
by many authors, the AGN infrared SED might be easily reshaped by these
extended dusty structures \citep[e.g.,][]{Sturm2005, Groves2006b, Honig2012,
Honig2013, Honig2017}.

Mid-IR interferometric observations have become available for some nearby
Seyfert nuclei, allowing a direct investigation of the geometry of their
nuclear IR structures at parsec scales. Interestingly, these studies show that
the mid-IR warm dust emission in some systems is largely distributed along the
AGN polar direction, instead of from an equatorial torus
\citep[e.g.,][]{Raban2009, Honig2012, Honig2013, Tristram2014,
Lopez-Gonzaga2016, Leftley2018}. These observations have motivated the
developments of increasingly sophisticated dust models to explain the few
best-studied cases \citep{Honig2017, Stalevski2017}. In the case of NGC 3783,
it is proposed that both the torus and the polar dust are composed of optically
thick clouds and that the polar dust is composed of large carbon grains
\citep{Honig2017}. In comparison, the Circinus Galaxy is modeled with a
parsec-scale optically-thick dusty disk and an IR optically-thin cone
following the structure of the narrow-line region (NLR) out to a distance
of $\sim$ 40 pc \citep{Stalevski2017}. The success of these different
approaches indicates that these complex models could be highly degenerate for
most AGNs due to the lack of detailed observational constraints.
We will address whether it is possible to  develop a much simpler model that
can be applied uniformly and is still of sufficient fidelity to provide useful
insights.

With the success of our intrinsic templates to reproduce the AGN IR emission of
bright quasars at different redshifts \citep{Lyu2017}, it is of considerable
interest to explore whether they also apply to relatively low-luminosity AGNs.
The possible existence of low-optical-depth dust in the vicinity of
the AGN nucleus, as outlined above, motivates us to develop a new library of
reddened AGN templates. We take the \citet{Lyu2017} empirical quasar templates
as givens for polar-dust-free AGNs and investigate the extent to which the
addition of a low-optical-depth but extended dust component with reasonable
assumptions for the dust grain properties and their large-scale distribution
can yield IR SED shapes consistent with those observed. In
Section~\ref{sec:model}, we introduce this model and validate it by fitting the
detailed observations of NGC 3783, the archetypical example of a type-1 AGN
with its mid-IR emission dominated by polar dust.

Although the extended dust distribution may have a range of morphologies, any
mid-infrared emission by low-optical-depth dust should be roughly isotropic and
hence detectable from any view angle. We therefore focus on whether the model
trained for NGC~3783 can be generally applied to match the infrared SEDs of
those AGNs where standard quasar templates fail, assuming the choice of
intrinsic AGN template and the optical depth of the obscuration as the only
free parameters for the SED shape. This analysis is carried out on 64 low-$z$
Seyfert nuclei with negligible mid-IR star formation contamination in
Section~\ref{sec:obs}. In addition, if the polar dust emission is a common
phenomenon for all populations of moderate-luminosity AGNs, a consequence is
that Seyfert-1 nuclei should be moderately obscured on a statistical
basis. We build composite SEDs of Seyfert-1 SEDs and confirm this prediction.

We find that the deviations from the quasar-like SED templates can indeed be
explained to first order by the combination of extinction and infrared emission
by polar dust. In Section~\ref{sec:pol_char}, we characterize the SED features
of the polar dust emission and discuss the prevalence of polar dust in a sample
of AGNs much larger than those that can currently be explored in any detail
through mid-infrared interferometry. We also demonstrate that one single
semi-empirical template can describe the influence on the AGN SED by the polar
dust emission for most objects. A consistency check of the results from
our SED analysis and those from morphology-based identification for AGN
polar dust emission is also carried out.

Various AGN populations with peculiar SED features that cannot be easily
matched by the classical AGN templates have been reported at high-$z$. Some
notable examples are extremely red quasars \citep{Ross2015, Hamann2017}, AGNs
with mid-IR warm-excess emission \citep[e.g.,][]{Xu2015a} and the hot
dust-obscured galaxies \citep[e.g.,][]{Eisenhardt2012, Wu2012}.  The success of
our model at low-$z$ encourages us to explore if these peculiar SED features
can be explained in a similar way. These studies are presented in
Section~\ref{sec:highz}.

Section~\ref{sec:dis} provides discussions on the implications of these results
for interpreting the AGN IR emission, the relation between X-ray obscuration
and polar dust extinction, and the AGN unification scheme. We propose a
tentative picture of the different circumnuclear dusty environments among AGNs
that leads to their diverse IR properties. Section~\ref{sec:summary} is a final
summary.

We adopt cosmology  \cosmology \citep{Bennett2003}~throughout this paper. 
We use the term {\it type-1} to describe AGNs showing broad emission lines
without distinguishing if the LOS is dust-obscured or not.  Since the word
{\it obscured} is frequently reserved to describe type-2 (or narrow-line)
AGNs, we adopt the name {\it reddened type-1 AGNs} to denote the
broad-line AGNs with some extinction along the LOS (aka the polar
direction), even if the real reddening might be insignificant because of a
very flat extinction curve \citep[e.g.,][]{Gaskell2004}. Lastly, the word
{\it unobscured} means no extinction, neither from the torus nor from any
extended dust distribution, along the LOS to the central engine.

\section{A Semi-empirical SED Model for Reddened Type-1 AGNs}\label{sec:model}

We introduce a relatively simple framework to produce a library of reddened
type-1 AGN templates, which will be used to fit the SEDs of Seyfert-1 nuclei in
Section \ref{sec:obs}. This model is based on two major assumptions: 
\begin{enumerate}
    \item Seyfert nuclei have a circumnuclear optically-thick torus whose SED
	variations from a face-on viewpoint can be described by the intrinsic
	AGN templates of unobscured quasars;
    
    \item Besides the torus, there could exist an extended dust component with
	some power-law density profile that is dominated by large dust grains
	heated by the AGN.
\end{enumerate} 
We describe the motivations as well as the details of these assumptions in
Section~\ref{sec:model-agn-temp}--\ref{sec:model-dust-obs}. The model and its
behavior are presented in Section~\ref{sec:model-model-pre}. In
Section~\ref{sec:ngc3783-model}, we test our approach by fitting the
observations of NGC~3783.

\subsection{Accretion Disk and Dusty Torus}\label{sec:model-agn-temp}

The continuum SED of an AGN is contributed mostly by the UV-optical emission
from the accretion disk around the black hole and the near-IR to mid-IR
emission emerges from the surrounding dusty structures. To reduce the
uncertainties, we adopt well-tested extinction-free empirical templates to
represent this AGN intrinsic emission.

The UV to mid-IR SEDs of most luminous type-1 quasars are well described by the
\citet{Elvis1994}-like AGN template, regardless of the redshift \citep[e.g.,
see discussion in][]{Lyu2017}. An AGN-heated dusty structure in type-1 quasars
is revealed by the broad IR emission bump at $\lambda\sim1.3$--40$~\mum$.  In
addition, broad emission lines have been detected in the optical polarized
spectra of type-2 quasars \citep{Zakamska2005}. Under the precepts of AGN
unification \citep{Antonucci1993a, Urry1995}, these observations support the
existence of some equatorial optically-thick dusty structures, which cause the
nuclear photons to preferentially escape along the polar direction. 

As demonstrated by \citet{Lyu2017}, the diversity of intrinsic IR emission among
type-1 quasars at $z\sim$0--6 can be characterized by three distinct templates
derived for (1) normal AGNs, (2) warm-dust-deficient (WDD) AGNs, and (3)
hot-dust-deficient (HDD) AGNs. These templates are unlikely to be affected
significantly by dust extinction since their derivations are based on the study
of optically-blue quasars that are not obscured. In other words, there should
be no significant dust distribution along the polar direction.  We suggest
these AGN templates describe the emission from the unobscured accretion disk
plus a face-on view of the dusty torus\footnote{We use the word {\it torus} to
describe the polar-dust-free obscuration structures as in optically-blue
quasars and do not make assumptions on the geometry or boundaries. }.

At $\lambda<0.1~\mum$, current observations do not give good constraints.
Following \citet{Stalevski2016}, we assume a broken power-law, where
\begin{equation}
    \nu F_\nu \propto 
     \begin{cases}
	 \lambda^{0}   & \quad  0.01~\mum <\lambda < 0.1~\mum  \\
	 \lambda^{1.2} & \quad  0.001~\mum <\lambda < 0.01~\mum 
     \end{cases}
\end{equation}
Nevertheless, our study will not be influenced by the assumed X-ray to
UV SED shape. In fact, considering the likely dominance of large dust grains
along the face-on direction (see Section~\ref{sec:model-grain}), the extinction
at these wavelengths is small and will not contribute to the IR SEDs.

\subsection{Extended Polar Dust Component}\label{sec:model-dust-obs}

Besides the torus component characterized by the intrinsic AGN templates
discussed above, we introduce another dust component to provide relatively
low-level obscuration for the nucleus as well as the additional IR emission
from the absorbed energy.  
 
We suggest that the dust size distribution in this component is dominated by
very large particles, which will be characterized by grain size cutoffs,
$a_{\rm max}$ and $a_{\rm min}$. In real situations, different grain species
sublimate at different temperatures and could have a broad range of dust
sublimation zones. However, we find that the calculated SEDs do not change
significantly between $T_{\rm sub} = 2000$ and $T_{\rm sub} = 1500$, indicating
that the introduction of separate values appropriate for carbon and silicates
would not change our results.  For simplicity, we adopt the same dust
sublimation temperature $T_{\rm sub}$ for all the grain compositions.

For the large scale structure, we assume a density profile parameterized as a
power-law in radius with slope $\alpha$:
\begin{equation}
    \rho(r)\propto r^{-\alpha},  ~~~~ r_{\rm in}<r < r_{\rm out}~,
\end{equation}
where the inner radius $r_{\rm in}$ is equal to the dust sublimation radius
$R_{\rm sub}$ set by $T_{\rm sub}$ as well as the light source luminosity, and
the outer radius $r_{\rm out}$ is a free-parameter. We introduce another
parameter, the outer-to-inner radius ratio $Y=r_{\rm out}/r_{\rm in}$, to
describe $r_{\rm out}$. Given the nature of this model, the geometry will not
influence the dust emission SED so that there is no need to introduce more free
parameters.

In the following, we outline the motivations behind these configurations.

\subsubsection{Grain Properties}\label{sec:model-grain}

There are strong reasons to suggest the classical dust properties are altered
by the harsh environment of the direct exposure to an AGN.  As suggested by
\citet{Aitken1985}, small grains ($a\sim10^{-3}~\mum$) can be easily destroyed
out to several hundred parsecs in a typical Seyfert-1 nucleus, on a timescale of
less than a few years. Physically, the majority of dust destruction mechanisms
around AGN are relatively less significant for large dust grains
\citep{Laor1993}. For example, large grains are expected to exist close to the
torus inner part since they have smaller $R_{\rm sub}$ than small grains. The
torus itself can be dynamically unstable and material exchanges with the
surrounding environments through various mechanisms are expected from
simulations \citep[e.g.,][]{Hopkins2012}. \cite{Baskin2018} have analyzed the
effects of sublimation on both carbon and silicate grains near an AGN. They
find that only large ($a > 0.1~\mum$) carbon grains can survive at the outer
edge of the BLR out to about 20 times this radius. They also conclude that the
silicate grain size distribution will be skewed toward large sizes to
significantly greater distances.

Conditions for grain growth may exist in the circumnuclear tori, where the
densities are high and grains are shielded from the X-ray and UV output of the
central engine \citep[e.g.,][]{Maiolino2001b}.  As suggested by e.g.,
\citet{Honig2012, Honig2013}, some dust in the torus can be uplifted into the
polar direction by AGN winds. It may also be possible for these large dust
grains to form in situ.  \citet{Elvis2002} suggested that dust can form in
AGN-driven winds where conditions are similar to those in the winds of
late-type stars. By these mechanisms, grains in the 0.1--1$~\mum$ range are
plausible \citep{Hofner2008}.  However, determining observational constraints
on the grain sizes around AGN can be quite difficult. This can be seen from the
diverse AGN UV-optical extinction curves reported that range from steeply
rising SMC-like laws \citep[e.g,][]{Hall2002, Richards2003} to flat or gray
laws \citep[e.g.,][]{Gaskell2004, Czerny2004, Gaskell2007}. As argued by e.g.,
\citet{Baskin2018}, it is possible the grain properties depend on the observing
angle.

Nonetheless, despite various uncertainties, there are observational indications
of relatively large dust grains around AGNs, including (1) the lower ratios
$A_V$/$N_H$ and $E(B-V)$/$N_H$ in intermediate-type Seyfert galaxies
(\citealt{Maiolino2001a, Maiolino2001b}; but see \citealt{Weingartner2002});
(2) the lower ratios between $A_V$ and the mid- IR silicate absorption
strength, $\Delta \tau_{9.7}$, in type-2 AGNs
\citep{Lyu2014, Shao2017}; (3) successful fittings of the silicate emission
profile in quasars and Seyfert galaxies with micron-sized grain models
\citep{Xie2017}; (4) the smaller observed torus inner radius from near-IR
interferometry of nearby Seyfert nuclei compared with the expectations for
classical dust grains (e.g.,\citealt{Kishimoto2007,Kishimoto2009, Honig2013,
Burtscher2013}; but see \citealt{Kawaguchi2010}).

As shown in Appendix~\ref{sec:model-thickness}, for classical ISM properties at
low optical thickness, the mid-IR silicate emission feature at
$\lambda\sim10~\mum$ would be very prominent with a sharp peak (see also, e.g.,
\citealt{Fritz2006, Nenkova2008a}) that is not commonly seen among Seyfert
nuclei \citep{Hao2007}. Instead, the lack of such detections could be an
expected consequence of large grains. For example, dust grains with size
$a\gtrsim0.3~\mum$ would reduce the strength of silicate features effectively
\citep{Laor1993}.

Due to the difficulties for setting direct constraints on the dust properties
around the AGN, we minimize departures from standard ISM grain models and
assume only the grain sizes, as characterized by the grain size cuts, $a_{\rm
max}$ and $a_{\rm min}$, are altered in the vicinity of an AGN.

\subsubsection{Large-scale Geometry}\label{sec:model-geometry}

Currently we do not have strong observational constrains about the geometry of
the dust responsible for the low-level obscuration in type-1 AGNs.
Nevertheless, AGN outflows \citep[e.g.,][]{Crenshaw2003, Piconcelli2005} could
be a natural mechanism to distribute the dust around the nucleus. Physically,
it has been found that the radiation pressure on resonant absorption lines
alone can not explain the outflow rates. The radiation feedback on dust within
the clouds could be an effective mechanism \citep[e.g.,][]{Roth2012}. Based on
a study of $\sim$3000 type-1 AGNs, \citet{Zhang2013} found that the relative
strength of the mid-IR to the optical flux of these objects is correlated with
the strength of outflows. 

Several teams have tried to explore the origin of the polar dust, showing the
outflow scenario is a promising solution. For example, \citet{Honig2012,
Honig2013} proposed that the dusty outflows could be launched from the surface
of the inner torus and the \citet{Honig2017} model motivated by this picture
successfully explained the behavior of NGC~3783, a Seyfert-1 nucleus with a
firm detection of polar dust emission. From an analysis of high-spatial mid-IR
images of 149 nearby Seyfert galaxies, \citet{Asmus2016} found that elongated
polar dust emission is co-spatial with the direction of AGN outflows for 18
objects.  

Little is known about the exact density profile of the gas outflows. As a
result, analytic analyses of self-similar solutions are typically pursued.  We
introduce a power-law density profile, $\rho(r)\propto r^{-\alpha}$, to
approximate the real situations.

Physically, we do not expect the outflow solution retains the memory of initial
conditions on large scales.  As suggested by \citet{Faucher-Giuere2012}, to
reach a finite free expansion radius, the gas density profile should have
profiles with $\alpha\lesssim2$.  Observationally, various values of $\alpha$
have been derived for materials in the outflows. For example, \citet{Behar2009}
derived $\alpha\sim$1.0--1.3 for five nearby Seyfert nuclei from analyzing the
X-ray absorption spectra. \citet{Feruglio2015} suggested a $r^{-2}$ profile for
the ultra-luminous IR galaxy Mrk 231. \citet{Revalski2018} derived the electron
density of the narrow-line regions in the Seyfert-2 nucleus Mrk 573, finding
$n_e\propto r^{-0.4}$--$r^{-0.6}$. Additionally, a constant density absorber
($\alpha$=0) is quite unlikely the real case. Assuming that the dust and gas
are well mixed with a constant dust-to-gas ratio, we suggest the dust density
profile should satisfy $0<\alpha\lesssim2$.

Based on mid-IR interferometry observations, the AGN polar dust emission is
found to be elongated \citep[e.g.,][]{Honig2012, Honig2013, Tristram2014,
Lopez-Gonzaga2016} and possibly distributed along the edges of the ionization
cone \citep{Stalevski2017}. In our model, however, the likely uneven
distribution of the polar dust component will not influence its IR emission
SED. This is a direct consequence of optically-thin dust emission in the IR,
especially for large grains \citep[e.g.,][]{Laor1993, Ivezic1997}. 

The modest levels of face-on extinction in type-1 AGNs correspond to a small
value of $\tau_{\rm V}$. Since the dust opacity is a strong function of
wavelength that decreases rapidly towards the infrared, the extinction for such
IR-reprocessed dust emission is likely to be close to zero.  In other words,
the IR emission of any dusty structures with a low $\tau_{\rm V}$ is highly
transparent: the emission from a single geometry element at some given
location, where the included dust grains can be considered in local
thermodynamic equilibrium (LTE), would share the same SED from different
viewing angles and this SED would transit through other surrounding dusty
structures without any notable changes.  Consequently, the total integrated IR
SED can be described as a summation of the dust emission from individual LTE
geometry elements at all possible locations. 

At the same distance $r$, the temperature of each LTE element would be the same
as is the SED, $B_\lambda(r)$. The total emission from all the dust at the same
distance is linearly scaled by the total numbers of LTE elements at the
corresponding radius, $\rho(r)$, and has little to do with their possible
uneven distribution.  The total integrated SED, $F_\lambda$, can be
approximated by adding the contributions of all the elements at various radii,
or
\begin{equation}
    F_\lambda \simeq \int_{r_{\rm in}}^{r_{\rm out}} \rho(r) B_\lambda(r) dr  ~ ~ .
\end{equation}
Thus if the average radial profile, $\rho(r)$, is similar, the dust
distribution at small scales, whether it is smooth, clumpy or filamentary, will
not influence the SED. 

As long as the integrated optical depth, $\tau_{\rm V}$, is low, the effects of
asymmetries in the dust large-scale structure and/or illumination along the
radial directions would proportionally change $\rho(r)$, only resulting in a
scaling down of the output polar dust emission. Note that $\tau_{\rm V}$ is
linearly scaled with $\rho(r)$ by 
\begin{equation}
    \tau_{\rm V} =  \int_{r_{\rm in}}^{r_{\rm out}} \rho(r) C_{\rm ext, V} dr 
                 = C_{\rm ext, V}  \int_{r_{\rm in}}^{r_{\rm out}} \rho(r) dr~ ~,
\end{equation}
where $C_{\rm ext, V}$ is the optical V-band extinction of all the grains
within a single LTE element. Thus these geometry effects can be modeled by
simply changing $\tau_{\rm V}$. Figure~\ref{fig:geo_cartoon} provides a
simple illustration. In Appendix~\ref{sec:model-thickness}, we provide some
simple demonstrations of the lack of influence of the dust geometry on the
derived SED with three-dimensional dust radiative transfer simulations.
Although the distribution of polar dust realistically could be highly
complicated, for the purpose of determining the resulting SED it is equivalent
to assume spherical symmetry. 

%%% Figure 1 %%%

\begin{figure}[!htb]
    \begin{center}
	\includegraphics[width=1.0\hsize]{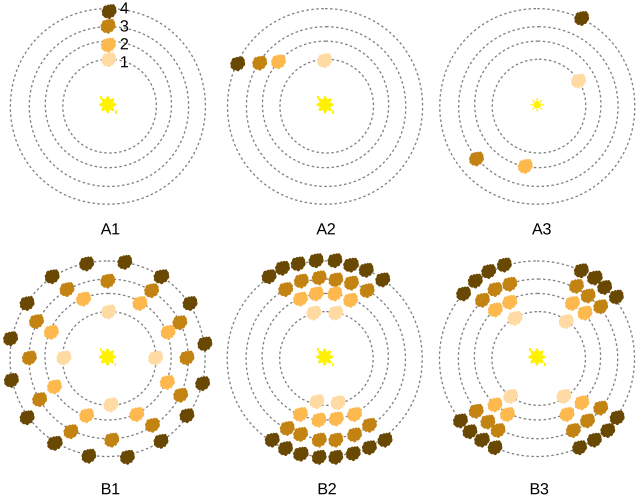}
		\caption{Cartoon illustrations of the optically-thin IR
		    emission for different geometry configurations. Imagine
		    that the dust structures are composed of similar small clouds
		    at four different radii $r_1$, $r_2$, $r_3$, $r_4$ to the
		    light source (the central yellow star in each panel). The
		    total emission of all the clouds at a given $r_i$ would
		    simply be $N_i F_i$, where $N_I$ is the number of clouds
		    and $F_i=F_i(\nu)$ is the emission SED of one single cloud.
		    Since the dust emission is assumed to be optically thin,
		    the integrated emission of all the dust clouds with
		    geometries described in panels A1, A2, A3 will be the same,
		    $F_{\rm tot, A}= F_1 + F_2 + F_3 + F_4$. Similarly, since
		    the number of clouds, $N_i$, in each radius, $r_i$, is the
		    same, the integrated SEDs in panels B1, B2, B3 are also
		    identical for optically-thin emission, and $F_{\rm tot, B}
		    = N_1 F_1 + N_2 F_2 + N_3 F_3 + N_4 F_4 = 2 F_1 + 4 F_2 + 6
		    F_3 + 8 F_4$, which is not dependent on how the clouds are
		    distributed.
		    }
	\label{fig:geo_cartoon}
    \end{center}
\end{figure}

In summary, when the $\tau_{\rm V}$ of the extended dust component is not very
high, its IR SED shape would be only dependent on the radial density profile.
Since the nuclear dust morphology cannot be constrained for most AGNs, we will
not introduce detailed geometry to match the very few observations.

\subsection{Reddened Type-1 SEDs from the Model}\label{sec:model-model-pre}

We use the latest version of the radiative transfer code DUSTY
\citep{dusty_new}\footnote{Accessible at
\url{https://github.com/ivezic/dusty}.} to obscure the three AGN templates
presented in \cite{Lyu2017}.  This code assumes a spherical symmetry and solves
the one-dimensional radiation transfer equations as described in
\citet{Ivezic1997}. Although a realistic model would place the emitting
clouds preferentially in the polar direction, we have just shown that the
output spectrum for the optically thin case would be identical. For a
centrally-heated spherical density distribution, the program needs the SED of
the radiation source, a density profile of the dust distribution and
information regarding the boundaries, the dust properties (chemical
composition, grain size distribution and the sublimation temperature), and a
range of required optical depths at some specific wavelength.

We calculate three sets of reddened templates separately for normal, HDD, and
WDD AGNs. The model parameters are summarized in Table~\ref{tab:dusty_model}.
We adopted optical properties for graphite and silicate grains from
\cite{Draine1984} with the standard Mathis-Rumpl-Nordsieck (MRN) power-law
grain size distribution distribution $dn/da\propto a^{-3.5}$ \citep{MRN1977}
but left the boundaries, $a_{\rm max}$ and $a_{\rm min}$, to be varied.

\capstartfalse
\begin{deluxetable}{ccc}
    %\tabletypesize{\scriptsize}
    \tabletypesize{\footnotesize}
    \tablewidth{1.0\hsize}
    \tablecolumns{3}
    \tablecaption{DUSTY model setup\label{tab:dusty_model}
    }
    \tablehead{
	\colhead{Parameter} & 
	\colhead{Label} & 
	\colhead{Value} 
}
\startdata
    $r_\text{in}$ temperature         & $T_\text{in}$ &    2000 K, 1500$^*$ K, 1000 K             \\
    density profile                   & $\alpha$  & 0, 0.5$^*$, 1, 1.5, 2  \\
    outer-to-inner radius             & $Y$  & $50$, $500^*$, $5000$         \\
    silicate:graphite mixture         &  & 0:1, 0.53:0.47$^*$, 1:0      \\
    maximum grain size  & $a_\text{max}$ &  0.25, 2.5, 10$^*$, 100       \\
    minimum grain size  & $a_\text{min}$ &  0.005, 0.01, 0.05$^*$, 0.1   \\
    \hline     
    input radiation SED               &  &  norm, WDD$^*$, HDD           \\
    optical depth        & $\tau_\text{V}$  &  0--10 with a step of 0.25
    \enddata
    \tablecomments{ We use * to indicate the reference parameters that
    adopted to demonstrate the influence of the output SEDs in
    Figure~\ref{fig:dusty_para}.
    }
\end{deluxetable}
\capstarttrue

%%% Figure 2 %%%

\begin{figure*}[!htb]
    \begin{center}
	\includegraphics[width=1.0\hsize]{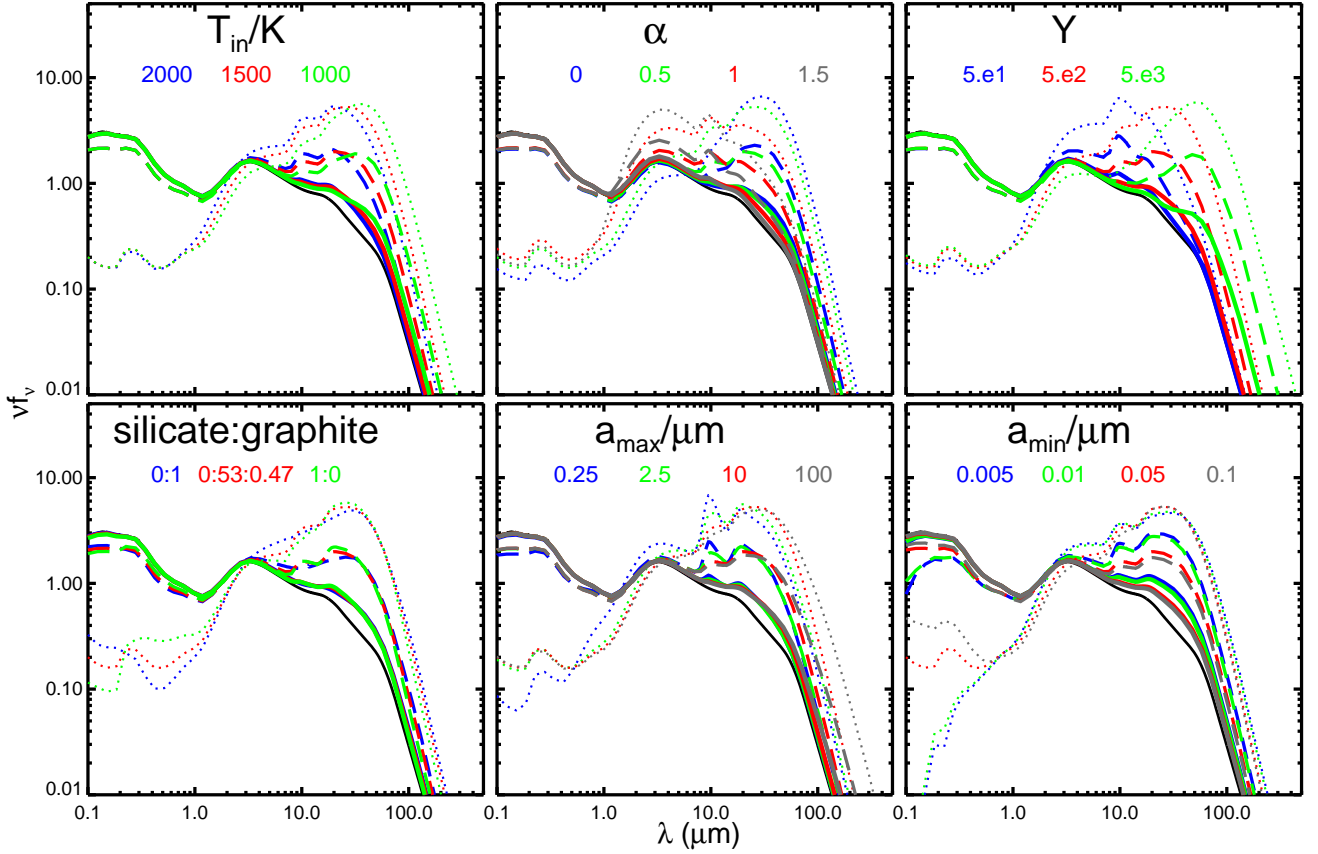}
		\caption{
		    The influence of model parameters on the output SEDs.  For
		    each model setup, we computed the output SEDs with
		    different levels of optical extinction $\tau_{\rm V}$=0.1
		    (solid lines), 0.63 (dashed lines) and 3.98 (dotted lines). 		    
		    }
	\label{fig:dusty_para}
    \end{center}
\end{figure*}

Figure~\ref{fig:dusty_para} presents the reddened WDD AGN SEDs as a function of
the optical extinction level $\tau_{\rm V}$ for different values of model
parameters. It is interesting that for large dust grains, the dust-reprocessed
SEDs are not sensitive to the mixtures of silicates and graphites. For a given
$\tau_{\rm V}$, the amount of UV-optical reddening is only sensitive to the
smallest grain size $a_\text{min}$ and has little to do with the dust geometry
configurations. The strength of the mid-IR silicate emission feature is
sensitive to the maximum grain size $a_\text{max}$ and the geometry parameters
$Y$ and $\alpha$.  The relative strength of the hot dust
emission is mainly determined by the compactness of the dust distribution
($\alpha$, $Y$).  Finally, the shape of the broad-band IR continuum is mainly
determined by the assumed geometry (as well as the input intrinsic SED).

Since the dust temperature at the inner radius ($T_{\rm in}=T_{\rm sub}$) and
relative fraction between silicate and graphite grains have limited effects on
the reddened SEDs, we decide to adopt $T_{\rm in}=1500$ K, and a normal
0.53:0.47 mixture of silicate and graphite grains.

\subsection{Model Validation with Observations of NGC~3783}\label{sec:ngc3783-model}

Our proposed simple model differs substantially from the approach of
\citet{Honig2017}, who fitted the SED and interferometric data for NGC 3783
with a specific system of very optically thick ($\tau_{\rm V} = 50$) small clouds. We
show here that our model can fit the SED of this object equally well.

We adopted the X-ray to mid-IR subarcsecond high-spatial-resolution (HSR)
nuclear photometry SED of NGC~3783 presented in \citet{Prieto2010}. To improve
the mid-IR continuum constraints, we collected its 2.5--5~$\mum$ {\it AKARI}
spectrum \citep{Kim2015} and 5.5--38~$\mum$ {\it Spitzer}/IRS spectrum
\citep{cassis}.  Since the spectral flux is consistent with the HSR photometry
at corresponding wavelengths (the {\it Spitzer}/IRS flux is also identical to
the spectrum obtained through a 0.75\arcsec$\times$0.53\arcsec~aperture; see
Figure 3 in \citealt{Honig2013}) and there is no obvious star-formation
feature, we believe these data apply to the AGN emission. To gauge the far-IR
emission strength from the nucleus, we used the 1~kpc aperture {\it Herschel}
70~$\mum$ and 100~$\mum$ photometry of NGC~3783 from
\citet{Garcia-Gonzalez2016}. For the foreground extinction by the Milky Way
(MW), we obtained $E(B-V)=0.10$ from the SFD dust map
\citep{Schlegel1998,Schlafy2011} and used the \cite{Fitzpatrick1999} extinction
law to correct the UV-optical data.

For NGC 3783, the torus component is believed to dominate the near-IR emission
but not the mid-IR \citep{Honig2013}. After comparing the observed SED of this
object with the \citet{Lyu2017} intrinsic AGN templates, we suggest its torus
emission should be WDD-like. The 0.1--30~$\mum$ SED of NGC 3783 is then fitted
by the reddened WDD AGN template with parameter ranges for the polar dust
component in Table~\ref{tab:dusty_model}. We use Markov Chain Monte Carlo
(MCMC) algorithms described by \citet{SATMC} to sample the large parameter
space and find the most likely parameter combination. The values of the
best-fit parameters and their 1-$\sigma$ uncertainties (68\% confidence levels)
are summarized in Table~\ref{tab:ngc3783_model}. The large fitted parameter
errors suggest a strong degeneracy. After considering the possible
contaminations by optical emission lines and the fact that the far-IR
photometry provides an upper limit to the AGN-heated dust emission, we finally
decide to adopt $\rho(r)\propto r^{-0.5}$, $Y=500$ and
$T_{\rm in}$=1500 K, and grain size cutoffs at $a_{\rm min}=0.04~\mum$ and
$a_{\rm max}=10~\mum$. Most of these values are picked near the mid-point of
the fitted ranges for illustration.

\capstartfalse
\begin{deluxetable}{cccc}
    \tabletypesize{\footnotesize}
    \tablewidth{1.0\hsize}
    \tablecolumns{4}
    \tablecaption{Suggested model parameters for NGC 3783\label{tab:ngc3783_model}
    }
    \tablehead{
	\colhead{Parameter} & 
	\colhead{Label} & 
	\colhead{Adopted Value} &
	\colhead{MCMC output}
}
\startdata
    $r_\text{in}$ temperature         & $T_\text{in}$   & {\bf 1500} K    &                                    \\
    density profile                   & $\alpha$        &  0.50           &  $0.47^{+0.50}_{-0.44}$          \\
    outer-to-inner radius             & $Y$             &  500            &  $475^{+513}_{-353}$              \\
    silicate:graphite mixture         &                 & {\bf 0.53:0.47} &                                    \\
    maximum grain size                & $a_\text{max}$  &  10            &  $20^{+79}_{-18}$                \\
    minimum grain size                & $a_\text{min}$  &  0.04           &  $0.05^{+0.05}_{-0.04}$       \\
    \hline                                                     
    input radiation SED               &                 & {\bf WDD}       &                              \\
    optical depth                     & $\tau_\text{V}$ &   1.4           &  $1.44^{+0.55}_{-0.60}$    
    \enddata
    \tablecomments{
	We use boldfaces to indicate assumed parameter values that do not go to
	MCMC parameter space sampling.
    } 
\end{deluxetable}
\capstarttrue

\begin{figure*}[!htb]
    \begin{center}
	\includegraphics[width=1.0\hsize]{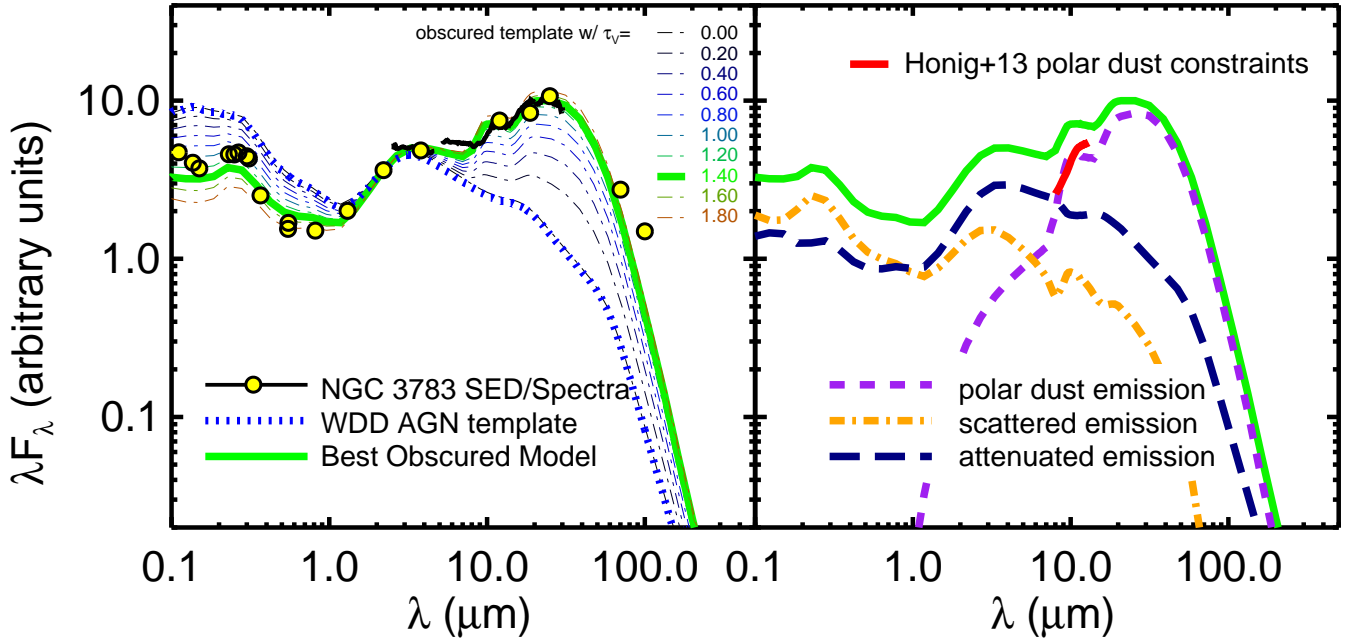}
		\caption{
		    {\bf Left}: simulated SEDs from the DUSTY model as a
		    function of optical depth $\tau_{\rm V}$ (thin lines).  We
		    assume the WDD AGN template (blue line) is obscured by a
		    dust distribution with density profile $\rho(r)\propto
		    r^{-0.5}$, outer-to-inner radius $Y=500$, and dust temperature at inner radius
		    $T_\text{in}$=1500~K. The dust grains have the standard ISM
		    mixture (53\% silicate and 47\% graphite) with a standard
		    distribution of grain sizes ($p$=3.5) but larger size
		    cutoffs with $a_\text{min} = 0.04~\mum$ and $a_\text{max} =
		    10~\mum$. The NGC 3783 SED is matched by the model with
		    $\tau_{\rm V}$=1.4 mag (thick green line).  {\bf Right}:
		    the contributions of various components of the best-fit
		    model as a function of wavelength. We also plot the
		    8--13~$\mum$ polar dust emission strength constrained by
		    \citet{Honig2013} for comparison.
		}
	\label{fig:rt_model}
    \end{center}
\end{figure*}

As shown in the left panel of Figure~\ref{fig:rt_model}, the UV-to-far-IR
nuclear SED of NGC~3783 is reasonably matched by the WDD AGN template obscured
by the suggested polar dust component with an optical depth $\tau_{\rm V}=1.4$.
In the right panel, we show the relative contributions of the attenuated and
scattered WDD emission, as well as the polar dust emission in the best model.
Besides the near-IR, the torus emission SED of the WDD AGN is not modified
by the polar dust obscuration, so we can linearly separate the torus emission
and the polar dust emission.  At $\sim$10~$\mum$, the optically-thin component
would contribute about 75\% of the total emission, which is in good agreement
with the polar dust emission strength constrained by interferometry
\citep{Honig2013, Lopez-Gonzaga2016}. Since the highly optically-thick torus
can block the accretion disk emission along the equatorial direction and the
dust along the polar direction is likely to be preferentially heated, some
elongated emission from the regions responsible for the low-level obscuration
could be observable if the system has an appropriate viewing angle. 

%%% Figure 3 %%%

Although our goal does not include reproducing any detailed dust morphology,
this model should yield similar sizes of the dusty structures to the
observations.  From fitting the mid-IR interferometry data of NGC~3783,
\citet{Honig2013} reported the nuclear dust is distributed over 20--70 $r_{\rm
in}$, where $r_{\rm in}\sim0.06$ pc is derived from near-IR reverberations. The
model geometry $r_{\rm out}/r_{\rm in}\sim500$ suggested from our SED fitting
is much larger than this observation. However, at a given wavelength, we only
observe dust with a narrow range of dust temperatures; the outer part of the
extended dust component would be too cold to observe in the mid-IR. Based on
the radial profile of dust temperature in our model, the observed geometry seen
at 8--13~$\mum$ would have a size of 20--40 $r_{\rm in}$, which in fact
qualitatively agrees with the \citet{Honig2013} observations (see more
discussion in Section~\ref{sec:size}).

\subsection{The Degeneracy of Models Constrained Only by the AGN SED}

The degeneracies of SED fitting - as shown here, equally good results from the
\cite{Honig2017} optically-thick clumpy dust model and our simple
low-optical-depth obscuration model based on empirical AGN templates - indicate
that the same observations can be explained through rather different
descriptions about the AGN circumnuclear extended dust. In general, we lack any
detailed structural information in the mid-infrared for most known AGNs.
Consequently, additional parameters to characterize the geometry of this
structure cannot be realistically constrained. As explained in
Section~\ref{sec:model-geometry}, at low optical depth, the behavior of polar
dust emission should not be sensitive to the shape of its large-scale
distribution, nor the observing angle. Thus our reddened AGN SED model is
consistent with the possible diverse polar dust morphologies.

\section{Reproducing the SEDs of Low-$z$ Seyfert Nuclei}\label{sec:obs}

\subsection{Sample and Data}

In view of the success in reproducing the observations of NGC~3783, we will
test if a similar model works for other Seyfert nuclei as well. To reduce the
ambiguity in interpreting the SED, we have searched the literature and archives
to find low-$z$ Seyfert-1 nuclei where the AGN mid-IR emission can be isolated
from the host star-formation contamination or the contribution from the latter
can be safely ignored. The details can be found in Appendix~\ref{sec:data}.
Below is a brief summary.

We first compiled AGNs with high spatial resolution (HSR) mid-IR SED
constraints.  About half of them are Seyfert 1-1.5 nuclei from
\citet{Asmus2014}; we required that the ground-based 12~$\mum$ flux
measurements be consistent with the much larger beam WISE W3 band observations. We
also included some very well-studied Seyfert-1 nuclei and all the type-1
objects with mid-IR interferometry observations. Due to these selections, the
near-IR stellar contamination is not very significant in this HSR sample. In
total, there are 32 nuclei selected. These objects span a redshift range of
0.002-0.16, with a median at $z=0.02$. IR photometry from the literature and
surveys like 2MASS, UKIDSS and WISE were collected to define the SEDs.

In addition, we looked for optically-selected broad-line AGNs from the Sloan
Digital Sky Survey (SDSS; \citealt{York2000}) that have {\it Spitzer}/IRS
spectra allowing isolation of the nuclear emission in the mid-IR. The
homogeneous datasets of this sample enable the derivation of composite SEDs
that minimize the effects of different observing angles and variability, and
can be used to test the complete range of  reddened AGN SEDs predicted from our
model.  For this SDSS/{\it Spitzer} sample, we ended up with 33 type-1
AGNs with a redshift range of 0.008--0.20, with the median at $z=0.06$.  Except
for NGC 4235, all of them are new objects with no duplication found in the HSR
sample.  Archival and literature X-ray, UV, optical and IR data from e.g.,
XMM-{\it Newton}, {\it Chandra}, {\it GALEX}, {\it Hubble}, {\it 2MASS},
{\it WISE}, {\it Spitzer}, {\it AKARI}, {\it Herschel} and other ground-based
optical and IR facilities are collected for these objects with a careful
selection of photometry apertures if possible.

To have some crude idea on the influence of variabilities on the IR SEDs
of these objects, we determined their mid-IR variability amplitudes in the
WISE W1 and W2 bands using the archival data from the {\it WISE} \citep{wise}
and the Near-Earth Object {\it WISE} Reactivation mission ({\it NEOWISE-R};
\citealt{neowise}) missions, covering a time period from 2010 to 2017.

Table~\ref{tab:sample} summarizes the information for the Seyfert-1 nuclei
studied in this work.  

%\input{table3.tex}
%%% Table 3 %%%%
%\startlongtable
\begin{longrotatetable}
\begin{deluxetable*}{cllccccccccccc}
%    \tabletypesize{\scriptsize}
    \tablewidth{1.0\hsize}
    \tablecolumns{14}
    \tablecaption{List of the Low-redshift Seyfert-1 Nuclei\label{tab:sample}
    }
    \tablehead{
	\colhead{\#}  &
	\colhead{Name} &
	\colhead{$z$}  & 
	\multicolumn{3}{c}{References for the IR SED data} &
	\colhead{$F_{\rm g}/F_{\rm s}$}  &
	\colhead{$f_{\rm MIR,SF}$} &
	\colhead{$EW_{\rm PAH}$}&
	\colhead{$\Delta{\rm W1}$} &
	\colhead{$\Delta{\rm W2}$} &
	\colhead{AGN type} &
	\colhead{$\tau_{\rm V, ext.}$} &
	\colhead{$f_{\rm pol, 10~\mum}$} 
	\\
	\colhead{} &
	\colhead{} &
	\colhead{} &
	\colhead{($\lambda\sim1$-$2.5~\mum$)}  &
	\colhead{($\lambda\sim2.5$-$40~\mum$)} & 
	\colhead{($\lambda>40~\mum$)} 	& 
	\colhead{} &
	\colhead{\%} &
	\colhead{$\mum$} &
	\colhead{mag.} &
	\colhead{mag.} &
	\colhead{} &
	\colhead{} &
	\colhead{} %&
	%\colhead{$\log(L/L_\odot)$}
	\\
	\colhead{(1)} & 
	\colhead{(2)} & 
	\colhead{(3)} & 
	\colhead{(4)} & 
	\colhead{(5)} & 
	\colhead{(6)} &
	\colhead{(7)} &
	\colhead{(8)} &
	\colhead{(9)} &
	\colhead{(10)} &
	\colhead{(11)} &
	\colhead{(12)} &
	\colhead{(13)} &
	\colhead{(14)} 
    }
    \startdata
\multicolumn{14}{c}{HSR AGN Sample} \\
       \hline                          
     1 &  PKS 1417-19      & 0.1195   & 2M  &  W  & \ldots          & 0.93   & \ldots   & \ldots    & 0.31 & 0.26  & WDD  & 1.00 & 0.65 \\
     2 &  ESO 141-55       & 0.0371   & 2M  &  W  & 1, 2, 3         & 0.94   & \ldots   & \ldots    & 0.26 & 0.23  & WDD  & 0.25 & 0.30 \\
     3 &  Mrk 509          & 0.0344   & 2M  &  W  & 2, 3            & 0.99   & \ldots   & \ldots    & 0.25 & 0.15  & WDD  & 0.50 & 0.47 \\  % 1.5
     4 &  Mrk 1239         & 0.0199   & 5,6 & 3,4,8,9,10,11,12 & 11 & 0.99   & \ldots   & \ldots    & 0.40 & 0.27  & NORM & 0.00 & 0.00 \\
     5 &  3C 382           & 0.0579   & 2M  & W, 1 & 4              & 1.02   & \ldots   & \ldots    & 0.27 & 0.23  & WDD  & 0.00 & 0.00 \\
     6 &  IRAS 09149-6206  & 0.0573   & 2M  & W, 2  & 5             & 1.09   & \ldots   & \ldots    & 0.28 & 0.20  & WDD  & 0.25 & 0.30 \\
     7 &  Ark 120          & 0.0327   & 2M  & W     & 2,3           & 1.13   & \ldots   & \ldots    & 0.38 & 0.37  & WDD  & 0.00 & 0.00 \\
     8 &  Fairall 51       & 0.0142   & 2M  & W     & 2,3           & 1.13   & \ldots   & \ldots    & 0.16 & 0.13  & WDD  & 0.50 & 0.47 \\
     9 &  IRAS 13349+2438  & 0.1076   & 2M  & W     & 1             & 1.14   & \ldots   & \ldots    & 0.09 & 0.05  & NORM & 2.50$^a$ & 0.83$^a$ \\
    10 &   I Zw 1          & 0.0589   & 2M  & W,13  &  12           & 1.16   & \ldots   & \ldots    & 0.29 & 0.20  & NORM & 0.50 & 0.33 \\
    11 &   H 0557-385      & 0.0339   & 2M  & W,1 & 2               & 1.18   & \ldots   & \ldots    & 0.24 & 0.15  & WDD  & 0.25 & 0.30 \\
    12 &   IC 4329A        & 0.0161   & 2M  & W,1 & 2,3             & 1.18   & \ldots   & \ldots    & 0.37 & 0.33  & WDD  & 0.75 & 0.58 \\
    13 &   3C 120          & 0.0330   & 2M  & W   &  2,3            & 1.20   & \ldots   & \ldots    & 0.33 & 0.26  & WDD  & 0.75 & 0.58 \\
    14 &   3C 390.3        & 0.0561   & 2M  & W,1 & 8               & 1.20   & \ldots   & \ldots    & 0.35 & 0.29  & WDD  & 0.50 & 0.47 \\
    15 &   Pictor A        & 0.0351   & 2M  & W   & 2,3             & 1.21   & \ldots   & \ldots    & 0.32 & 0.30  & WDD  & 0.50 & 0.47 \\
    16 &   MR 2251-178     & 0.0640   & 1   & W,1,3   & 9           & 1.21   & \ldots   & \ldots    & ??   & ??    & WDD  & 0.25 & 0.30 \\
    17 &   3C 227          & 0.0858   & 2M  & W,3,6   & 10          & 1.28   & \ldots   & \ldots    & 1.02 & 0.52  & WDD  & 0.25 & 0.30 \\
    18 &   Mrk 1014        & 0.1631   & 2M  & W   &   11,12,13      & 1.28   & \ldots   & \ldots    & 0.24 & 0.18  & NORM & 10.00 & 0.93 \\
    19 &   3C 445          & 0.0559   & 2M  & W   &   1             & 1.29   & \ldots   & \ldots    & 0.13 & 0.08  & NORM & 0.00 & 0.00 \\
    20 &   3C 93           & 0.3571   & 2M  & W   & \ldots          & 1.54   & \ldots   & \ldots    & 0.33 & 0.41  & WDD  & 0.50 & 0.47 \\
    21 &   NGC 3783        &  0.0097  & P10 & P10 &  2,3, 19        & 1.03   & \ldots   & \ldots    & 1.91 & 2.37  & NORM & 1.00 & 0.51 \\ %???
    22 &   NGC 4507        &  0.0118  & 2M  & W  &   2, 3           & 1.15   & \ldots   & \ldots    & 0.16 & 0.18  & WDD  & 1.75 & 0.78 \\
    23 &   ESO 323-77      &  0.0150  & 2M  & 1, 3, 7 & 2,3, 19     & 0.67   & \ldots   & \ldots    & 0.31 & 0.26  & WDD  & 0.25 & 0.30 \\
    24 &   NGC 4151        &  0.0033  & AH13 & AH13 &   19          & 0.78   & \ldots   & \ldots    & 1.15 & 0.78  & NORM & 0.75 & 0.44 \\
    25 &   NGC 7469        &  0.0163  & P10, AH13 & P10, AH13 & 2,3 & 0.48   & \ldots   & \ldots    & 0.34 & 0.31  & NORM & 7.75 & 0.91 \\ % 1.5 Fuller
    26 &   NGC 1566        &  0.0050  & P10 & P10 &                 & 0.36   & \ldots   & \ldots    & 0.11 & 0.32  & NORM & 9.50 & 0.93 \\
    27 &   NGC 4593        &  0.0090  & 2M  &  W, 3 &  11           & 0.85   & \ldots   & \ldots    & 0.35 & 0.52  & WDD  & 0.75 & 0.58 \\  % no good data
    28 &   NGC 3227        &  0.0039  & F16 & F16 &   2,3           & 0.40   & \ldots   & \ldots    & 0.25 & 0.25  & HDD  & 4.50 & 0.94 \\
    29 &   NGC 4235        &  0.0080  & AH13 & AH13&  2,14          & 0.78   & \ldots   & \ldots    & 0.23 & 0.26  & NORM & 0.00 & 0.00 \\  % duplicated in saga
    30 &   NGC 4015        &  0.0023  & 2M & W &     9              & \ldots & \ldots   & \ldots    & 0.41 & 0.43  & NORM? & 1.75? & 0.66? \\
    31 &   Fairall 9       &  0.0470  & 2M & W &     2,3            & 1.11   & \ldots   & \ldots    & 0.09 & 0.04  & NORM & 0.00 & 0.00 \\
    32 &   NGC 3516        &  0.0088  & AH13 & AH13& 2,3            & \ldots & \ldots   & \ldots    & 0.66 & 0.79  & NORM & 1.00? & 0.51? \\
       \hline                          
\multicolumn{14}{c}{SDSS/{\it Spitzer} Sample} \\
       \hline                          
   33  &   Mrk 1393                       &  0.0543  & 2M & W  & 15     & \ldots & 0.2   & 0.01   & 0.22 & 0.23 & NORM & 0.75 & 0.44 \\
   34  &   Mrk 506                        &  0.0431  & 2M & W  & \ldots & \ldots & 2.2   & 0.03   & ??   &  ??  & WDD  & 0.25 & 0.30 \\
   35  &   Mrk 926                        &  0.0470  & 2M & W  & 2,3    & \ldots & 4.6   & 0.01   & 0.73 & 0.61 & NORM & 0.25 & 0.20 \\
   36  &   NGC 4074                       &  0.0226  & 2M & W  & 2,3    & \ldots & 7.3   & 0.02   & ??   &  ??  & NORM & 1.50 & 0.62 \\
   37  &   Mrk 1392                       &  0.0359  & 2M & W  & 2,3    & \ldots & 1.9   & 0.01   & ??   &  ??  & NORM & 0.75 & 0.44 \\
   38  &   [VV2006c] J020823.8-002000     &  0.0741  & 2M & W  & \ldots & \ldots & 4.5   & 0.00   & 0.19 & 0.21 & NORM & 1.00 & 0.51 \\
   39  &   NGC 2484                       &  0.0408  & 2M & W  & \ldots & \ldots & 1.7   & 0.00   & 0.08 & 0.15 & NORM? & 10.00? & 0.93? \\
   40  &   2MASX J14510879+2709272        &  0.0645  & 2M & W  & 12     & \ldots & 1.3   & 0.05   & ??   &  ??  & NORM & 0.50 & 0.33 \\
   41  &   2MASX J16164729+3716209        &  0.1518  & 2M & W  & 16     & \ldots & 0.0   & 0.03   & 0.03 & 0.08 & NORM & 10.00 & 0.93 \\
   42  &   NGC 863                        &  0.0261  & 2M & W  & 2,3    & 1.43   & 2.2   & 0.03   & 0.13 & 0.34 & NORM & 4.00 & 0.84 \\   % duplicated  Mrk 590
   43  &   Mrk 1018                       &  0.0430  & 2M & W  & 2,3    & \ldots & 4.4   & 0.00   & 0.77 & 1.11 & NORM & 0.00 & 0.00 \\
   44  &   3C 15                          &  0.0735  & 2M & W  & \ldots & \ldots & 1.2   & 0.06   & 0.13 & 0.20 & WDD  & 0.25 & 0.30  \\
   45  &   2dFGRS TGN254Z050              &  0.0888  & 2M & W  & 16     & \ldots & 0.0   & 0.04   & ??   & ??   & NORM & 10.00 & 0.93 \\
   46  &   Ton 730                        &  0.0864  & 2M & W  & 12     & \ldots & 0.6   & 0.01   & ??   & ??   & HDD  & 0.25 & 0.42 \\
   47  &   Mrk 110                        &  0.0355  & 2M & W  & 12     & \ldots & 6.7   & 0.00   & ??   & ??   & NORM & 0.00 & 0.00 \\
   48  &   2MASX J09191322+5527552        &  0.0489  & 2M & W  & 2,3    & \ldots & 6.1   & 0.00   & 0.22 & 0.20 & NORM & 1.00 & 0.51 \\
   49  &   2MASX J14492067+4221013        &  0.1786  & 2M & W  & 16     & \ldots & 0.4   & 0.03   & 0.06 & 0.10 & NORM & 7.75$^a$ & 0.95$^a$ \\
   50  &   NGC 5252                       &  0.0229  & 2M & W  & 2,3    & \ldots & 2.2   & 0.01   & 0.87 & 0.98 & WDD  & 0.00 & 0.00 \\
   51  &   3C 219                         &  0.1746  & 2M & W  & \ldots & \ldots & 2.8   & 0.02   & 0.10 & 0.12 & WDD  & 0.25 & 0.30 \\
   52  &   SDSS J095504.55+170556.3       &  0.1378  & 2M & W  & 12     & \ldots & 1.5   & 0.00   & 0.32 & 0.22 & NORM & 0.00 & 0.00 \\   % not identified by simBAD or NED,  2MASS J09550456+1705561        
   53  &   Mrk 176                        &  0.0265  & 2M & W  & 17     & \ldots & 8.9   & 0.04   & 0.16 & 0.12 & NORM & 1.00 & 0.51 \\
   54  &   2MASX J14054117+4026326        &  0.0806  & 2M & W  & 9      & \ldots & 0.2   & 0.08   & 0.06 & 0.09 & NORM & 10.00 & 0.93 \\
   55  &   2MASX J14482512+3559462        &  0.1133  & 2M & W  & 16     & \ldots & 3.9   & 0.08   & ??   & ??   & HDD  & 0.50 & 0.60 \\
   56  &   [GH2004] 9                     &  0.1952  & 2M & W  & \ldots & \ldots & 0.0   & 0.00   & ??   & ??   & NORM & 2.50 & 0.75 \\
   57  &   Mrk 417                        &  0.0328  & 2M & W  & 2,3    & \ldots & 6.7   & 0.01   & 0.36 & 0.30 & NORM & 1.00 & 0.51  \\
   58  &   Mrk 668                        &  0.0770  & 2M & W  & 1      & \ldots & 2.8   & 0.08   & 0.78 & 0.95 & NORM & 0.50 & 0.33  \\   % not identified by simBAD or NED, QSO B1404+2863007      
   59  &   NGC 4235$^b$                       &  0.0075  & 2M & W  & 2,14   & \ldots & 7.1   & 0.10   & 0.23 & 0.26 & NORM? & 3.25? & 0.80? \\ 
   60  &   2MASX J12384342+0927362        &  0.0829  & 2M & W  & \ldots & \ldots & 1.4   & 0.01   & 0.14 & 0.15 & HDD  & 3.00 & 0.92 \\
   61  &   2MASS J10405880+5817034        &  0.0712  & 2M & W  & 18     & \ldots & 7.0   & 0.07   & 0.16 & 0.15 & NORM & 1.25 & 0.58 \\
   62  &   SDSS J164019.66+403744.4       &  0.1512  & 2M & W  & 18     & \ldots & 3.3   & 0.10   & 0.38 & 0.49 & NORM & 1.50 & 0.62 \\   % no detection in herschel bands, 3 sigma upper limits are calcuated.
   63  &   Mrk 50                         &  0.0239  & 2M & W  & 2,3    & \ldots & 3.2   & 0.04   & ??   & ??   & HDD  & 0.00 & 0.00 \\
   64  &   2dFGRS TGN404Z026              &  0.0329  & 2M & W  & 2,3    & \ldots & 5.1   & 0.00   & 0.44 & 0.35 & NORM & 0.50 & 0.33 \\
   65  &   Mrk 771                        &  0.0636  & 2M & W  & 12     & \ldots & 5.5   & 0.00   & 0.45 & 0.34 & NORM & 0.75 & 0.44
\enddata                                    
    \tablecomments{ 
    Col. (4)-(6): references for the SED data, P10 -
    \citet{Prieto2010}, AH03 - \citet{Alonso-Herrero2003},
    RA09-\citet{RamosAlmeida2009}, F16 -\citet{Fuller2016}, and the photometry data,
    NIR: 2M- 2MASS \citep{2MASS},
         1- \citet{Elvis1994},
         2- \citet{Peng2006},
         3- \citet{Abrahamyan2015},
         4- \citet{Scoville2000},
	 5- \citet{Spinoglio1995},
	 6- \citet{Rudy1982};
    MIR: W- WISE \citep{wise}, 
         1- Spitzer Science Center Source list \citep{Teplitz2010},
         2- \citet{Matsuta2012},
	 4- \citet{Gallimore2010},
	 3- \citet{Asmus2014},
	 5- \citet{Shi2010},
	 6- \citet{Dicken2008},
	 7- \citet{Honig2010a},
	 8- \citet{Spinoglio1995},
	 9- \citet{Rudy1982},
	 10- \citet{Reunanen2010},
	 11- \citet{Haas2007},
	 12- \citet{Gorjian2004};
	 13 - \citet{Shi2014};
    FIR: 1- \citet{Hernan-Caballero2011}, 
         3-\citet{Schimizu2016},
	 2- \citet{Melendez2014},
	 4-\citet{Dicken2010},
	 5- \citet{Matsuta2012},
	 6- \citet{Pollo2010},
	 7- \citet{Sargsyan2011},
	 8- \citet{Landt2010},
	 9- IRSA Faint Source Catalog \citep{Moshir1990},
	 10- \citet{Dicken2008},
	 11- AKARI/FIS All-Sky Survey Point Source Catalogues \citep{Yamamura2010},
	 12- \citet{Shi2014},
	 13- \citet{Ma2015}, 
	 14- \citet{Auld2013},
	 15- \citet{Hanish2015},
	 16- \citet{Abrahamyan2015},
	 17- \citet{Bitsakis2014},
	 18- \citet{Oliver2012},
	 19- \citet{Garcia-Gonzalez2016}
    Col. (7): the 12-$\mum$ flux ratio between the ground-based observations
    from \cite{Asmus2014} and WISE band 3 data;
    Col. (8): the flux contribution of star-forming component in the {\it
    Spitzer}/IRS mid-IR spectra using the decomposition method proposed by
    \cite{Hernan-Caballero2015};
    Col. (9): the Equivalent Width of the aromatic feature at 11.3 $\mum$;
    Col. (10)-(11): the maximum variations among different observing epochs for
    WISE W1 ($\sim3.4~\mum$) and W2 ($\sim4.6~\mum$) bands. Questions marks
    `??' indicate cases where there are not enough good single-epoch photometry
    data to establish a meaningful extinction curve;
    Col. (12): the intrinsic AGN template suggested from our best-fits, `?'
    indicates highly uncertain results; Col. (13)-(14): the derived V-band
    optical depth and relative 10~$\mum$ emission strength of the extended
    polar dust component, `?' indicates highly uncertain results.
    \\
    $^a$ these values are based on the hot-dust-obscured AGN model introduced
    in Section~\ref{sec:dog}; $^b$ the duplication of NGC 4235 serves as an
    example to demonstrate the uncertainties caused by strong host galaxy
    contamination on the SED fitting results.
    }
\end{deluxetable*}
\end{longrotatetable}

\subsection{SED Fitting Method}

\subsubsection{Stellar Emission Template}

While the galaxy star formation contribution to the mid-IR has been minimized
by our sample selections, stellar contamination in the near-IR can be
significant, especially for the SDSS/{\it Spitzer} AGNs. Luckily, the near-IR
bands of low-$z$ galaxies are dominated by old stellar populations that share
nearly identical broad-band SEDs.  Despite the likely multiple stellar
populations with different star formation histories,  the stellar photospheric
SEDs of these galaxies peak at $\sim$1~$\mum$ and drop quickly following a
Rayleigh-Jeans tail toward the mid-IR \citep[e.g.,][]{Polletta2007}. It is
known that dust around evolved stars could have additional emission that
supplements the quickly dropping photospheric SED at $\lambda\gtrsim7~\mum$,
particularly in early-type galaxies \citep[e.g.,][]{Bressan2006, Rampazzo2013}.
For some low-luminosity AGNs, where the stellar contamination can even
contribute to the mid-IR, we need more accurate templates.

\begin{figure}[!htb]
    \begin{center}
	\includegraphics[width=1.0\hsize]{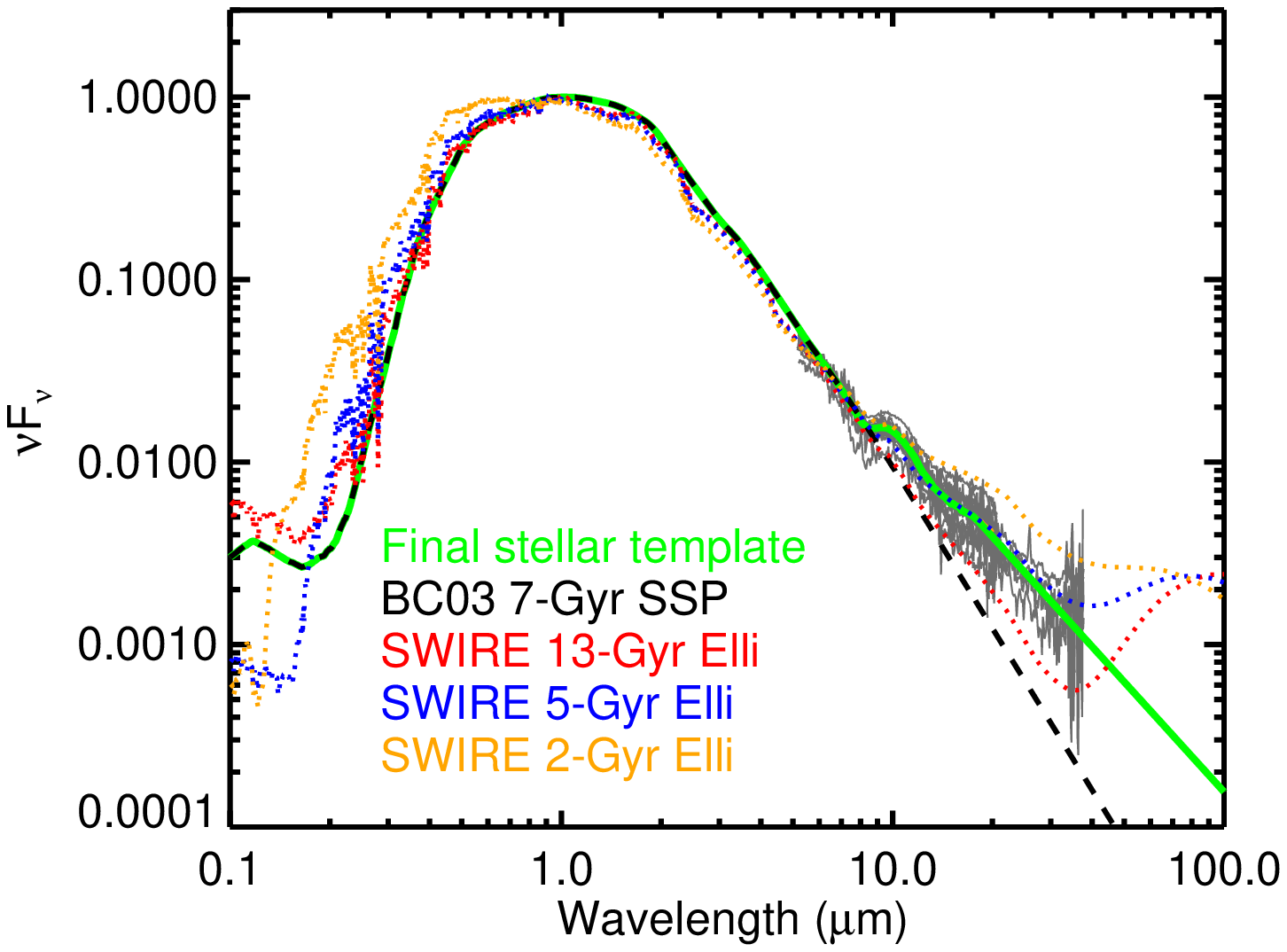}
		\caption{
		    Comparisons of different stellar templates. We derive an
		    empirical template by replacing a \citet{bc03} 7 Gyr SSP
		    SED template with the mid-IR (6--20~$\mum$) stellar
		    template and a power-law SED ($\lambda>$20~$\mum$). We plot the
		    individual {\it Spitzer}/IRS spectra used to derived the
		    mid-IR template as grey lines. The original \citet{bc03} and
		    the elliptical galaxy templates with age 2, 5 and 13 Gyr
		    from the SWIRE library \citep{Polletta2007} are also
		    presented.
		    }
	\label{fig:sta_temp}
    \end{center}
\end{figure}

There are several stellar population synthesis models that include the effect
of circumstellar dust shells from evolved stars \citep[e.g.,][]{Bressan1998,
Silva1998, Piovan2003, Gonzalez-Lopezlira2010, Cassara2013, Villaume2015}.
However, due to the lack of constraints on the metallicity, mass loss rates or
properties of the dust shells around the evolved stars in the AGN host
galaxies, we decided to derive an empirical template. Following
\citet{Hernan-Caballero2015}, we used the {\it Spitzer}/IRS spectra of 18 local
early-type galaxies with negligible indication of star formation
activities\footnote{This early-type galaxy sample includes: NGC 4474, NGC 4377,
NGC 4564, NGC 4570, NGC 4660, M 85, NGC 4473, NGC 5812, NGC 1700, NGC 1374, NGC
0821, NGC 5831, NGC 1297, NGC 1366, NGC 3818, NGC 7332, NGC 1549, NGC 3904. }
to derive a mean mid-IR stellar template at 6--20~$\mum$ with the normalization
at 7~$\mum$. Only a few of them have LL2 (19.9--39.9~$\mum$) observations but
the spectral slope can be roughly described by a power-law $f_\nu \propto
\nu^{1.0}$ at $\lambda>20~\mum$. We continue this mid-IR stellar template as a
power-law at longer wavelengths and join it to the 7~Gyr single stellar
population (SSP) template from \citet{bc03} at 7.0~$\mum$.  The results can be
seen in Figure~\ref{fig:sta_temp}, together with three elliptical galaxy
templates generated with the GRASIL code \citep{Silva1998} by
\citet{Polletta2007}.  Compared with the original dust-free \citet{bc03}
template, the star-heated dust features increase the flux by a factor of
$\gtrsim1.5-3$ at $\lambda>10~\mum$. At $\lambda\sim$1--8~$\mum$, the SEDs of
different templates are remarkably similar, so the choice of the \citet{bc03}
templates has little effect for the IR SED modeling.

%%% Figure 4 %%%

\subsubsection{AGN templates}

For the AGN component, a series of semi-empirical SEDs is produced with the
model introduced in Section~\ref{sec:model}.  Because our major goal is to test
if Seyfert nuclei behave similarly to NGC~3783, the radial density profile and
dust grain properties of the extended dust component are fixed to the values in
table~\ref{tab:ngc3783_model}. All three AGN intrinsic templates presented in
\citet{Lyu2017} are obscured with an increment of 0.25 in optical depth
($\tau_{\rm V}$). We end up with 123 AGN templates that cover
$\tau_\text{V}$=0--10. 

\subsubsection{Other Far-infrared Components}

More than 80\% of our sample have constraints on the far-IR SEDs. We used
\cite{Rieke2009} star-forming galaxy (SFG) templates to describe
the host galaxy dust emission. This library was developed based on {\it
Spitzer} observations of local galaxies and each template is parameterized by a
different IR luminosity $\log(L_{\rm SF, temp}/L_\odot)$. We decide to adopt a
default value of $\log(L_{\rm SF, temp}/L_\odot)$=10.75 for most objects.

Some radio-loud AGNs can have strong synchrotron radiation in the far-IR bands.
We assume a broken power-law model \citep[see, e.g.,][]{Peer2014} to represent
this contribution
\begin{equation}
    F_\nu \propto 
     \begin{cases}
	 \nu^{-0.7}:   &  \nu < 10^{13}~\text{Hz, or } \lambda > 30~\mum  \\
	 \nu^{-1.3}:   &  10^{13}~\text{Hz} < \nu < 10^{14}~\text{Hz, or} \\
	               & \quad \quad  3~\mum <\lambda < 30~\mum \\
	 0    :        &   \nu > 10^{14}~\text{Hz, or } \lambda< 3~\mum
     \end{cases}
\end{equation}

\subsubsection{SED Models}

Our model would predict obscuration in the UV-optical band for many AGNs.
However, this effect will be dependent on our line of sight (LOS); any
clumpiness in the distribution of the polar dust will result in variations in
the extinction. Indeed, it is possible to imagine a system where the polar dust
covers 99\% of the sky as seen by the central engine and hence provides an
important infrared SED component, but where our LOS by chance falls along the
1\% of directions without dust and hence the central engine appears to be
unobscured. In light of this, we only focus on the IR SEDs for individual
objects. To reduce the influence of host galaxy star-formation contamination,
the fittings are limited to $\lambda\sim$1--30~$\mum$.

We combined the (reddened) AGN templates with the stellar templates and used
$\chi^2$ minimization to fit the SEDs. Since the data have a range of
non-statistical errors, a uniform weight for all bands is assumed instead of
using the quoted flux uncertainties. There are only four free parameters in
our model: the normalizations of the stellar and the AGN templates, the type of
the AGN template and its optical extinction $\tau_\text{V}$. The best AGN
template is selected by searching the combination with the minimum $\chi^2$
value among all three sets of reddened AGN libraries (normal, WDD, HDD).
However, if $\chi^2$ with the best reddened normal AGN template is less than
1.5 times that with the best reddened dust-deficient template, the normal AGN
template is selected.

For objects with at least two photometry data points at $\lambda>50~\mum$, a
second-round SED fitting at $\lambda\sim$1--500~$\mum$ was carried out by
combining the AGN template selected above, a stellar template and an SFG far-IR
template. For eight objects, we changed the default SFG template to another
\citep{Rieke2009} template with a lower or higher $\log(L_{\rm SF,
temp}/L_\odot)$ in a range of 9.75--10.50 to improve the fitting of the far-IR
SED peak. For two objects, Pictor A and 3C 120, the synchrotron emission
template was added to reproduce their far-IR and submm SEDs.

\subsection{Fitting the IR SEDs of Individual Objects}\label{sec:ind_fit}

The best-fit results for type-1 nuclei can be seen in
Figure~\ref{fig:agn1_sed1}. The SED shape of the AGN component is only
determined by two free parameters: the type of the intrinsic template (normal,
WDD, or HDD) and the optical depth, $\tau_{\rm V}$, of the polar dust
component. Surprisingly, this simple model fits the broad-band IR SEDs of most
objects reasonably well. The only notable exceptions are IRAS 13349+2438 and
2MASX J14492067+4221013, whose SEDs feature strong hot dust excess emission. As
shown later (Section~\ref{sec:highz}), similar characteristics can be found in
the so-called hot dust-obscured galaxies, whose SEDs our model can fit by
adjusting the density profile of the polar dust.

%%% Figure 5a %%%
\begin{figure*}[!htb]
    \begin{center}
	\includegraphics[width=1.0\hsize]{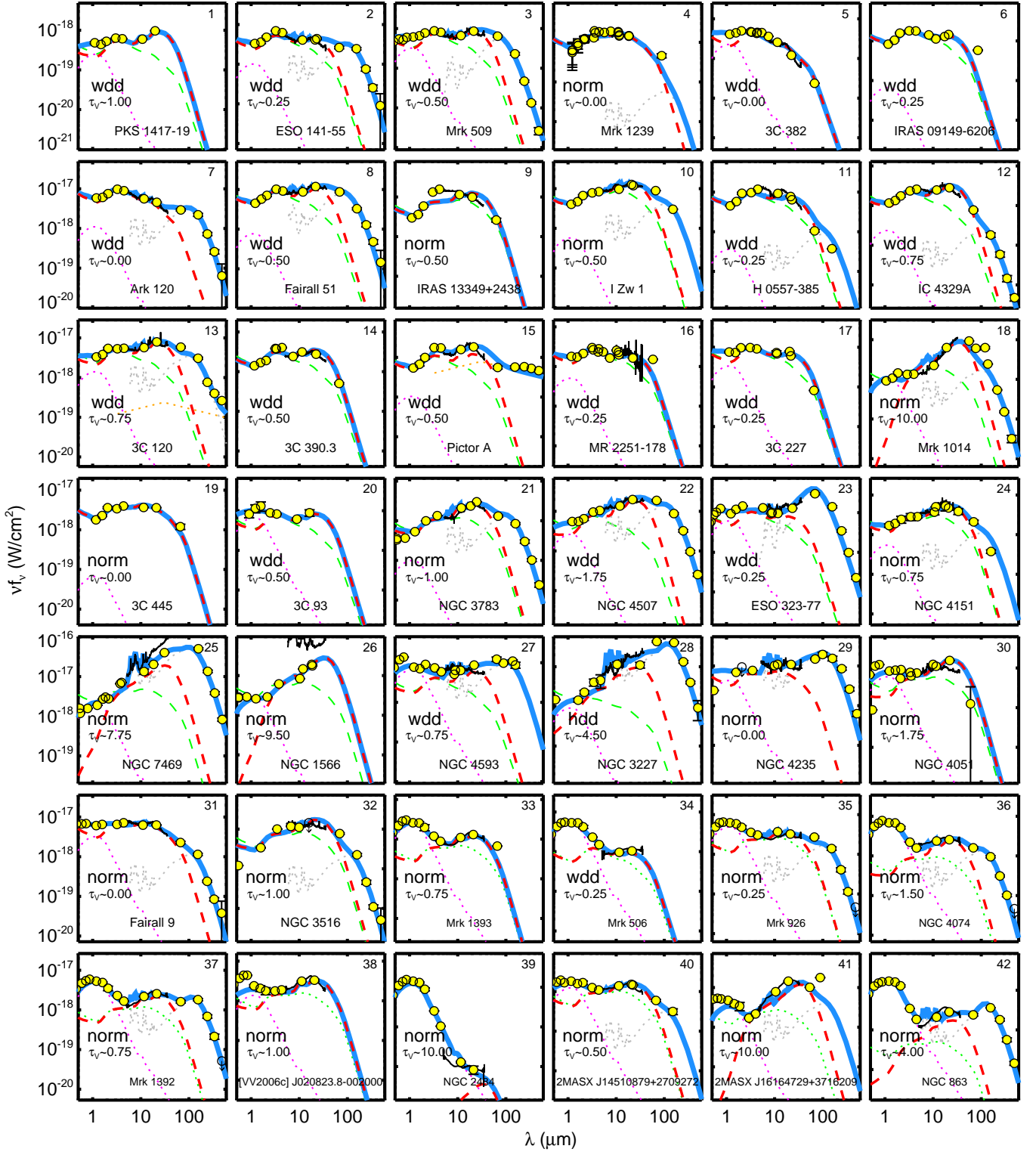}
		\caption{
		    Best-fit results for low-$z$ Seyfert-1 nuclei with the 
		    reddened AGN model trained for NGC~3783.  Photometric data
		    points are shown as yellow dots and the {\it Spitzer/IRS}
		    mid-IR spectra are plotted as black solid lines. The SED
		    model (blue thick solid lines) is composed of the AGN
		    component (red dashed lines), the stellar component
		    (magenta dotted lines), and the far-IR star formation
		    component (grey dotted lines). For 3C 120 and Pictor A, the
		    sychrotrom emission component (orange dotted lines) is
		    added into the fittings. We also plot the suggested
		    intrinsic AGN template (green dashed lines) for each object
		    from our SED fittings to compare the observed SED.
		    }
	\label{fig:agn1_sed1}
    \end{center}
\end{figure*}

\addtocounter{figure}{-1}

%%% Figure 5b %%%
\begin{figure*}[!htb]
    \begin{center}
	\includegraphics[width=1.0\hsize]{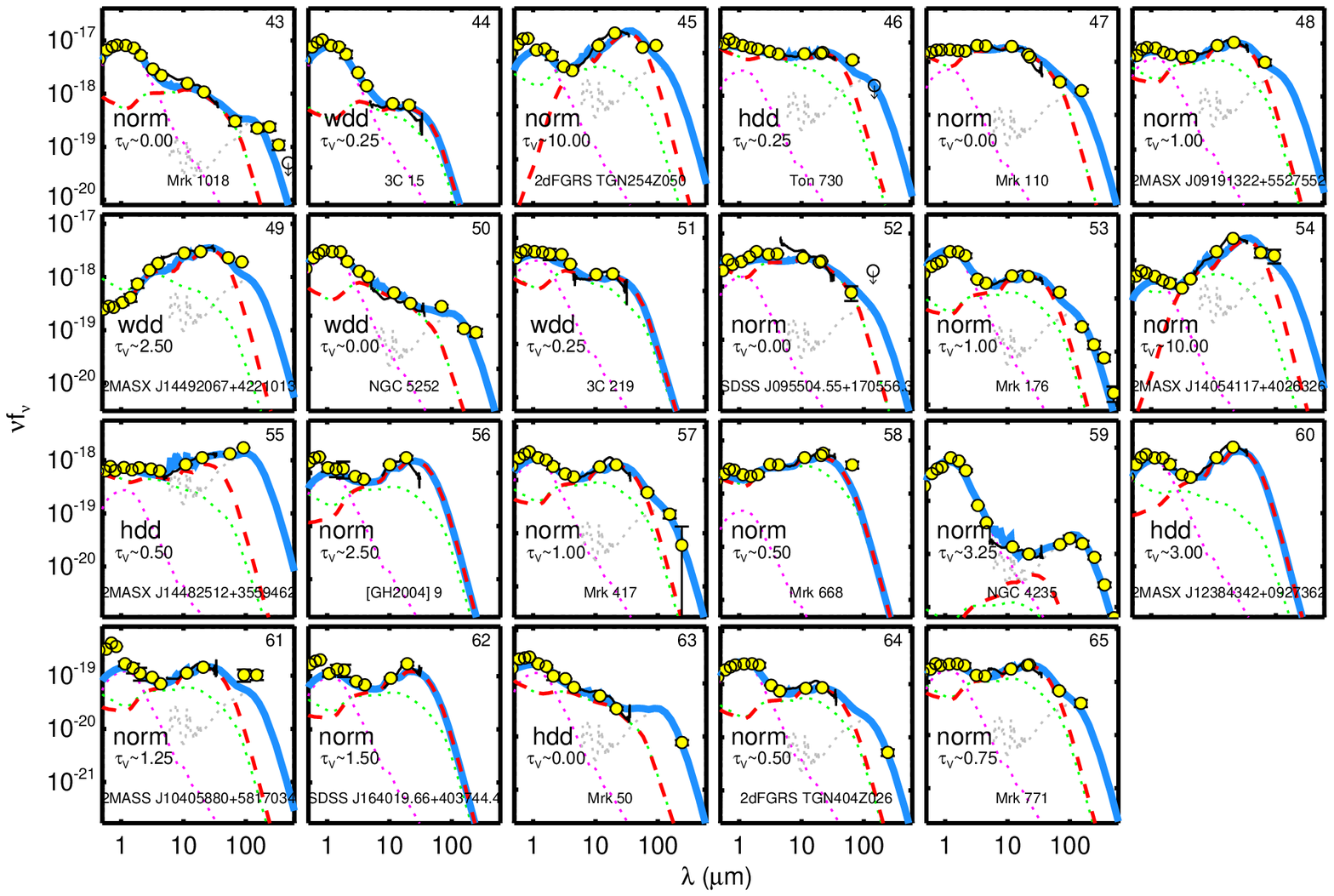}
		\caption{
		    (continued.) Best-fit results for low-$z$ Seyfert-1 AGNs
		    with the reddened AGN model trained for NGC~3783.
		    }
	\label{fig:agn1_sed2}
    \end{center}
\end{figure*}

We want to mention several complications that could influence the quality of
the SED fitting. First, AGNs show near- and mid-IR variability that could
produce a gap between different datasets. In fact, $\gtrsim40\%$ of the
Seyfert sample have strong mid-IR variability with $\Delta W1>0.3$ mag during
the available WISE epochs, in contrast with $\sim$13\% for the PG quasar sample
in our previous study \citep{Lyu2017}.  Secondly, it is quite likely that the
dust has different compositions or distributions among different objects.
Indeed, we can improve the fittings by sampling more parameter space with
similar MCMC methods used for NGC~3783 (see Section~\ref{sec:ngc3783-model}).
However, the exact situations for these properties cannot be observationally
constrained.  Our goal is to reproduce a wide range of observations in a way
that minimizes the uncertainties due to model degeneracy, not to produce
another complicated model that may fit a few observations perfectly. 

In summary, our four-free-parameter simple model provides reasonably good fits
in the near-IR and mid-IR to nearly the entire sample of AGNs.

%\clearpage

\subsection{Testing the reddened AGN Templates with Composite SEDs} \label{sec:sed}

We now work with SEDs averaged over the SDSS-{\it Spitzer}/IRS sample to test
the prediction that objects with excess IR emission due to polar dust should
have obscuration in some directions. By averaging over a sample of AGNs,
evidence of such obscuration should be obvious even if it is very non-uniform
for individual sources.

\subsubsection{Composite SEDs of the SDSS Seyfert-1 Nuclei}

For each object, we corrected the flux for the foreground Galactic extinction
in the UV-optical bands according to the SRD dust map
\citep{Schlegel1998,Schlafy2011}. The observed flux is converted into
luminosity and the SED is smoothed with a $\Delta \log(\nu)=0.2$ boxcar in the
$\lambda-\lambda L_\lambda$ space. The median of all these SEDs is then
computed to produce a composite SED. To reduce the uncertainties in the mid-IR
due to the low resolution of photometry data points, we also derive a median
mid-IR spectrum of the sample using the {\it Spitzer}/IRS data in a similar
fashion.  In the last step, we replace the 6-35~$\mum$ low-resolution SED
template based on photometric data with the median mid-IR spectrum of the
sample.

\begin{figure*}[!htb]
    \begin{center}
	\includegraphics[width=1.0\hsize]{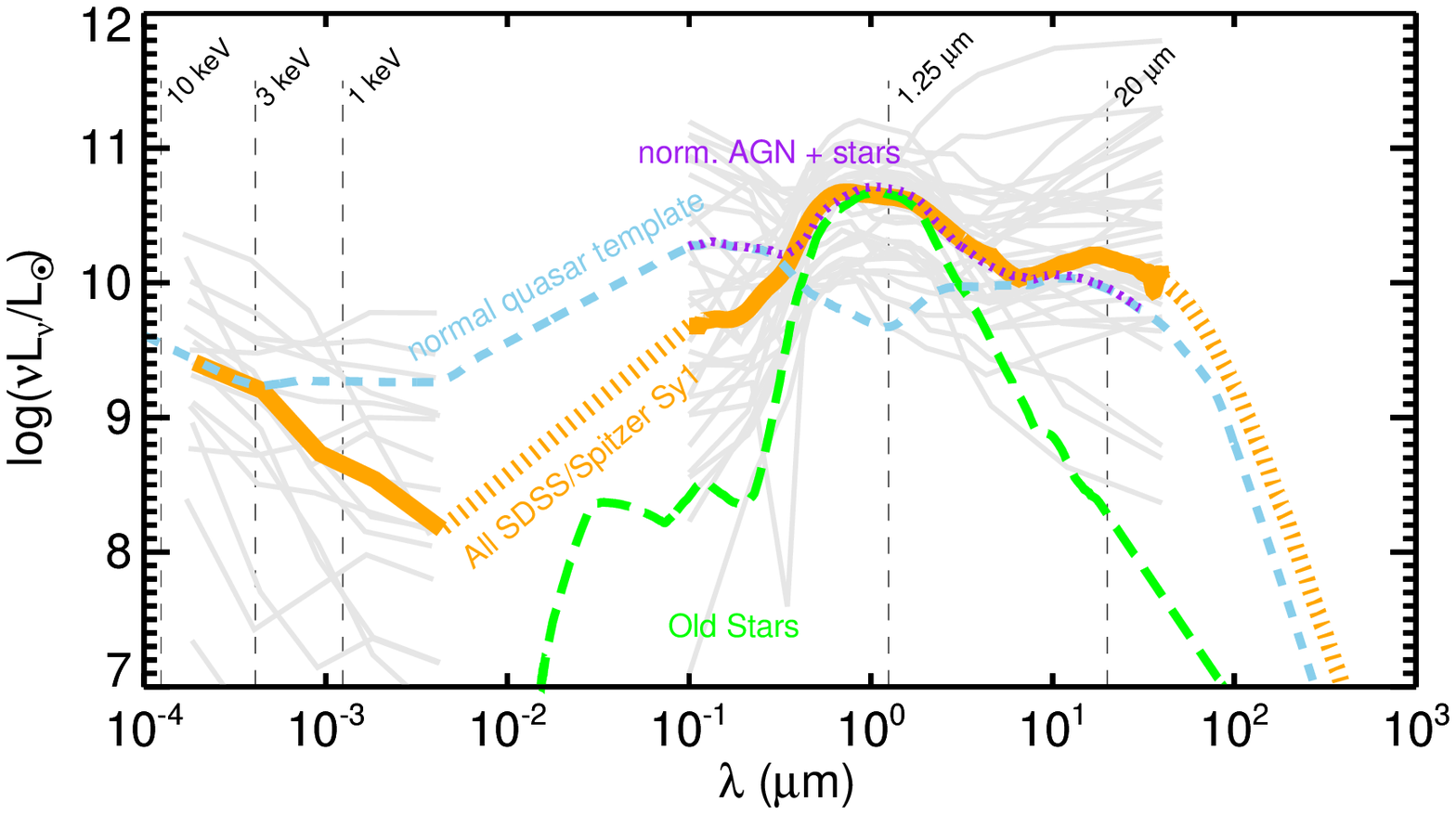}
		\caption{
		    The individual SEDs (grey thin lines) and the median
		    composite SEDs (orange thick line) of the SDSS-{\it
		    Spitzer}/IRS Seyfert-1 AGN sample.  As a comparison, we
		    also plot the intrinsic AGN template for normal quasars
		    (dashed blue line) from \citet{Xu2015a}. We used the
		    modified \citet{bc03} stellar template with age $\sim$
		    7~Gyr (green long-dashed line) and the normal AGN template
		    (blue dashed line) to reproduce the composite SED at
		    0.5--3.0~$\mum$ with the model as a magenta dotted line.
		    }
	\label{fig:Sey_comp}
    \end{center}
\end{figure*}

Figure~\ref{fig:Sey_comp} presents the median SED for the whole SDSS-{\it
Spitzer}/IRS Seyfert-1 sample. Since we do not remove the near-IR host galaxy
contamination, a near-IR SED bump due to emission from an old stellar
population is expected. In the mid-IR, this Seyfert-1 median SED presents
stronger emission at $\lambda\gtrsim10~\mum$ compared with the near-IR
normalized normal Elvis-like template. Since any objects with strong host
galaxy contamination in the mid-IR have been dropped, such excess emission has
to be AGN-dominated. On the other hand, the UV and soft X-ray (to $\sim$ 3 keV)
composite SED of this Seyfert-1 sample is suppressed relative to the quasar
template. Beyond 3 keV, the two templates are similar. This UV-optical dip is
the expected signature of obscuration in these Seyfert-1 nuclei. It is an
indication that the Seyfert-1 mid-IR excess is from the reradiation of the
absorbed energy by the dust that also provides the short-wavelength
obscuration.

%%% Figure 6 %%%

With the SDSS/{\it Spitzer} sample, we also demonstrate that both normal AGNs
and dust-deficient AGNs exist in the Seyfert-1 population, which supports the
idea that similar intrinsic AGN templates are also valid for relatively
low-luminosity AGNs. Based on the mid-IR continuum shape, objects with similar
IR SED characteristics are grouped together and a composite SED is derived
similarly to the whole sample median SED. We classify these AGNs into three
categories and show that their composite SEDs have the following
characteristics compared with the \citet{Elvis1994}-like templates for normal
AGNs:

\begin{itemize}
    \item (C1, N=5) normal AGNs. 
	
	These objects have SEDs well-reproduced by the normal AGN template with
	some stellar contribution in the near-IR. Some of them have a little
	far-IR excess emission, which might be associated with the host galaxy.

    \item (C2, N=7) dust-deficient AGNs. 
	
	These objects present a deficiency of the dust emission at
	$\lambda\gtrsim5~\mum$. However, since the stellar contribution in the
	near-IR can not be calibrated, it is hard to argue this SED is
	hot-dust-deficient or warm-dust-deficient. It is likely a combination
	of both populations.

    \item (C3, N=20) warm-excess AGNs. 
	
	These objects feature a mid-IR bump peaking around
	$\lambda\sim20~\mum$, in contrast with the flat
	$\sim$3--20~\mum~continuum of normal quasars. A similar mid-IR feature
	is also present in the SED of NGC 3783 as well as the composite SEDs of
	the whole sample.

\end{itemize}

\begin{figure*}[!htb]
    \begin{center}
	\includegraphics[width=1.0\hsize]{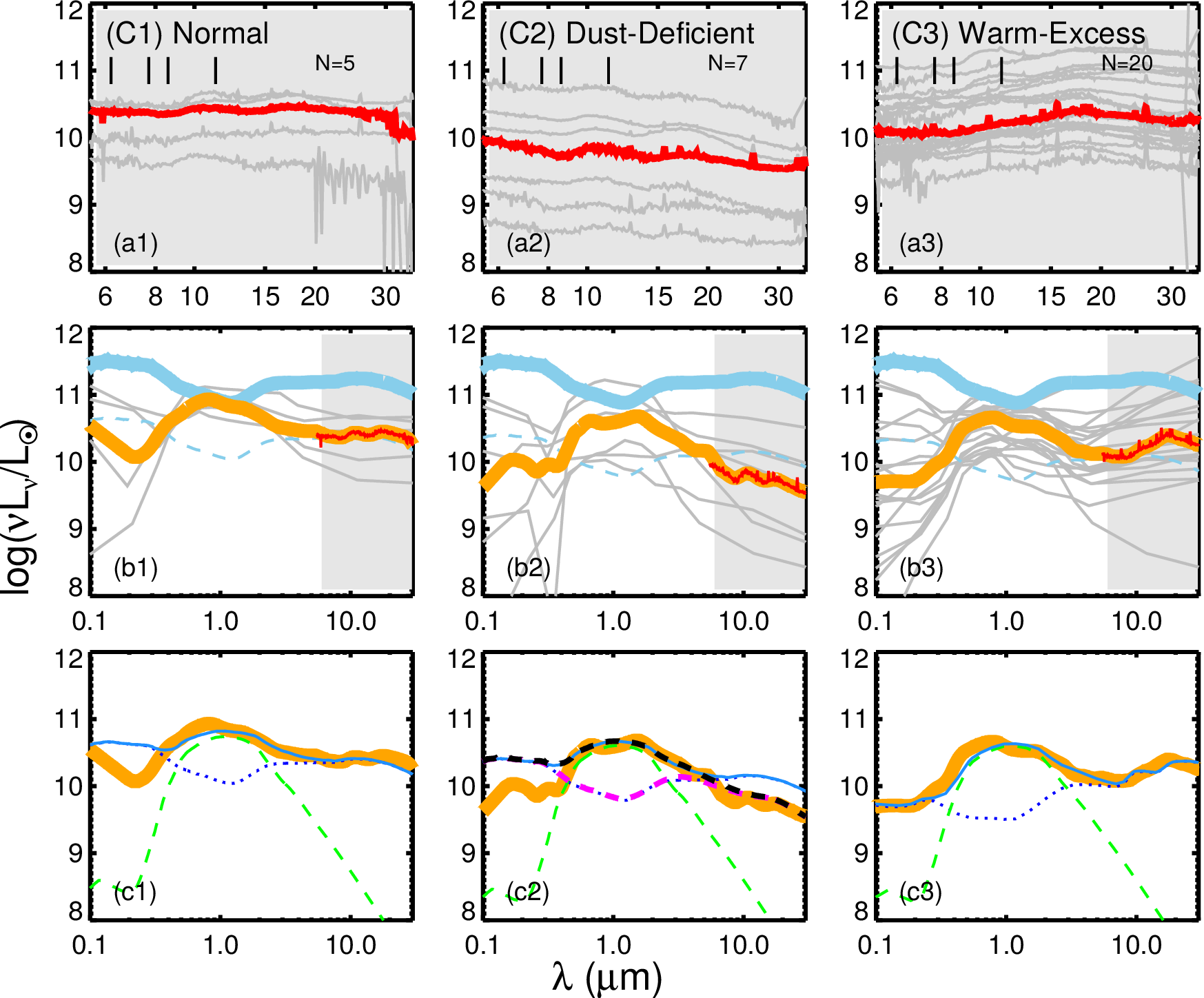}
		\caption{
		    (a1-a3): Mid-IR spectra of the SDSS-{\it Spitzer}/IRS
		    Seyfert-1 sample. Based on the continuum shape, we grouped
		    them into normal AGNs, dust-deficient AGNs and warm-excess
		    AGNs. Spectra of individual objects are plotted as grey lines and the
		    composites are shown as red lines. The black vertical lines on the
		    top panels label the location of the aromatic features at
		    6.2, 7.7, 8.6 and 11.3~$\mum$.  
		    (b1-b3): Composite (orange) and individual UV-to-MIR SEDs
		    (grey lines) for different groups of the SDSS-{\it
		    Spitzer}/IRS Seyfert-1 sample. We also show templates of a
		    normal AGN (blue lines) and the Seyfert-1 mid-IR spectral
		    composites (red lines).  
		    (c1-c3): Reproductions of the composite SEDs of Seyfert-1
		    Nuclei.  The model SED (blue solid lines) is composed of an
		    AGN component (blue dotted lines) and a stellar component
		    (green dashed lines). For (c2), we also present the
		    fittings with the WDD intrinsic template (purple dashed
		    line for the AGN component and black dashed line for the
		    SED model).
		    }
	\label{fig:Sey_group}
    \end{center}
\end{figure*}

These results are presented in Figure~\ref{fig:Sey_group}: (1) In the top
panels we demonstrate using the {\it Spitzer}/IRS spectra to construct
composite spectra in the 6--30 $\mum$ range. (2) In the middle panels we merge
the mid-IR spectral templates with composite SEDs at the shorter wavelengths
built primarily using photometry. The middle right panel shows the mid-IR bump
clearly when compared with the left and center panels. (3) The bottom panels
present our model fits (see below).

%%% Figure 7 %%%

\subsubsection{Reproducing the Composite SEDs}

For the composite templates (C1) and (C2), due to the relatively small sample
sizes ($N<10$), the UV-optical SEDs may not be representative of the overall
extinction and suffer the possibility of variability. However, either the
normal AGN template or the dust-deficient AGN template plus the stellar
template can reproduce the Seyfert IR SED.

For (C3), we apply the reddened AGN templates to fit the average composite SEDs
together with the stellar template. This UV-to-mid-IR composite SED is matched
by the reddened WDD template with $\tau_{\rm V}=1.5$ and all other parameters
of the polar dust component fixed the same for NGC~3783.

We conclude that the unobscured quasar templates can be directly applied to
some Seyfert-1 nuclei ($\sim$37\% in our case).  Both normal AGNs and
dust-deficient AGNs are seen among the Seyfert-1 population.  Meanwhile, most
Seyfert-1 nuclei present evidence for low- to moderate-level obscuration from
the soft X-ray to the optical with some dust excess emission starting from the
mid-IR, which can be reproduced by our polar-dust-obscured AGN template. These
results support our model assumptions introduced in Section~\ref{sec:model}.

\section{Characterizing the AGN Polar Dust Emission}\label{sec:pol_char}

In Section~\ref{sec:ngc3783-model}, the SED features and the polar dust
emission strength of NGC 3783 were accurately reproduced by the reddened AGN
model proposed in this study.  With the same parameters used for this single
object, our reddened AGN templates, combined with the host galaxy templates,
successfully fit the IR SEDs of another 64 Seyfert-1 nuclei and the UV to
mid-IR composite SEDs of the SDSS/{\it Spitzer} type-1 AGNs
(Section~\ref{sec:ind_fit}). In Section~\ref{sec:sed}, we found that on average
Type-1 AGN SEDs show indications of extinction as predicted by our approach.
These results support our proposal that the SED differences between quasars and
Seyfert-1 nuclei are due to the IR reprocessed emission from the extended
distribution of dust that is also responsible for the common low-level
obscuration of the Seyfert nuclei.

Now we characterize the possible SED features of the polar dust emission and
explore its prevalence among the Seyfert-1 nuclei studied in this work.

\subsection{SED Features}

\begin{figure*}[!htb]
    \begin{center}
	\includegraphics[width=1.0\hsize]{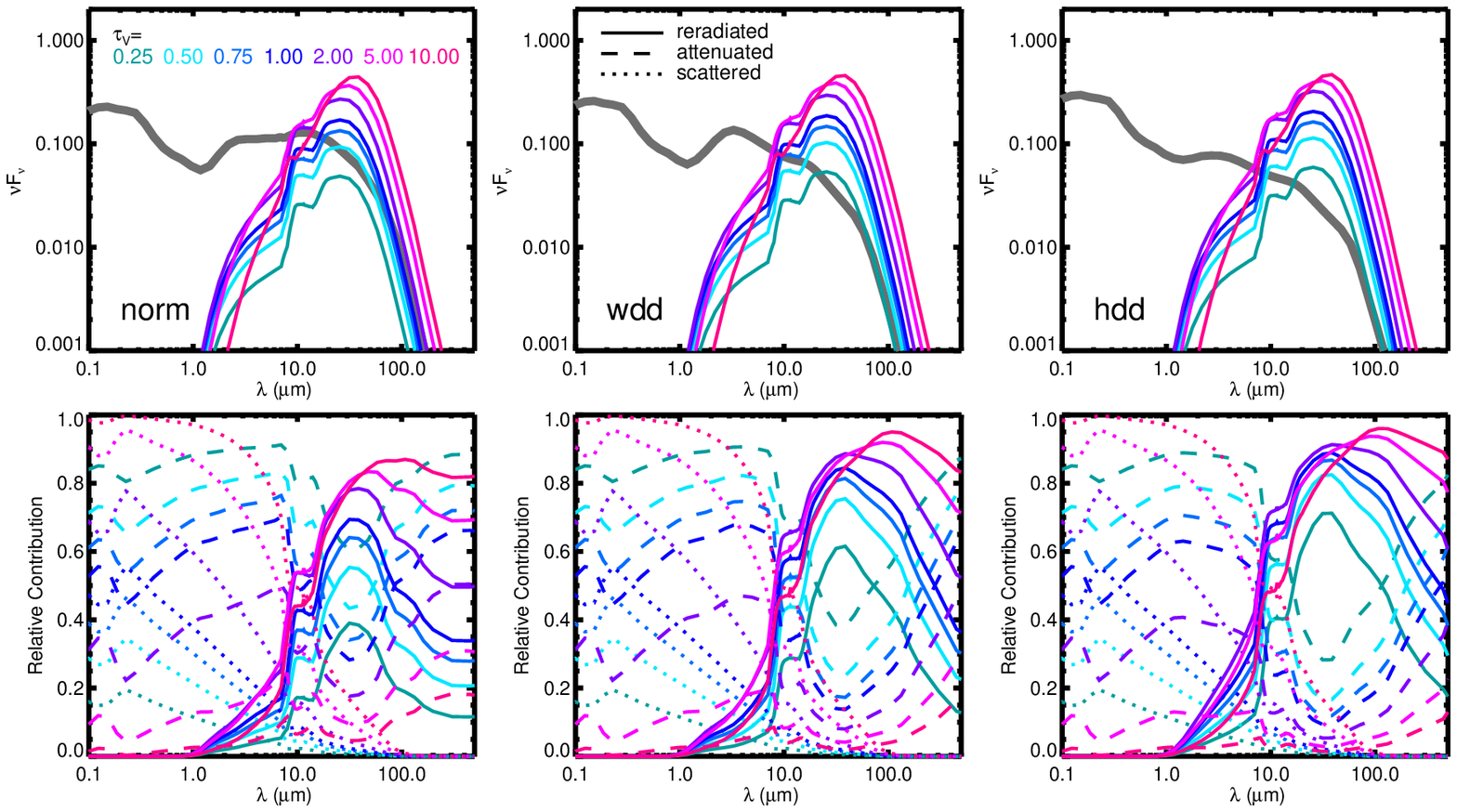}
	\caption{
	    The IR emission SED of the polar dust component (top panels) and
	    the relative contributions of the IR-reradiated emission from polar
	    dust, the scattered emission of accretion disk plus torus (by the
	    polar dust), and the attenuated emission of accretion disk plus
	    torus by the polar dust obscuration, in our reddened AGN
	    templates (bottom panels).
	}
	\label{fig:pol_decomp}
    \end{center}
\end{figure*}

In the top panels of Figure~\ref{fig:pol_decomp}, we show the SEDs of the
IR-reprocessed emission from the extended polar dust component as a function of
$\tau_{\rm V}$ for normal, WDD and HDD AGN.  In the bottom panels, the relative
contributions of the dust IR-reprocessed emission, dust scattered emission and
the attenuated emission of the intrinsic AGN templates are compared.

%%% Figure 8 %%%

There are several interesting results. Firstly, at a given $\tau_{\rm V}$,
regardless of the intrinsic AGN template, the strength and the SED shape of the
polar dust emission is almost identical. This is easily understood since the
configurations of the dust obscuration properties and the UV to optical SED
inputs, where the emission mostly heats the dust, are the same. This result
also underlies the possibility to fit a broad variety of observed SEDs with
$\tau_{\rm V}$ as the only free parameter to characterize the contribution of
the polar dust component.

In addition, at relatively low optical depth ($\tau_{\rm V}<5$), although the
overall strength of the polar dust emission changes with $\tau_{\rm V}$, its
SED shape is identical at the wavelengths where this component matters.  As
discussed in Section~\ref{sec:model-geometry}, different dust covering factors
would only change the relative scaling of the dust emission SED. Thus we can
always change the value of $\tau_{\rm V}$ to match the effect caused by the
dust covering factor on the final SED shape, as long as $\tau_{\rm V}$ is not
so large that the polar dust emission becomes optically thick in the mid-IR. 

Thus, we can describe the AGN-heated polar dust emission with one single
template when the integrated optical depth is modest ($\tau_{\rm V}\lesssim5$).
In fact, $\sim$90\% of our low-$z$ Seyfert-1 AGN sample are fitted by the
reddened AGN templates with $\tau_{\rm V}\leq5$. We suggest that this polar
dust template can be generally used to further reduce the number of free
parameters in the SED fittings.

\begin{figure}[!htb]
    \begin{center}
	\includegraphics[width=1.0\hsize]{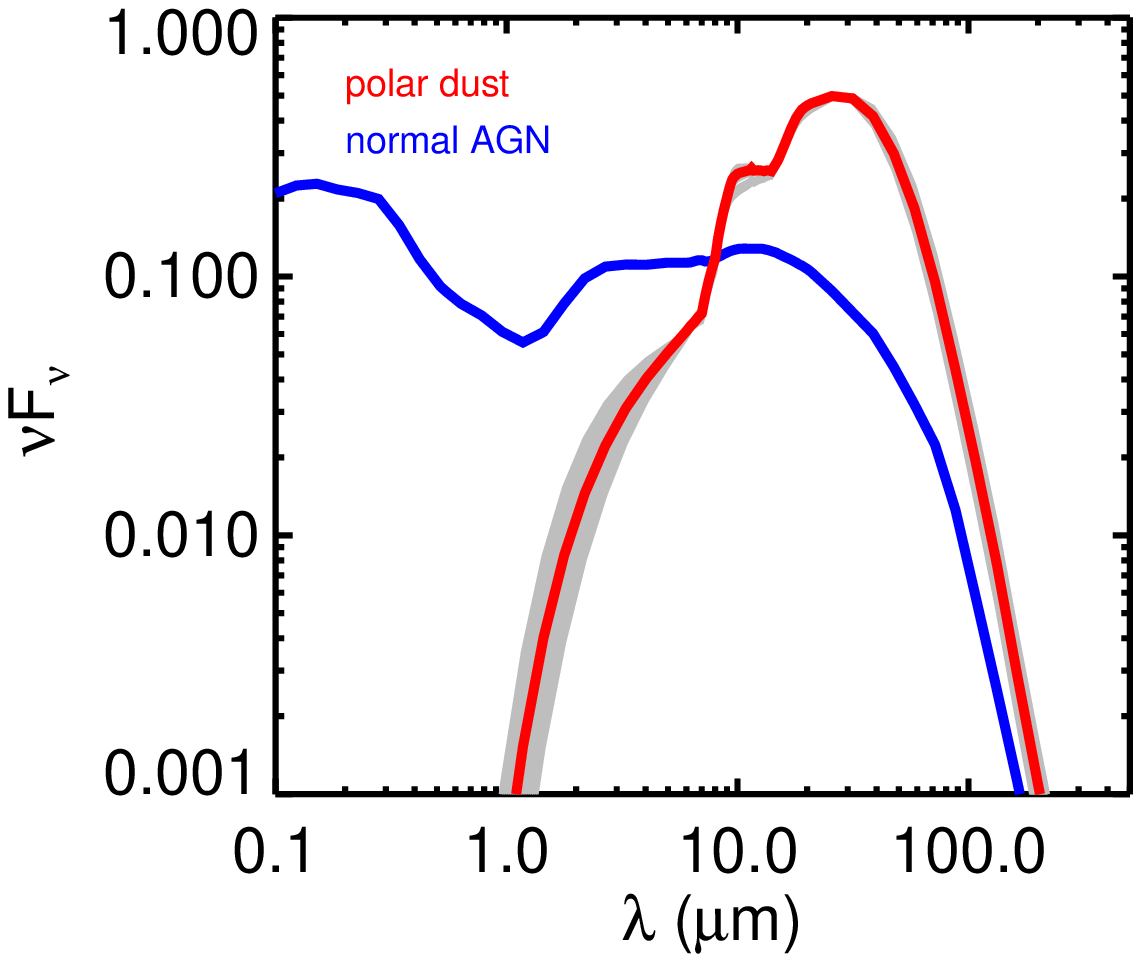}
	\caption{
	    IR SED template for AGN polar dust emission (red). We also plot the
	    intrinsic normal AGN template as a comparison (blue). The grey
	    region represents the range of SED variations of polar dust
	    emission with the normalizations at 26.5 $\mum$ and
	    $\tau_V\sim$0--5.
	}
	\label{fig:pol_sed}
    \end{center}
\end{figure}

\begin{figure}[!htb]
    \begin{center}
	\includegraphics[width=1.0\hsize]{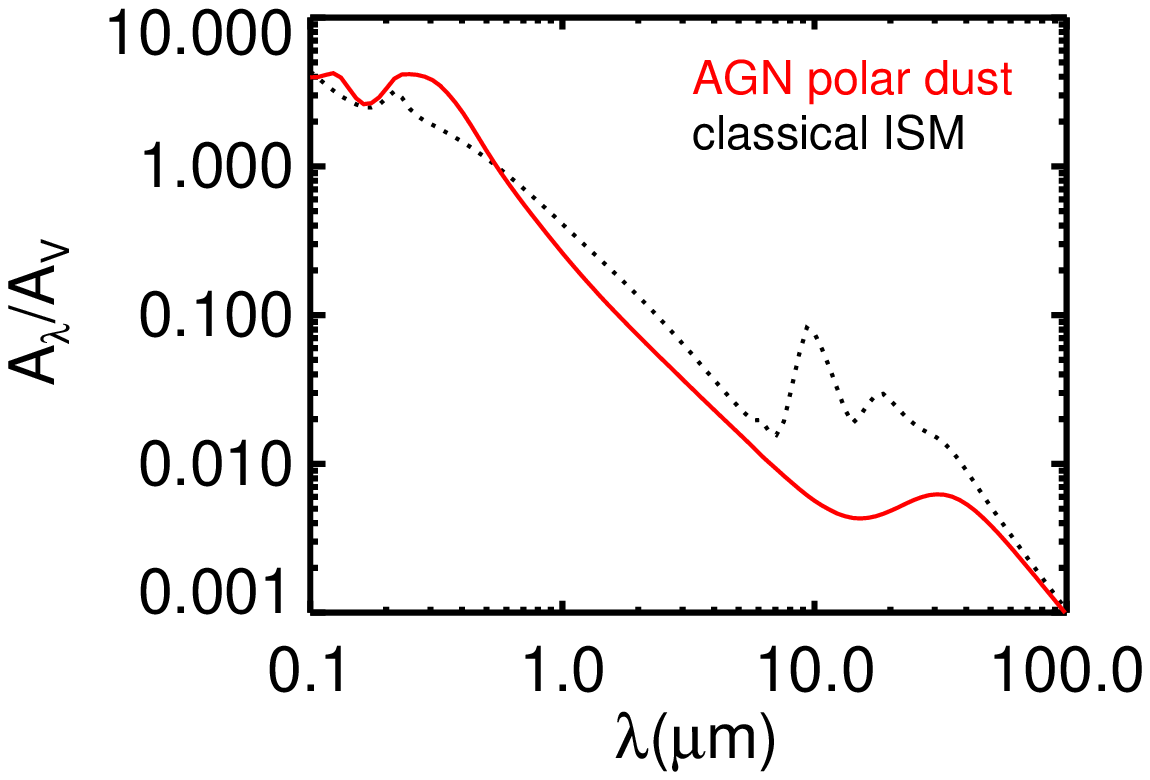}
		\caption{
		    Extinction curves of suggested AGN polar dust (red solid
		    line) and classical ISM dust (black dashed line).
		}
	\label{fig:ext_curve}
    \end{center}
\end{figure}

Figure~\ref{fig:pol_sed} shows the SED template for the polar dust emission.
Its emission is peaked at $\sim25.6~\mum$, corresponding to a characteristic
dust temperature $\sim$113 K. In other words, most of the energy from the
central engine absorbed by the polar dust will be radiatively transferred into
the mid- to far-IR bands. As we discussed in Section~\ref{sec:model}, the
polar dust should consist of large grains, e.g., $a_{\rm min}\sim0.04~\mum$ and
$a_{\rm max}\sim10~\mum$. We use the SKIRT code \citep{Baes2003, Baes2011} to
compute the optical properties of the polar dust by calculating the total
extinction cross section averaged over such a standard grain size distribution
with 20 bins for silicate and graphite with a mixture 0.53:0.47. Then an
extinction is derived by normalizing the polar dust optical properties at
V-band (0.55~$\mum$) and shown in Figure~\ref{fig:ext_curve}.  Due to the
presence of $a\gtrsim0.3\mum$ grains, the mid-IR extinction, e.g.,
$A_{10\mum}$, is at least two orders of magnitude lower relative to the optical
extinction, $A_{\rm V}$.  Since the majority of our sample yields $A_{\rm
V}\lesssim5$, the possible extinction at the wavelengths where the polar dust
emits the most energy is extremely low.  In this case, the shape of its
emission SED would not be influenced by the geometry of the polar dust
component but only depend on its radial density profile, as explained in
Section~\ref{sec:model-geometry}. 

%%% Figure 9 %%%

%%% Figure 10 %%%

\subsection{Observed Size}\label{sec:size}

\begin{figure}[!htb]
    \begin{center}
	\includegraphics[width=1.0\hsize]{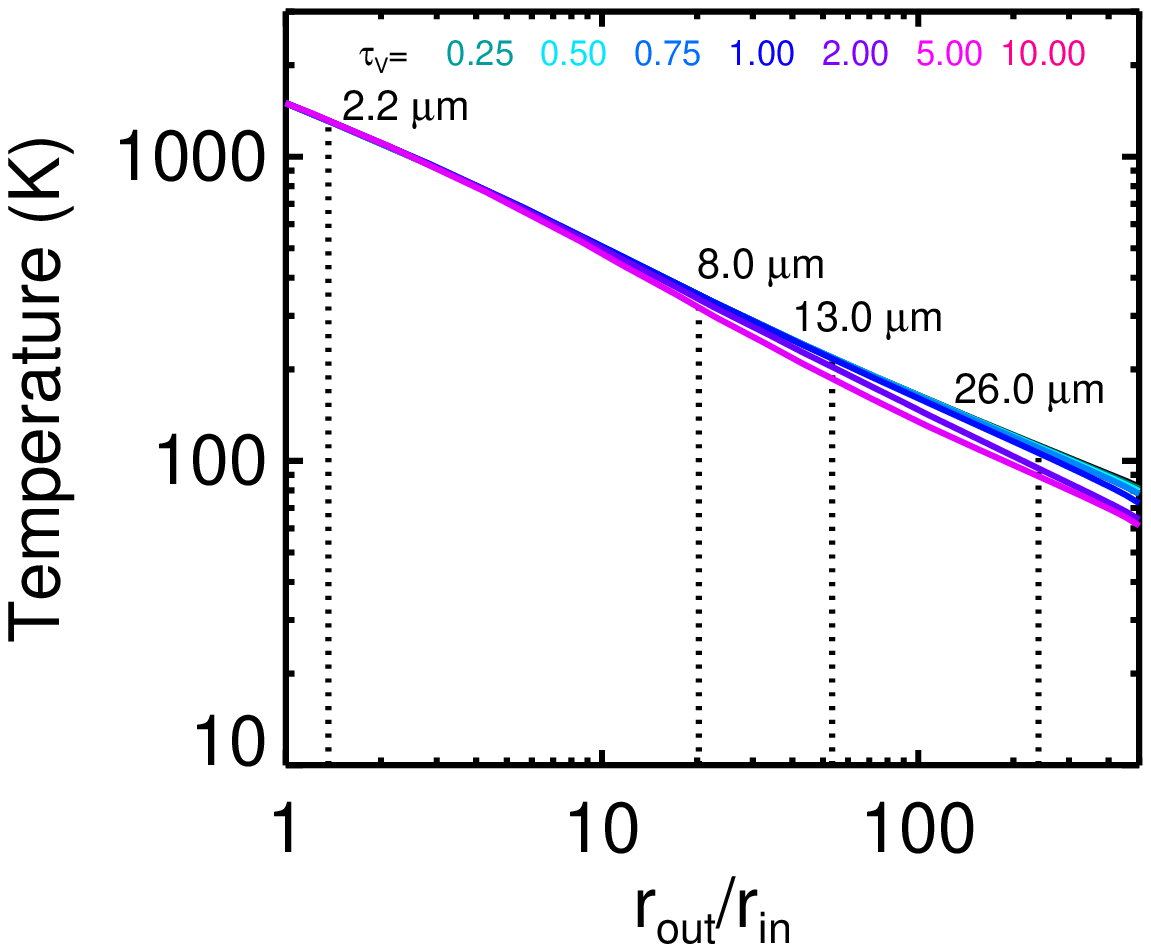}
	\caption{
	    Radial profiles of the dust temperature of the extended dust
	    component with $\tau_{\rm V}$ = 0.25, 0.50, 0.75, 1.00, 2, 5, 9.75.
	    The geometry structure and grain properties are the same as for
	    NGC~3783 (see Table~\ref{tab:ngc3783_model}). We denote the radii
	    that dominate the emission at four typical wavelengths by the
	    vertical dotted lines.
	}
	\label{fig:pol_temp}
    \end{center}
\end{figure}

Figure~\ref{fig:pol_temp} presents the radial profiles of the temperature
distribution for the extended polar dust component in our model. Similar to the
IR-reprocessed emission SEDs, the temperature profiles also have little
dependence on the optical depth ($\tau_{\rm V}$). With the Wien displacement
law, we relate some typical observed wavelengths to the dust temperature and
estimate the corresponding physical size. In the near-IR ($\sim2.2~\mum$), we
can only see the very inner part ($r_{\rm out,obs}\sim 1.3 r_{\rm in}$) of the
extended dust component. In the mid-IR ($\sim$8--13~$\mum$), the observed size
increases to $r_{\rm out,obs}\sim$ 20--41$r_{\rm in}$. At $\lambda\sim26~\mum$,
the polar dust emission reaches its peak, corresponding to an observed size
$r_{\rm out,obs}\sim$ 200--300$r_{\rm in}$. At even longer wavelengths, the
observed size of dust distribution is expected to change slowly.  However,
since its SED drops quickly as a power-law, the polar dust emission would be
very faint and spreads out in a relatively extended area, making any detections
challenging.

%%% Figure 11 %%%

Considering the AGN polar dust distribution could have some dust-covering
factor and its morphology will depend on the observing angle, the estimations
provided here should be considered only as order-of-magnitude values.

\subsection{The Prevalence of AGNs with Strong Polar Dust Emission}\label{sec:prevalence}

As presented in the bottom panels of Figure~\ref{fig:pol_decomp}, the relative
strength of the reprocessed and scattered emission by the polar dust is a
function of the optical depth $\tau_{\rm V}$. We take the relative contribution
of the these two components at 10~$\mum$, $f_{\rm pol, 10~\mum}\gtrsim0.5$ as
the threshold for AGNs with significant polar dust emission. However, the exact
values can not be definitive since there are several sources of uncertainties.
First, the intrinsic IR emission of the AGNs presents variations and there
could exist objects with intermediate SEDs that cannot be unambiguously grouped
into normal, WDD or HDD AGNs \citep{Lyu2017}. In such cases, the SED model
would not make a decisive selection of the intrinsic AGN type, making the value
of $f_{pol,10~\mum}$ uncertain.  In addition, despite the model simplicity,
fitting degeneracy still exists. A good example is NGC 3783, where our SED
fittings in Section~\ref{sec:ind_fit} suggest the intrinsic AGN emission can be
also represented by a normal AGN template with $f_{\rm pol, 10~\mum}\sim0.50$,
which is much lower than the \citet{Honig2013} mid-IR interferometry
measurements. Last but not the least, as shown by the example of NGC 4235
(see the fittings \#29 and \#59 in Figure~\ref{fig:agn1_sed1}), if the SED were
strongly contaminated by the galaxy emission, the derived AGN component could
be highly uncertain, making the identification of polar dust emission
specious.

To get a conservative number of significant polar dust emitters, we remove
objects with significant near-IR stellar contamination and count the fittings
with the reddened normal AGN templates only. In this way, the fraction of
objects with significant polar dust emission among the HSR Seyfert-1 sample is
13\%--31\%. For the SDSS/{\it Spitzer} sample, this value is 36\%--45\%. In
Table~\ref{tab:pol_candidate}, we list the candidate objects with significant
polar dust emission. This behavior appears to be characteristic for at least
25--38\% of relatively low-luminosity type-1 AGNs in Seyfert galaxies.

\capstartfalse
\begin{deluxetable}{clccc}
    %\tabletypesize{\scriptsize}
    \tabletypesize{\footnotesize}
    \tablewidth{1.0\hsize}
    \tablecolumns{5}
    \tablecaption{Candidate AGNs with Significant Polar Dust Emission  \label{tab:pol_candidate}
    }
    \tablehead{
	\colhead{ID} & 
	\colhead{Name} &
	\colhead{Type} &
	\colhead{$f_{pol,10~\mum}$} &
	\colhead{Morph.}
}
\startdata
\multicolumn{5}{c}{HSR AGN sample} \\
\hline 
 1 & PKS 1417-19  &  WDD     &     0.65   & p    \\
12 & IC 4329A     &  WDD     &     0.58   & p, U    \\
13 & 3C 120       &  WDD     &     0.58   & u    \\
18 & Mrk 1014     &  NORM    &     0.93   & p    \\
21 & NGC 3783     &  NORM    &     0.51   & u, E  \\
22 & NGC 4507     &  WDD     &     0.78   & n, C \\
25 & NGC 7469     &  NORM     &     0.91   & y, U   \\
26 & NGC 1566     &  NORM     &     0.93   & u   \\
27 & NGC 4593     &  WDD      &     0.58   & y, U   \\
28 & NGC 3227     &  HDD      &     0.94   & y   \\
\hline 
\multicolumn{5}{c}{SDSS/{\it Spitzer} AGN sample} \\
\hline 
36 & NGC 4074                 & NORM   &   0.62 & p  \\
38 & [VV2006c] J020823.8-002000  & NORM   &   0.51 & \ldots\\
41 & 2MASX J16164729+3716209  & NORM   &   0.93 & \ldots  \\
42 & NGC 863 (aka Mrk 590)    & NORM   &   0.84 &  n      \\
45 & 2dFGRS TGN254Z050        & NORM   &   0.93 & \ldots  \\
48 & 2MASX J09191322+5527552  & NORM   &   0.51 & \ldots \\
49 & 2MASX J14492067+4221013  & WDD    &   0.84 & \ldots  \\
53 & Mrk 176                  & NORM   &   0.51 & \ldots\\ 
54 & 2MASX J14054117+4026326  & NORM   &   0.93 & \ldots  \\
55 & 2MASX J14482512+3559462  & HDD    &   0.60 & \ldots  \\
56 & [GH2004] 9               & NORM   &   0.75 & \ldots\\
57 & Mrk 417                  & NORM   &   0.51 & \ldots\\
60 & 2MASX J12384342+0927362  & HDD    &   0.92 & \ldots \\ 
61 & 2MASS J10405880+5817034  & NORM   &   0.58 & \ldots\\
62 & SDSS J164019.66+403744.4 & NORM   &   0.63 & \ldots\\
    \enddata
    \tablecomments{
	These candiates are selected from our SED modelings.  Morph. - nuclear
	mid-IR extension derived from single-disk images by \citet{Asmus2014}:
	n - not resolved, p - possibly extended, u - unknown extension, y -
	extended, and interferometry observations by \citet{Lopez-Gonzaga2016}:
	C - circular, E - elongated, U - marginally resolved or unresolved~12
    $\mum$ emission. }
\end{deluxetable}
\capstarttrue

Based on the SED analysis, we can also put an upper limit on the fraction of
objects with polar dust emission. As argued in
Section~\ref{sec:model-agn-temp}, an AGN whose SED can be directly matched by
any of the \citet{Lyu2017} intrinsic templates (i.e., $\tau_{\rm V,ext}\sim0$)
is unlikely to have strong polar dust emission. There are 11 such objects in
our whole sample, making the polar-dust-free sample fraction about $\sim20\%$.
There are another 11 objects with $\tau_{\rm V, ext}\sim0.25$, or $f_{\rm pol,
10\mum}\lesssim30\%$.  Accordingly, the proportion in our Seyfert-1 sample with
significant polar dust emission should be no more than $\sim70\%$, assuming
that the values of $f_{\rm pol,10~\mum}$ were underestimated among half of the
samples with moderate polar dust emission.

Lastly, note that our intrinsic AGN templates are the descriptions of
common features and the IR SEDs of polar-dust-free AGNs are expected to have
additional variations among individual objects (e.g., see Figure 10 in
\citealt{Lyu2017}). Bearing this in mind, we may conclude that about 1/3 of the
type-1 nuclei studied in this work show evidence for strong emission by polar
dust, about 1/3 show no evidence, and the remaining 1/3 have weaker or absent
emission - they are ambiguous.

\subsection{Comparison with Mid-IR Morphology-based Polar Dust Identification}

Among the type-1 AGNs studied in this work, there are thirteen objects whose
mid-IR morphology has been studied via interferometry by
\citet{Lopez-Gonzaga2016}. Ten of them, Mrk 1239,  IRAS 09149-6206,  IRAS
13349+2438,  I Zw 1, H 0557-385,  IC 4329A,  ESO 323-77,  NGC 4151, NGC 7469,
NGC 4593, are marginally resolved or unresolved at $\sim$12~$\mum$. For one
object, NGC 1566, the result is uncertain. Conclusive arguments on the
existence of their polar dust emission cannot be reached due to limited UV
coverages and signal-to-noise ratios.  From our SED analysis, NGC 7469 and
NGC 1566 may present very preminant polar dust emission with $f_{\rm pol,
10~\mum}\gtrsim0.9$; IC 4329 A, NGC 4151, NGC 4593, and I Zw 1 have $f_{\rm
pol, 10~\mum}\sim$0.4--0.6; IRAS 09149-6206 and ESO 323-77 could be
moderate polar dust emitter with $f_{\rm pol, 10~\mum}\sim0.3$; Mrk 1239
has an IR SED best-desribed by the intrinsic normal AGN template so it is
unlikely to have much polar dust emission.  Besides NGC 3783,
\citet{Lopez-Gonzaga2016} found another type-1 AGN, NGC 4507, has evidence for
polar dust emission but with a nearly circular morphology. The strong polar
dust emission of NGC 4507 is also identified by us with $f_{\rm pol,
10~\mum}\sim 78\%$. 

\citet{Asmus2016} studied the mid-IR emission extension for a large number of
low-$z$ Seyfert nuclei at subarcsecond scales. Among their objects with
extended mid-IR emission that are likely associated with polar dust emission,
four type-1 nuclei have been studied with SED analysis in this work: NGC 3227,
NGC 4593, NGC 7469 and ESO 323-77. The first three objects have been
successfully identified by us in Table~\ref{tab:pol_candidate} with $f_{\rm
pol, 10~\mum}\gtrsim 0.6$. The SED of ESO 323-77 is best fitted by the 
reddened WDD AGN template with $\tau_V\sim0.25$ and $f_{\rm pol, 10~\mum}\sim0.30$,
suggesting the likelihood of moderate mid-IR polar dust emission. In fact,
    very recently, \cite{Leftley2018} reported interferometric observations of
ESO 323-77, arguing that $\sim$35\% of its flux at 8--13$~\mum$ is polar
extended. This is in good agreement with our SED analysis.

These consistent results demonstrate that the infrared SED analysis could be a
promising and low-budget method to look for AGNs with polar dust emission.

\subsection{Type-2 AGNs}

Since most AGNs with evidence for the extended polar dust emission are type-2
objects \citep[e.g.,][]{Lopez-Gonzaga2016, Asmus2016}, it is ideal to carry out
some similar SED analysis of such objects. However, we do not have a robust
understanding of the SEDs of the circumnuclear tori for these sources. Although
current radiative transfer models can produce a wide range of SED features,
they have too many degeneracies to be tested robustly particularly given the
likelihood of contamination by the emission of star forming regions and
AGN-heated polar dust. Consequently, we have had to leave type-2 AGN out of
our study.

\section{High-$z$ Type-1 AGNs with Peculiar SED Features}\label{sec:highz}

Recent observations have identified some high-$z$ type-1 AGNs with peculiar
SEDs that cannot be easily reproduced by the classical AGN template with simple
UV-optical reddening. Although we lack the detailed understanding of these
types of object that has been accumulated for Seyfert galaxies, we show in this
section how their SEDs are consistent with being shaped by polar dust that can
be fitted by our model.

\subsection{Extremely Red Quasars}

From the Baryon Oscillation Sky Survey, \citet{Ross2015} identified a
population of extremely red quasars (ERQs), using SDSS and WISE photometry.
These objects are very luminous with AGN bolometric luminosities
$\gtrsim10^{13}~L_\odot$, so that any host galaxy contamination at longer
wavelengths can be ignored. In addition, outflows are commonly revealed by the
UV to optical emission line profiles in this type of objects
\citep{Zakamska2016, Hamann2017}. These characteristics of ERQs match our model
assumptions, offering a unique test for the validity of our reddened templates.

\begin{figure}[!htb]
    \begin{center}
	\includegraphics[width=1.0\hsize]{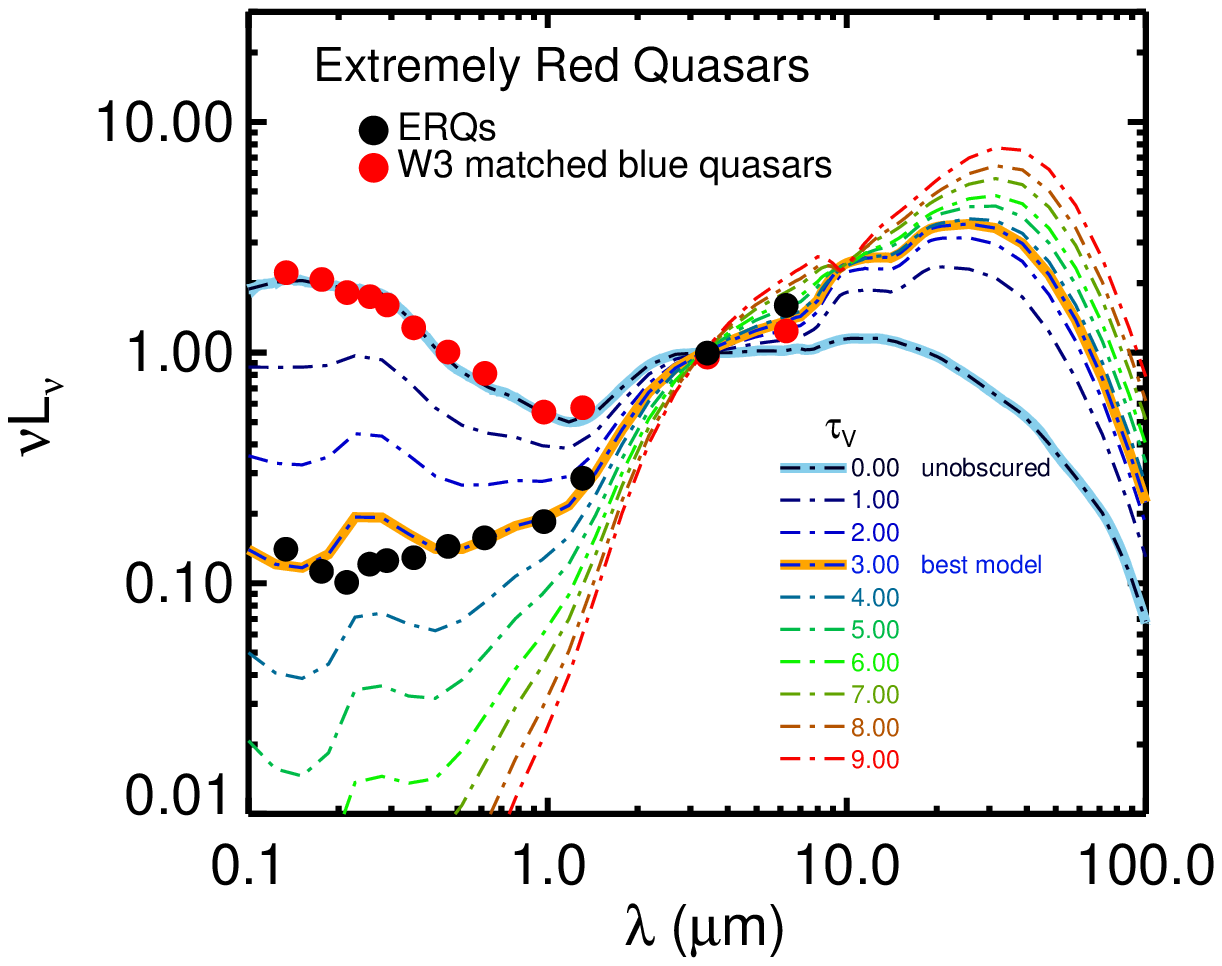}
		\caption{
		    Reproduction of the median SEDs of extremely red
		    quasars and blue quasars matched to the ERQ sample in the
		    WISE W3 bands from \citet{Hamann2017}. The polar dust
		    component is configured with the same parameters as
		    NGC~3783. The ERQ SED is best matched by the reddened
		    normal AGN template with $\tau_{\rm V}\sim3.0$.
		    }
	\label{fig:erq}
    \end{center}
\end{figure}

\citet{Hamann2017} built the median SEDs of Type-1 ERQs and showed that they
were inconsistent with the simple reddening of the UV-optical SED of the normal
quasar template, assuming a SMC-like extinction curve (see their Section 5.5).
In Figure~\ref{fig:erq}, we compare their median ERQ SED for the non-BAL
core-sample with our model templates used for low-$z$ Seyfert-1 nuclei.
Without any fine-tunings of the dust geometry and grain properties, this
composite SED can be matched by the reddened normal AGN template with
$\tau_{\rm V}=3.0$.

%%% Figure 12 %%%

The success of our model for these extremely red quasars indicates that similar
dust obscuration structures as well as the grain properties of polar dust might
be shared among AGNs with a very wide range of luminosities ($L_{\rm AGN,
bol}\sim10^8$--$10^{13}~L_\odot$). 

\subsection{AGNs with Mid-IR Warm-excess Emission}

From a study of 24~$\mum$-selected AGNs in the Local Cluster Substructure
Survey, \citet{Xu2015a} showed that the UV-to-IR SEDs of most Type-1 AGNs at
$z\sim$0.3--2.5 can be reasonably reproduced by combining three empirical
templates that describe AGN, stellar and star formation components. However, an
additional warm dust emission component ($T\gtrsim$50 K) was found for eight
type-1 AGNs whose SEDs can not be fitted by combining the Elvis-like intrinsic
AGN template and any SFG galaxy template. Similar SED behavior has also been
reported in other high-$z$ samples \citep[e.g.,][]{Kirkpatrick2015} as well as
a few quasars in the Palomar-Green sample \citep{Lyu2017}.

\begin{figure}[!htb]
    \begin{center}
	\includegraphics[width=1.0\hsize]{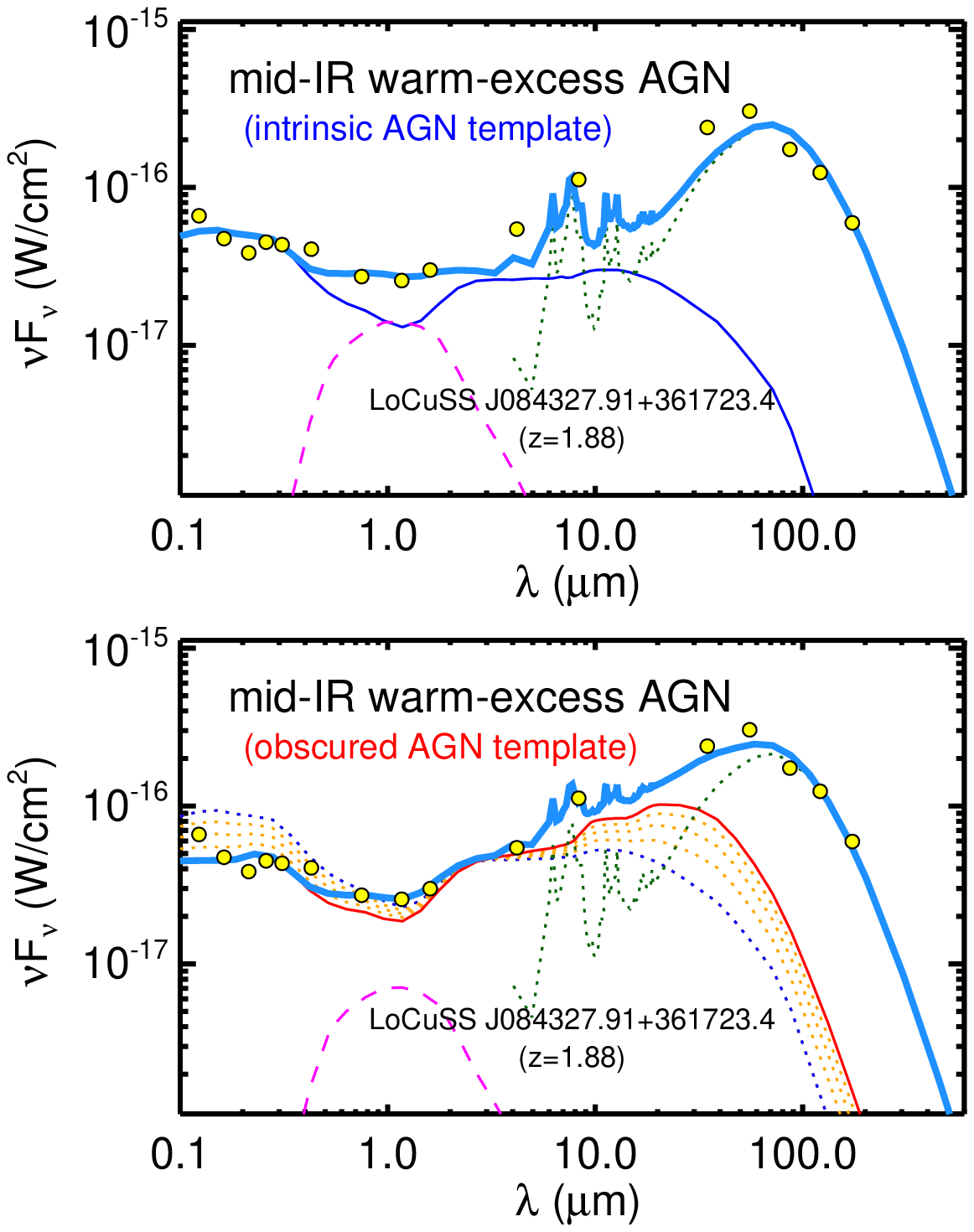}
		\caption{
		    Best-fit results for a $z=1.883$ type-1 AGN, LIRAS
		    J084327.91+361723.4, in \citet{Xu2015a}. Top: the results
		    from fitting the SED with the intrinsic normal AGN template
		    (blue solid thin line), the stellar template (magenta dashed
		    line) and the star-forming template (green dotted line);
		    the observed SED (yellow points) presents significant
		    warm-excess emission in the mid-IR and far-IR compared with
		    the final best-fit model (sky blue solid thick line).
		    Bottom: the same fit as above, but with the addition of a
		    standard template for emission by polar dust (red solid
		    line).
		    }
	\label{fig:xu}
    \end{center}
\end{figure}

Figure~\ref{fig:xu} shows an example for high-$z$ warm-excess AGNs in
\citet{Xu2015a}. The contribution of the warm-excess emission to the total IR
luminosity of this object is estimated to be 56\% using the original Elvis-like
AGN template \citep{Xu2015a} and the \citet{Rieke2009} star-forming galaxy
template. As demonstrated in this work, warm excess emission above the
intrinsic AGN templates can be easily produced if there is polar dust around
the nucleus. After allowing the AGN template to be obscured by our model, we
made excellent fittings of the rest-frame 0.1--500~$\mum$ SEDs for all the
warm-excess AGNs reported in \citet{Xu2015a}.

%%% Figure 13 %%%

The real origin of the warm-excess emission is not clear. \citet{Xu2015a}
demonstrated that this warm component can be fitted by the spectrum of a
parsec-scale starburst disk \citep{Thompson2005, Ballantyne2008}. In contrast,
the success of our reddened AGN model suggests that this feature could be caused by
the AGN heating of the surrounding polar dust, a possibly common phenomenon in
many objects.

\subsection{Hot Dust-obscured Galaxies}\label{sec:dog}

An important discovery made by the {\it WISE} survey is the identification of
very luminous hot dust-obscured galaxies, or hot DOGs
\citep{Eisenhardt2012,Wu2012}. Compared with typical IR luminous galaxies, the
SEDs of hot DOGs have very prominent hot dust emission with characteristic
temperatures around 60--100K \citep[e.g.,][]{Wu2012, Fan2016}. From
investigations of X-ray observations and rest-frame optical spectra, these
objects are found to be powered by luminous but heavily reddened Type-1 AGNs
\citep[e.g.,][]{Stern2014, Ricci2017b, Wu2018}. 

\citet{Fan2016} presented the IR SEDs of 22 submm-detected Hot DOGs and
suggested that they can be described by the CLUMPY torus model
\citep{Nenkova2008a, Nenkova2008b} plus a cold dust component to represent the
host galaxy star formation. Given the large number of free parameters of their
models, we would like to see if our semi-empirical templates can provide an
alternative solution. 

\begin{figure}[!htb]
    \begin{center}
	\includegraphics[width=1.0\hsize]{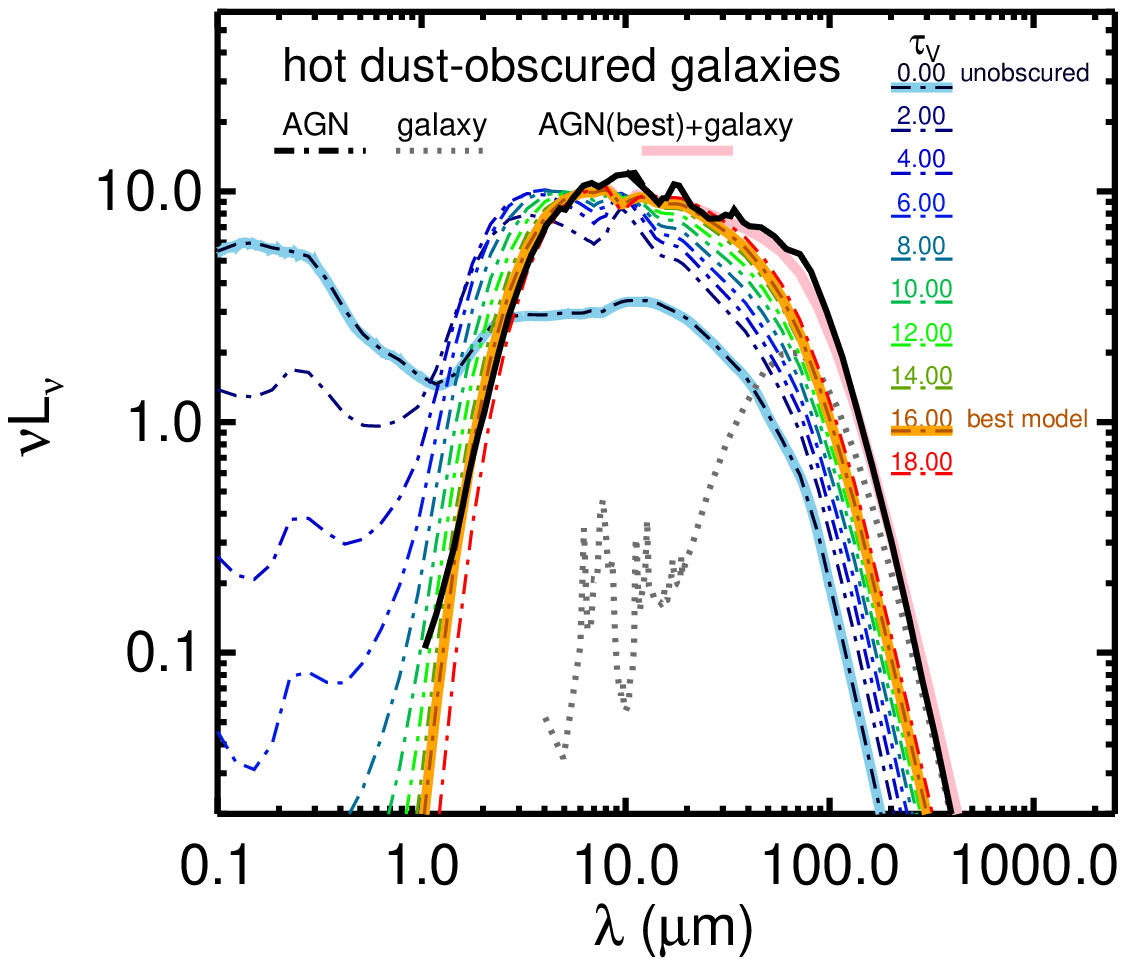}
		\caption{ 
		    Reproduction of the median SED of hot dust-obscured AGNs
		    from \citet{Fan2016}. This SED is roughly matched with the
		    reddened AGN template with $\tau_{\rm V}\sim16.0$ (orange thick
		    line) plus a moderate star-formation template (grey dotted
		    line) in the far-IR.  Compared with other objects, the
		    polar dust geometry to reproduced the Hot DOG SEDs features
		    a steeply decreasing density profile ($n\sim r^{-1.5}$)
		    and more extended structure ($Y\gtrsim1000$).
		    }
	\label{fig:dog}
    \end{center}
\end{figure}

Figure~\ref{fig:dog} presents a median IR SED of hot DOGs \citep{Fan2016}. To
represent the galaxy far-IR emission, we adopted the empirical star-forming
galaxy SED library derived by \citet{Rieke2009}. Since the reddened AGN
templates trained for NGC~3783 do not contain such strong hot dust emission, we
allowed the geometry of the extended dust component to be variable but left the
dust grain properties unchanged.  Similar to the case for NGC~3783 (see
Section~\ref{sec:ngc3783-model}), we used the MCMC algorithms to find the
best-fit parameters. The results are summarized in Table~\ref{tab:dog_model}.
Reasonable fittings can be achieved with a broad range of parameters, such as a
dust density profile $r^{-1.5}$, and outer-to-inner radius $Y=5000$. Compared
with most Seyfert nuclei, the fitted polar dust component for AGNs in
hot-dust-obscured galaxies has a steeper density profile ($n \propto
r^{-0.5}\rightarrow r^{-1.5}$) and more extended distribution ($Y\sim
500\rightarrow 5000$).

%%% Figure 14 %%%

\capstartfalse
\begin{deluxetable}{cccc}
    %\tabletypesize{\scriptsize}
    \tabletypesize{\footnotesize}
    \tablewidth{1.0\hsize}
    \tablecolumns{4}
    \tablecaption{Suggested Model Parameters for Hot Dust-Obscured Galaxies\label{tab:dog_model}
    }
    \tablehead{
	\colhead{Parameter} & 
	\colhead{Label} & 
	\colhead{Adopted Value} &
	\colhead{MCMC output}
}
\startdata
    $r_\text{in}$ temperature         & $T_\text{in}$   & {\bf 1500} K       &                                    \\
    density profile                   & $\alpha$        &  0.50              &  $1.49^{+0.46}_{-0.45}$          \\
    outer-to-inner radius             & $Y$             &  5000               &  $4778^{+4762}_{-3323}$              \\
    silicate:graphite mixture         &                 & {\bf 0.53:0.47}    &                                    \\
    maximum grain size                & $a_\text{max}$  & {\bf 10}~$\mum$    &                               \\
    minimum grain size                & $a_\text{min}$  & {\bf 0.04}~$\mum$  &                                \\
    \hline                                                     
    input radiation SED               &                 & {\bf normal}       &                              \\
    optical depth                     & $\tau_\text{V}$ &   16               &  $17.16^{+1.64}_{-1.97}$    \\
    $\log(L_{\rm SF, temp}/L_\odot)$  &                 &    11.5            &  $11.49^{+0.46}_{-0.44}$ 
    \enddata
    \tablecomments{     
	We use boldfaces to indicate assumed parameter values that do not go to
	MCMC parameter space sampling.
    } 
\end{deluxetable}
\capstarttrue

Are the best-fit parameters of our model to match the Hot DOG SED physical?
Assuming the central engine has $L_{\rm AGN, bol}=10^{14} L_\odot$, the
suggested extended dust would extend to kpc scales.  Interestingly, ALMA
observations of [CII] emission in the most luminous hot DOG W2246$-$0526 show a
uniform and highly turbulent ISM, suggesting isotropic galaxy-scale outflows
\citep{Diaz-Santos2016} which is consistent with our argument that the extended
dust distribution might be related to AGN outflows. Hot DOGs are
suggested to live in overdense environments based on submm \citep{Jones2014,
Jones2015} and near-IR observations \citep{Assef2015}. It is possible that
these objects represent a special phase of galaxy evolution, e.g., galaxy
mergers, that could change the gas and dust distribution around the AGNs
compared with the relatively undisturbed environment for typical Seyfert
nuclei.  Indeed, complex velocity structures based on CO(4-3) emission lines
are reported in hot DOGs, suggesting a violent environment in such systems
\citep{Fan2018}. Lastly, the SED model selected the \citet{Rieke2009}
star-forming galaxy template with $\log(L_{\rm SF, temp}/L_\odot)\sim11.5$,
consistent with expectations for IR luminous galaxies at the hot DOG redshifts
\citep{Rujopakarn2013}.

\begin{figure}[!htb]
    \begin{center}
	\includegraphics[width=1.0\hsize]{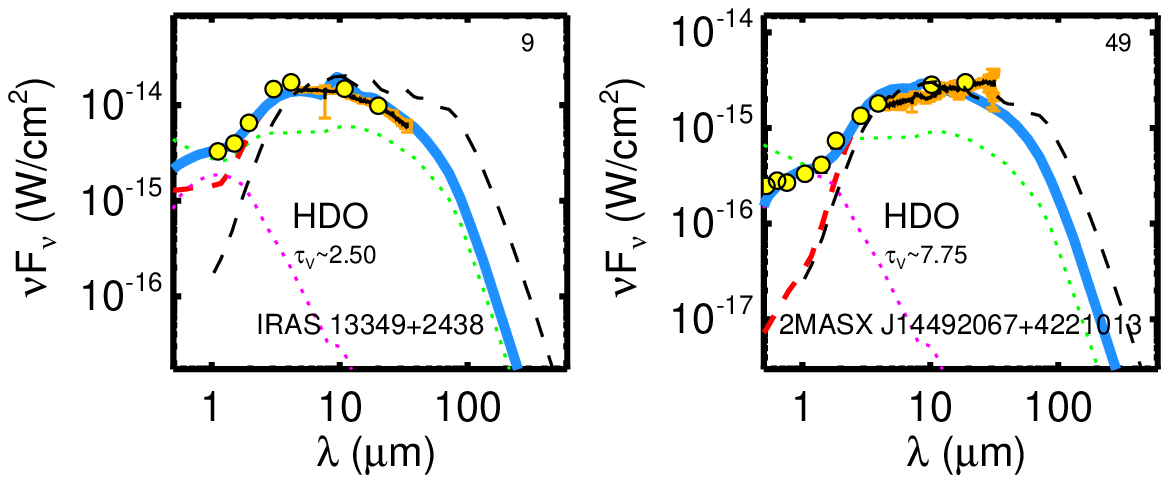}
	\caption{
	    SEDs and the best-fit models of low-$z$ candidate hot dust-obscured
	    (HDO) AGNs.  The legend follows Figure~\ref{fig:agn1_sed1}. In each
	    panel, we plot the composite SED template for hot DOGs from
	    \citet{Fan2016} as a black dashed line.
	}
	\label{fig:lowz_dog}
    \end{center}
\end{figure}

Although hot DOGs are claimed to exist only at high-$z$
\citep[e.g.,][]{Eisenhardt2012,Wu2012}, type-1 AGNs with similar SED features
could exist at low-$z$. In Section~\ref{sec:ind_fit}, we note that two low-$z$
Seyfert-1 nuclei, IRAS 13349+2438 ($z=0.11$), 2MASX J14492067+4221013
($z=0.18$), present strong hot dust emission for which our model with
parameters set for NGC 3783 did not produce reasonable fits. In
Figure~\ref{fig:lowz_dog}, we present the SEDs of these objects and the
best-fit model with the reddened AGN templates developed for hot DOGs. Due to
their relatively low AGN luminosities, the stellar contamination in the near-IR
cannot be ignored so that the apparent hot dust emission is not as strong as
the case of high-$z$ hot DOGs. However, these objects have much stronger hot
dust emission compared with the normal AGN template. Together with the stellar
template, our hot dust-obscured AGN templates explain these SEDs well. Thus, we
suggest AGNs with some extreme SEDs similar to high-$z$ hot DOGs do exist at
low-$z$.

%%% Figure 15 %%%

%\clearpage

\section{Discussion}\label{sec:dis}

We have shown that the IR SEDs of most type-1 AGNs regardless of luminosity can
be fitted by a small set of templates for the accretion disk and torus,
supplemented by an extended component to represent emission by polar dust. This
result indicates that the differences in the SEDs between Seyfert-1 nuclei and
type-1 quasars are largely the result of the absorption of UV energy by polar
dust and its reradiation in the IR.  As argued by e.g., \citet{Stalevski2016},
the IR SEDs of the optically-thick torus would stay almost the same for any
lines of sight not blocked by the torus. Our intrinsic AGN templates should be
faithful descriptions of the torus emission in type-1 systems with a range of
possible observing angles. The suggested polar dust for most objects can be
modeled with the similar density profiles and grain properties. Therefore, the
diverse broad-band IR SED features of many type-1 AGNs are only determined by
two free parameters: the AGN intrinsic IR SED type and the level of surrounding
obscuration.  These results have the following implications.

\subsection{Interpreting the AGN Mid-IR Emission}

Previous AGN-related studies mostly only consider the effects of the extinction
in the UV-optical bands, leaving the IR SED unchanged
\citep[e.g.,][]{Assef2010, Xu2015a}. An embedded assumption for such an
approach is that the dust is distributed very far away from the AGN and the
reprocessed emission is weak and possibly mixed with the host galaxy emission
heated by stars. However, as suggested in this work, if the obscuration occurs
in the vicinity of the AGN, it could naturally explain the polar dust emission
in nearby AGNs as a result of energy balance and reconcile the diverse
broad-band IR SED features of various AGN populations. This IR-reprocessed
emission can be an important contributor to the AGN-heated IR features besides
the torus.

Various radiative transfer models have been proposed to reproduce the AGN IR
emission. For simplicity, the geometric structure of the nuclear dust is
commonly assumed to be doughnut-like and numerous dusty torus radiative
transfer models have been developed to produce the AGN-heated IR SEDs
\citep[e.g.,][]{Fritz2006, Nenkova2008a, Alonso-Herrero2011, Stalevski2012,
Siebenmorgen2015}. Although they have matched the SEDs or spectral observations
\citep[e.g.,][]{Fritz2006, Nenkova2008b, Alonso-Herrero2011, Siebenmorgen2015},
most such comparisons are limited to the mid-IR bands and strong parameter
degeneracy exists (e.g., see discussions in \citealt{Hoenig2013_proc} and
\citealt{Netzer2015}).  Additionally, some work found that the torus models
alone are typically not enough to reproduce the complete AGN IR SEDs of bright
quasars \citep[e.g.,][]{Mor2009, Leipski2014}.

An important support for the clumpy torus models comes from their ability to
reproduce the behaviors of AGN mid-IR silicate features, especially the
silicate emission seen in some type-2 AGNs (e.g., \citealt{Nikutta2009}, but
see \citealt{Feltre2012}). However, the possible existence of AGN-heated polar
dust makes the interpretation of the origin of the mid-IR spectral features
ambiguous. In fact, as suggested by, e.g., \citet{Sturm2005}, the silicate
emission feature observed in Seyfert nuclei could come from low-optical-depth
dust located in the AGN narrow-line regions. The success of our model supports
this possibility (also see e.g., \citealt{Efstathiou2006},
\citealt{Schweitzer2008}).

For a single object, the AGN-heated IR spectral features are a result of a
mixture of intrinsic variations of the torus properties, possible
IR-reprocessed emission by the polar dust and some specific observing angle.
With these complications as well as the torus model degeneracies, the fittings
of individual observations provide little information on the credibility of the
model. Instead, we suggest that the best test is to see if the torus model can
reproduce the empirical unobscured AGN templates, such as the three in
\citet{Lyu2017}. The objects used to derive the templates do not have evidence
of obscuration, so that (1) the possibility of the polar dust emission is
minimized and (2) these templates represent a face-on view of the system. In
addition, by constructing average templates, variations of individual objects
are smoothed out. In other words, the model degeneracies and observational
uncertainties are greatly reduced.

\subsection{Dust-covering Factor of Type-1 AGNs}

A common tool to study the AGN dust environment is the so-called dust-covering
factor, typically measured by the relative flux ratios between the near- to
mid-IR band and the optical bands \citep[e.g.,][]{Maiolino2007, Treister2008,
Mor2011, Roseboom2013, Lusso2013}. Many authors have tried to use this
parameter to explore the possible evolution of the torus, assuming the AGN
near- to mid-IR emission comes from the torus and the optical emission
originates from the accretion disk \citep[e.g.,][]{Lawrence1991, Simpson2005,
Assef2013}.  Given the direct detections of the polar dust \citep{Braatz1993,
Cameron1993, Raban2009, Honig2012, Honig2013} and its possibly frequent
occurrence inferred in this work as well as e.g., \citet{Asmus2016}, we should
be cautious about the interpretations of the AGN IR-to-optical flux ratios.

\begin{figure}[!htb] 
    \begin{center}
	\includegraphics[width=1.0\hsize]{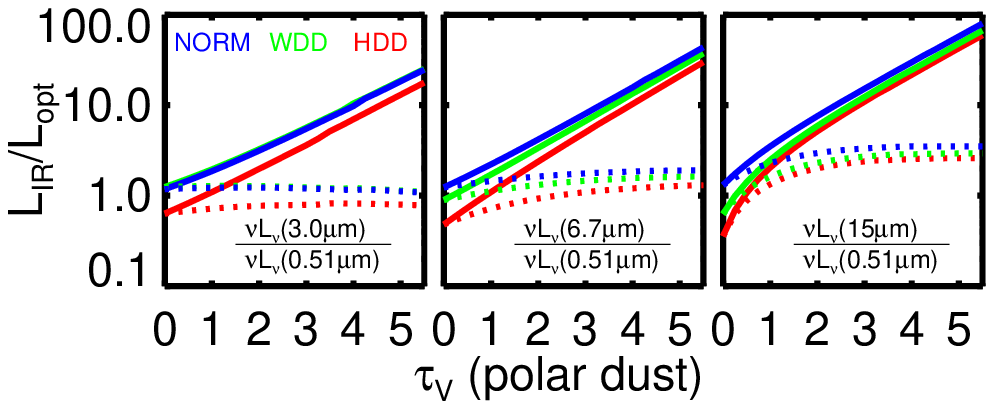} 
	\caption
	{ 
	    MIR-to-optical luminosity ratios ($L_{\rm IR}/L_{\rm opt}$) as a
	    function of optical depth of the extended dust component
	    ($\tau_{\rm V}$) from our model for normal (blue), WDD (green) and
	    HDD (red) AGNs.  We compute these relations with the apparent
	    (obscured) and intrinsic (unobscured) values of $L_{\rm opt}$ and
	    show them separately as solid and dotted lines.
    }
	    \label{fig:polar_cf} 
    \end{center} 
\end{figure}

We can obtain some rough idea on how significantly the AGN IR-to-optical flux
ratio can be changed by the polar dust.  Figure~\ref{fig:polar_cf} presents
some characteristic tracers of dust covering factors calculated from our
reddened AGN model as a function of optical depth $\tau_{\rm V}$. For the
direct observed values, we divided the IR luminosity, $L_{\rm IR}$, at
$\lambda$= 3, 6.7, 15~$\mum$ by the apparent AGN optical luminosity, $L_{\rm
opt}$, at 5100~\AA~  of reddened templates for normal, WDD and HDD AGNs. In all
cases, the IR-to-optical flux is found to increase exponentially with the
optical depth. With moderate extinction ($\tau_{\rm V}\lesssim$1--2), the
change caused by polar-dust obscuration and emission would easily exceed the
intrinsic variations of the torus in a type-1 AGN. Conversely, if we use the
intrinsic (unobscured) luminosity at 5100~\AA~ to represent the $L_{\rm opt}$,
the integrated optical depth of the polar dust component would have limited
influence on $L_{\rm IR}/L_{\rm opt}$, especially at shorter wavelengths. With
$\tau_{\rm V}$ ranges from 0 to 5, the values of $L_{\rm 3.4~\mum}/L_{\rm
0.51~\mum}$ are changed by a factor of 0.94, 0.91 and 1.26 for normal, WDD, HDD
AGNs. For $L_{\rm 6.7~\mum}/L_{\rm 0.51~\mum}$, the changes are 1.53, 1.81,
2.67. For $L_{\rm 15~\mum}/L_{\rm 0.51~\mum}$, the corresponding values are
2.68, 4.67, 7.28.  Since the SED of polar dust emission peaks at
$\lambda\sim26~\mum$ (Section~\ref{sec:pol_char}), its effects would be only
important for the dust covering factors of warm and cold dust.

%%% Figure 16 %%%

Nevertheless, as discussed in Section~\ref{sec:model-geometry}, the
optically-thin IR emission from a dust component with different dust-covering
factors can be easily reproduced by changing its overall optical depth $\tau_V$
with an assumption of a spherical dust distribution. Given this degeneracy, it
is impossible to get a definite estimation of the dust geometry, like its
covering factor, from the SED alone when the IR emission is not
optically-thick.

Besides the possible existence of polar dust, there are many other factors that
can change the value of $L_{\rm IR}/L_{\rm opt}$. In the optical band, the AGN
emission can be modified by the host galaxy contamination, dust obscuration at
different physical scales with uncertain extinction curves, or even some
short-term variability. Based on a 3D radiative transfer model,
\cite{Stalevski2016} explored how the anisotropic emission of the dusty torus
and the accretion disk could influnence the estimation of real dust covering
factor from $L_{\rm IR}/L_{\rm opt}$, showing that their relation could be
non-linear and depend strongly on the assumed torus optical depth. Considering
these complications, it is difficult to make conclusive arguments on the
meaning of $L_{\rm IR}/L_{\rm opt}$ traced by the simple colors derived from a
few photometric bands.

\subsection{AGN X-ray Obscuration and Polar Dust Optical Depth}

\begin{figure}[!htb] 
    \begin{center}
	\includegraphics[width=1.0\hsize]{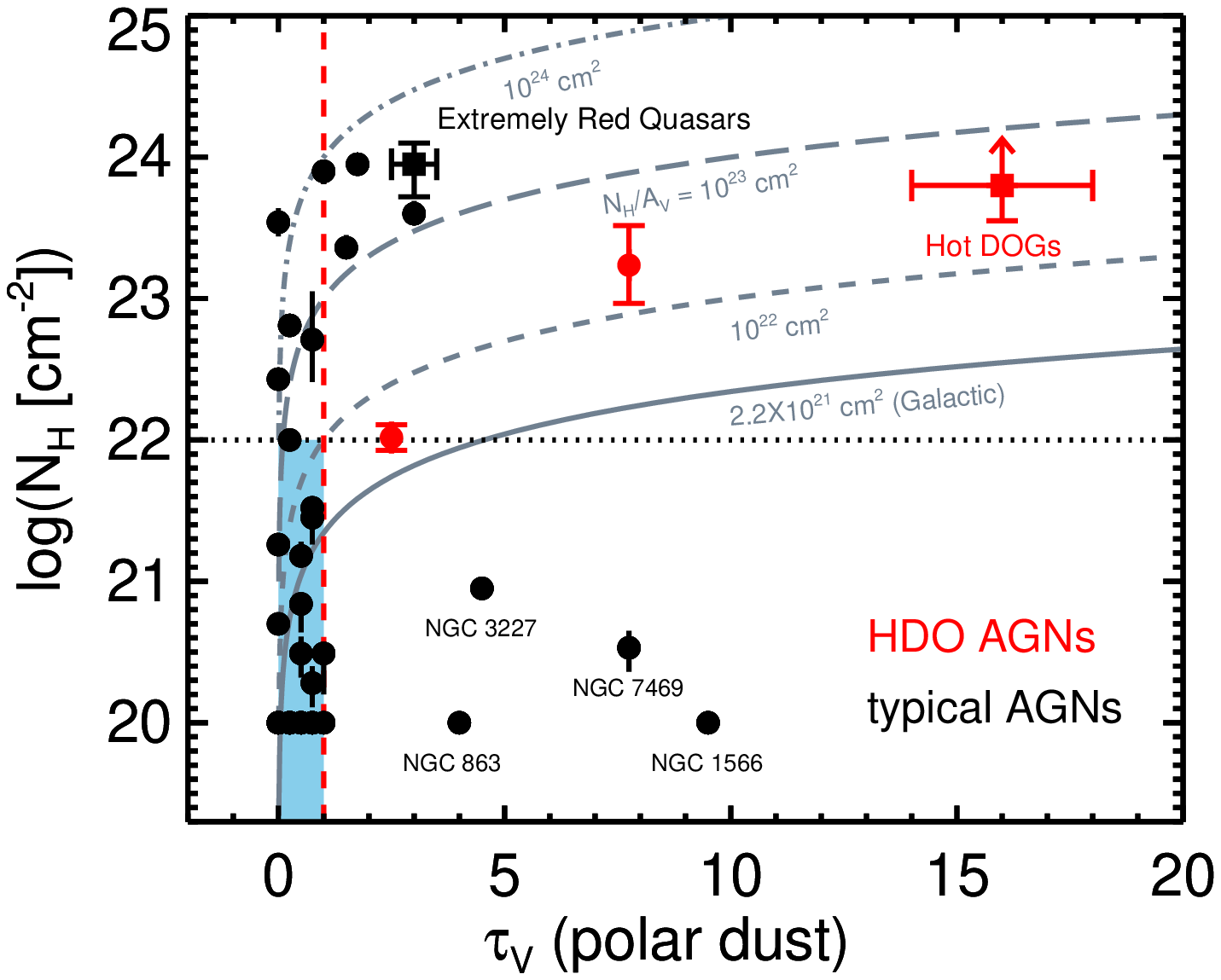} 
	\caption
	{ The distribution of gas column density and polar dust optical depth
	    of the low-$z$ Seyfert-1 (dots) and the high-z reddened populations
	    (squares). We use the black color to indicate objects whose IR SEDs
	    are best described by the NGC~3783-like model and red colors to
	    indicate the HDO AGNs. The blue shaded region represents
	    low-extinction type-1 AGNs with $A_{\rm V}\lesssim1$~mag and $N_{\rm
	    H}\lesssim10^{22}~cm^{2}$.
    }
	    \label{fig:x_ray} 
    \end{center} 
\end{figure}

The extinction caused by the polar dust component in many AGNs is likely
associated with the behavior of their X-ray obscuration. To explore this
possibility, we collected the literature measurements of the absorbing column
density, $N_{\bf H}$, and compared them to the derived integrated optical depth
of the polar dust component (derived purely from IR SED fittings) in
Figure~\ref{fig:x_ray}. 41 out of 64 nearby Seyfert-1 nuclei in our sample are
found to have good measurements. Among them, besides 3C 219 (taken from
\citealt{Comastri2003}) and Mrk 1239 (taken from from \citealt{Corral2011}),
the measurements of all others are collected from \cite{Ricci2017}.

%%% Figure 17 %%%

First we check if the X-ray obscuration and the possible optical extinction
caused by polar dust are consistent. As argued by \cite{Shimizu2018}, most
type-1 AGNs are unobscured in the X-ray, i.e., $N_{\bf H}\lesssim10^{22}
~{\bf cm}^{-2}$.  If we adopt the AGN value of $N_{\bf H}/A_{\bf V}$ from
\cite{Maiolino2001a}, this means the corresponding optical extinction should
satisfy $A_V\lesssim1$.  By this criteria, 28 out of the 41 objects
($\sim68\%$) have low extinction both in the X-ray and optical. In addition, 9
out of 41 ($\sim22\%$) are X-ray absorbed without much polar dust extinction.
For most X-ray obscured Seyfert-1 nuclei for which the NGC 3783-like model
works well, a value of $N_{\bf H}/A_{\bf V}\gtrsim10^{23}~{\bf cm^{-2}/mag}$
seems favored, which is much higher than the typical values found for the
Galactic ISM (e.g., $\sim2.2\times10^{21}~{\bf cm^{-2}/mag}$;
\citealt{Guver2009}).  This nicely fits the picture that small grains, which
are most efficient carriers for the extinction in the optical band, can be
easily destroyed by their direct exposures to the AGN radiation. The effect of
large grains agrees with the behavior of the nine X-ray-absorbed objects with
small $A_{\bf V}$.  Meanwhile, there are four outstanding outliers with high
$\tau_{\bf V}$ but very low $N_{\bf H}$.  However, this is also not a surprise
since $\tau_{\bf V}$ describes the {\it average} polar dust optical depth along
all possible LOSs.  If the polar dusty clouds distribute close to the edge of
the ionization core or are simply clumpy, the AGN X-ray corona could be
unobscured along some lucky observing angles while the amount of polar dust is
still significant.  Consequently, we conclude that there is no obvious tension
between the literature results on the X-ray obscuration and the optical depth
of the polar dust component inferred by our SED model.

We can also compare the behaviors of the high-$z$ extremely red quasars and hot
dust-obscured galaxies discussed in Section~\ref{sec:highz} with these low-z
Seyfert-1 nuclei to further check if they are also similar (or not) in the
X-ray.  Taking the gas column density measured for the stacked X-ray images
from \cite{Goulding2018}, a typical value of $N_{\bf H}/A_{\bf V}$ for
extremely red quasars is $\sim3\times10^{23}~{\bf cm^{-2}/mag}$, consistent
with the trend of most X-ray obscured Seyfert-1 AGNs in our sample. In other
words, besides a polar dust configuration resembling NGC~3783, these high-$z$
extremely red quasars also share the relation between X-ray gas obscuration and
optical dust extinction of typical reddened Seyfert-1 nuclei.

In contrast, hot DOGs might be different. We computed an average $N_{\bf
H}\gtrsim 6\times 10^{23}~{\bf cm}^{-2}$ for high-$z$ hot DOGs with reported
measurements from the literature (see \citealt{Vito2018} and the references
therein). Combined with the optical depth of polar dust component estimated
from our SED modeling, this suggests a $N_{\bf H}/A_{\bf V}\gtrsim 4.0\times
10^{22}~{\bf cm^{-2}}$.  Together with the measurements of two low-$z$
candidates, these HDO objects have $N_{\bf H}/A_{\bf V}$ values of
$\sim10^{22}$--$10^{23}~{\bf cm^{-2}/mag}$, which are lower than those of
typical Seyfert-1 nuclei (as well as the extremely red quasars) but still
higher than the Galactic values.  Perhaps the hot DOGs do represent a different
phase of AGN-host evolution compared to typical Seyfert-1 nuclei (as well as
extremely red quasars) so that their polar dust grains are not removed as
significantly as the latter. For example, the host galaxies of HDO AGNs could
be very obscured so that the AGN radiation is not efficient enough to destroy
most dust distant from the nucleus.  Alternatively, considering the large
extent of the outflows in these cases, there could be some channels for dust
production, e.g., (post-)AGB stars and/or supernovas, that mitigate the effects
of dust destruction by the AGN.

\subsection{AGN Structures and Unification}

An anisotropic obscuration structure composed of optically-thick dust,
typically pictured as a torus, provides a simple solution to unify the
behaviors of different types of AGNs \citep[][]{Antonucci1993a, Urry1995}.
Besides the optically-thick torus, this work suggests the presence of an
extended dust distribution that modifies both UV-optical and IR properties of
the AGN.  We can get some rough ideas on the physical scales associated with
the extended dust distribution and discuss its relation with other known AGN
components.

Based on our results, the observations of many type-1 AGNs can be reproduced by
assuming the same polar dust model configuration featuring a density profile
$r^{-0.5}$ , outer-to-inner radius $Y=500$ and temperature at the inner
boundary $T\sim1500$ K.  For an $L_\odot=10^{11}$ AGN, the dust responsible for
the extended obscuration has a maximum size $r_{\rm out}\sim0.1$~kpc, which is
about one order of magnitude larger than the size of cold dust in the AGN torus
($r_{\rm torus, cold}\sim10$ pc; see Section 4.2 in \citealt{Lyu2017b}).
AGN-driven winds or outflows can easily distribute dust around the torus to
such scales.

One likely location for the polar dust grains is the narrow-line-region clouds.
In fact, the required $A_{\rm V}\lesssim$ 5~mag for most type-1 AGNs can be
easily reproduced by combining the typical NLR column density
($N_\text{H}\sim10^{20}$--$10^{21}$~cm$^{-2}$) with the assumption of the
classical value of $N_{\rm H}$/$A_V\sim2.2\times10^{21}$ cm$^{-2}$/mag in the
Galactic ISM \citep[][]{Guver2009}. Alternatively, if we adopt $N_{\rm
H}/A_V\gtrsim10^{22}$ cm$^{-2}$/mag for Seyfert galaxies \citep{Maiolino2001a}
and assume the NLR cloud density $\sim100$ cm$^{-3}$, the required physical
scale of the line-of-sight dusty clouds would span $\sim$150 pc, which is also
realistic and consistent with the value estimated above. 

\begin{figure}[!htb] 
    \begin{center}
	\includegraphics[width=1.0\hsize]{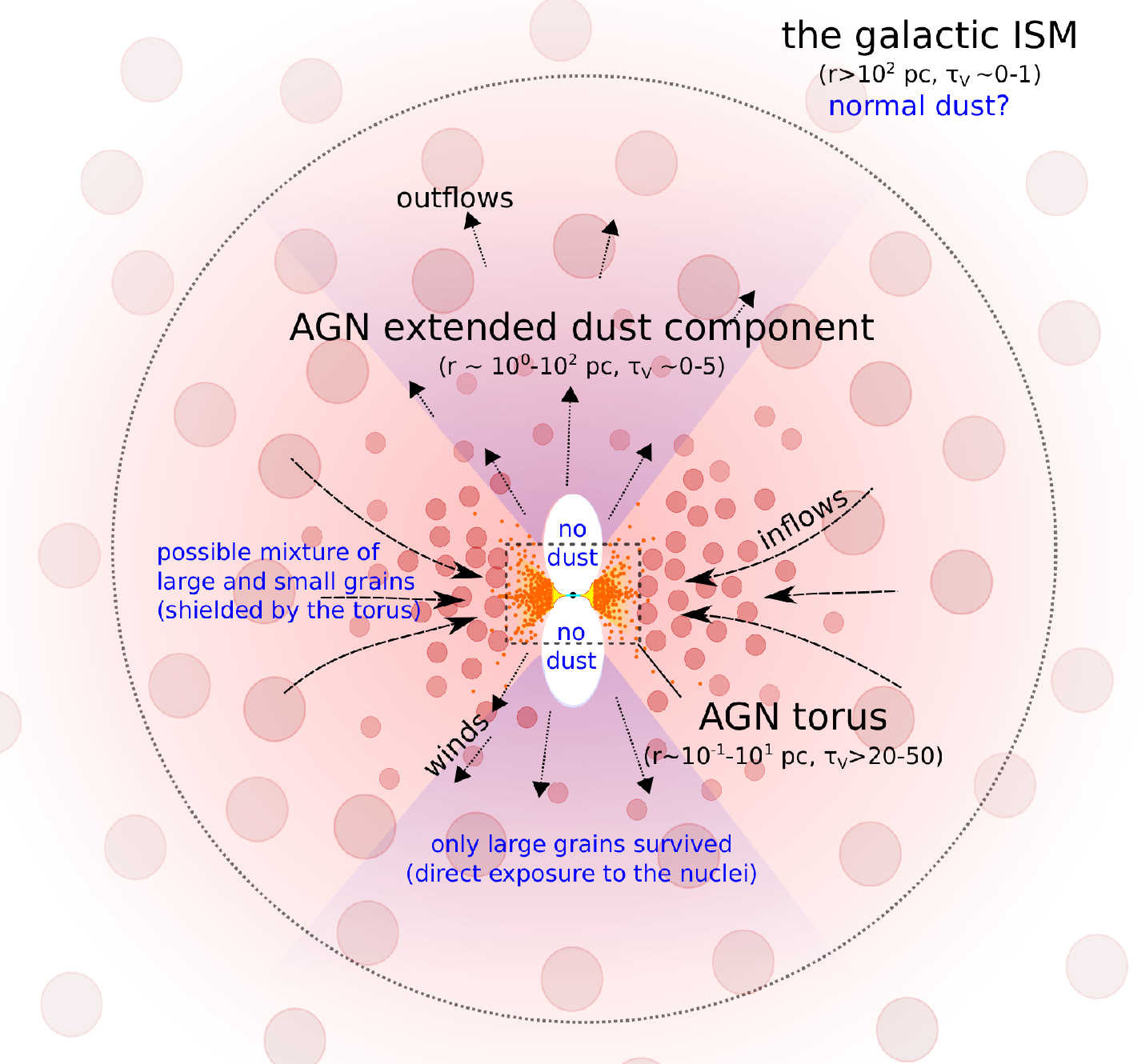} 
	\caption
	{ 
	Schematic drawing for the dust environment around a typical
	Seyfert-like nuclei. See the text for details. 	     }
	    \label{fig:picture} 
    \end{center} 
\end{figure}

In Figure~\ref{fig:picture}, we illustrate the various dust components
surrounding a typical Seyfert nucleus. Very close to the accretion disk, dust
could not survive due to evaporation at high temperatures. Since the accretion
disk emission is not isotropic, some concave shape of dust-free regions is
expected above and below the accretion disk. In the equatorial direction, we
expect a torus-like component. Thanks to the shielding against the direct AGN
emission by very optically-thick ($\tau_{\rm V}\gtrsim$20--50) clouds, a range
of dust grain properties is expected within the torus. Along the polar
direction, as discussed in Section~\ref{sec:model-grain}, only large dust
grains might survive. In other words, we would expect the dust properties are
changing along different observing angles \citep[see][]{Baskin2018}, which may
lead to diverse AGN extinction curves as reported in the literature
\citep[e.g.,][]{Hall2002, Richards2003, Hopkins2004, Gaskell2004, Czerny2004,
Gaskell2007}. In regions far away from the nucleus, e.g., the galactic ISM at
(sub-)kpc scales, the influence of the AGN is minimal so that classical dust
properties are expected. Between the torus and the galactic ISM, there exists
an extended dust component with low optical depth ($\tau_{\rm V}\lesssim$5)
which causes the LOS obscuration and the polar dust emission seen in type-1
AGNs. In reality, there should no clear boundaries between these components and
they can exchange dust by various feedback mechanisms, e.g., inflows or
outflows. Although our sketch shows a clumpy environment, the actual
configuration could have other forms, such as filaments \citep{Wada2009,
Wada2012}. In addition, the shape of the polar dust distribution could deviate
strongly from isotropic symmetry, e.g., having some dust covering factor.

%%% Figure 18 %%%

The intrinsic IR emission of AGNs at wide ranges of luminosity ($L_{\rm AGN,
bol}\sim10^{8}$--$10^{14}~L_\odot$) and redshift ($z\sim$0--6) appears to be
described well by a limited family of SED shapes. In \citet{Lyu2017}, we
demonstrated that the intrinsic IR emission of unobscured quasars can be
represented by a similar set of templates independent of redshift. Based on the
low-$z$ Palomar-Green sample, complete IR SED templates have been built and
tested for AGNs with $L_{\rm AGN, bol}\gtrsim10^{11}~L_\odot$. With the results
obtained in this work, their validity is extended to Seyfert-1 nuclei and some
peculiar AGN populations. Besides the torus SED variations, the IR SED
differences are caused by an extended dust component that might reside in NLR
clouds. Consequently, it is not necessary to invoke any new component that is
only found in some special groups of AGNs.

\subsection{The Diverse Dust Environments of AGNs}

With our previous study of unobscured quasars \citep{Lyu2017, Lyu2017b} and the
results on other type-1 AGN populations obtained here, we have found that the
IR properties of AGNs are not only determined by the observing angles, but also
caused by the intrinsic variations of the torus and the different contributions
of polar dust. 

As illustrated in Figure~\ref{fig:evoluation}, dust environments of AGNs can be
grouped into three different scenarios, characterized by the obscuration along
the face-on direction:

%%% Figure 19 %%%
\begin{figure*}[!htb] 
    \begin{center}
	\includegraphics[width=1.0\hsize]{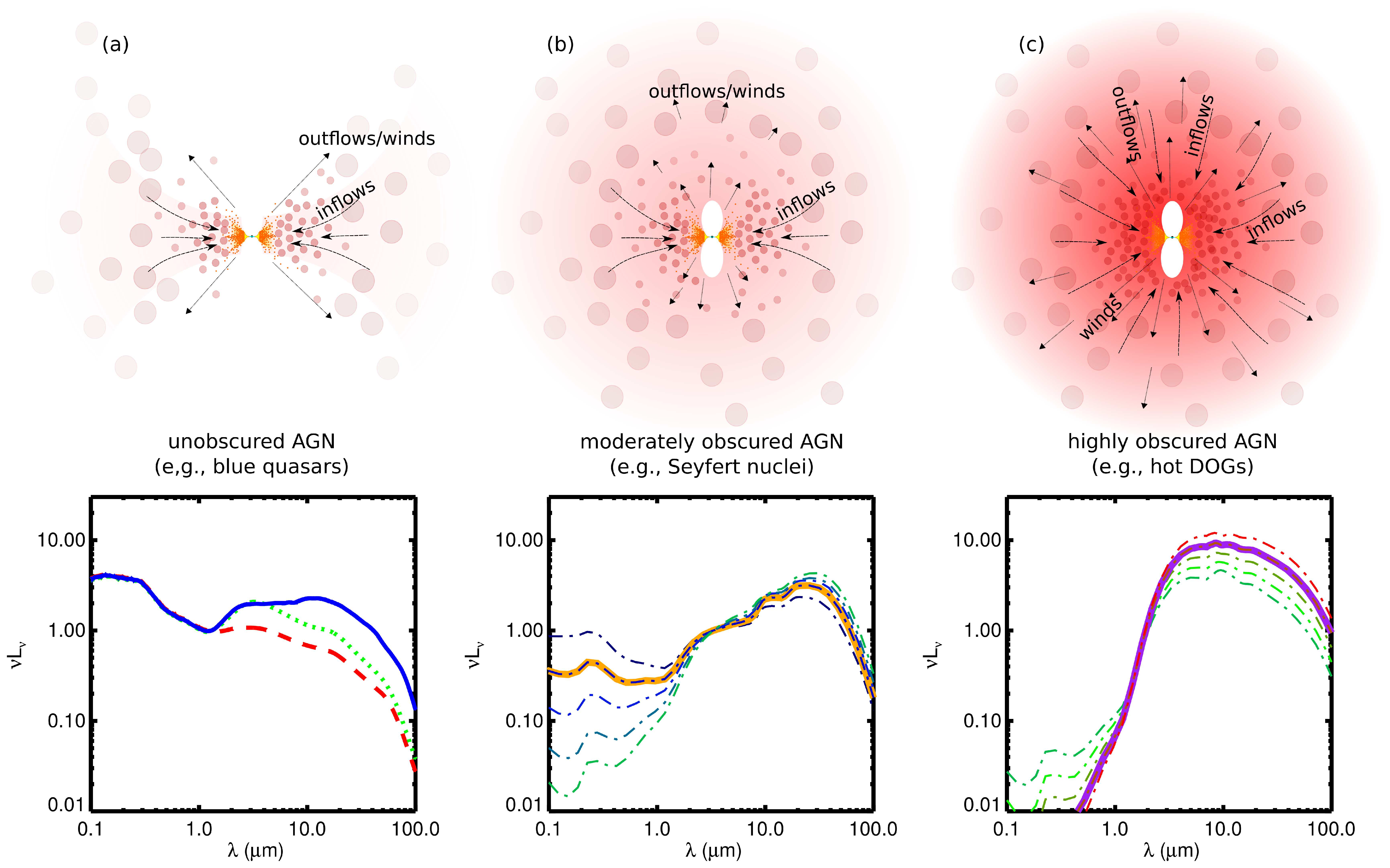} 
	\caption
	{ 
	Schematic drawings for the dust environment around an AGN (upper panels)
	and typical SEDs (lower panels). In column (a), we present the
	suggested dust environment for an unobscured AGN and example SEDs for
	optically-blue quasars from \citet{Lyu2017} (blue solid line: normal
	AGN, green dotted line: WDD AGN, red dashed line: HDD AGN). Column (b)
	shows the corresponding picture and SEDs for a moderately obscured AGN.
	We present the model SEDs for typical Seyfert-1 nuclei with $\tau_{\rm
	V}\sim$1--5.  Column (c) is the description of a highly obscured AGN.
	The SEDs are from our model for the hot dust-obscured AGNs with
	$\tau_{\rm V}\sim$5--20. See the text for details.
	     }
	    \label{fig:evoluation} 
    \end{center} 
\end{figure*}

\begin{itemize}
    \item unobscured AGNs: 

	The IR properties of such systems are dominated by a torus-like
	structure without much dust distributed along the polar direction. It
	is likely that AGN activity is extremely strong, blowing out the dusty
	gas or simply destroying most dust along the polar direction. The
	photons from the central engine can be directly seen from a face-on
	observing angle. Typical examples are luminous blue quasars, where
	there is little extinction in the UV-optical bands.

	Due to its relatively compact size, the behavior of the torus is mainly
	determined by accretion parameters.  As shown in \citet{Lyu2017}, the
	intrinsic IR SEDs of unobscured quasars present clear variations and
	might be related to different AGN properties (e.g., luminosity and
	Eddington ratios). Among Seyfert nuclei, we have also seen the
	appearance of both normal and dust-deficient AGNs, suggesting the torus
	structures among Seyfert-1 populations are not exactly the same. 

    \item  moderately-obscured AGNs: 
	
	Besides the torus, there exists an extended dust component ($r\sim10^2$
	pc) with low-optical depth ($\tau_{\rm V}\gtrsim$0--5), which could be
	AGN dusty outflows, winds or dusty narrow line regions. Along the polar
	direction, the dust obscures the central engine, resulting in mid-IR
	excess emission and some moderate level of optical extinction if the
	observation is not made through some lucky dust-free LOS.  

	Most AGNs, i.e., those in Seyfert galaxies, will fall into this
	category. In relation to unobscured quasars, our SED analysis suggested
	that Seyfert nuclei frequently have strong polar dust emission,
	indicating an extended dusty component that can be heated by the
	central engine. In fact, there are observational suggestions that the
	AGN NLR could disappear at very high luminosities \citep{Netzer2004}.
	If we believe these NLR clouds are dusty, a decreasing frequency of
	AGNs with polar dust emission with increasing luminosity should be
	expected. 
   
    \item  highly-obscured AGNs: 
	
	The AGN-heated dust in such systems can be very extended ($r\sim10^3$
	pc) with high optical depth ($\tau_{\rm V}\sim$5--20), which might be a
	result of, e.g., galaxy mergers. The gravitational torque of
	large-scale interaction can bring a substantial amount of galactic ISM
	towards the nucleus which can be collide with the AGN outflows,
	resulting in a very turbulent environment that changes the circumnuclar
	dust density profile. These objects present clear evidence of
	UV-optical obscuration and strong near- to mid-IR dust emission.
	Typical examples are AGNs in hot DOGs.

\end{itemize}

In this picture, the torus is formed and maintained during the black hole
accretion \citep[e.g.,][ and references therein]{Hopkins2012}. An extended dust
component surrounding the AGN nucleus naturally bridges the torus and the host
galaxy ISM. Both AGN and galaxy feedbacks may modify the properties of this
component, resulting in the changing obscuration behaviors at different phases
of galaxy evolution as suggested by many previous studies
\citep[e.g.,][]{Sanders1988a, Sanders1988b, King2003, Hopkins2010,
Hopkins2016}.

\section{Summary and Conclusions}\label{sec:summary}

Motivated by the commonly seen UV-optical obscuration and the discoveries of
polar dust emission in a few nearby Seyfert nuclei, we have proposed an
empirically-driven model to produce the SEDs of reddened type-1 AGNs. The
intrinsic AGN templates in \citet{Lyu2017} have been assumed as reasonable
choices for polar-dust-free type-1 AGNs. With radiation transfer calculations,
we obscured them by an extended dusty structure with a power-law density
profile and large dust grains. This model naturally:

\begin{enumerate}
    \item reproduced the nuclear UV to mid-IR SED \citep{Prieto2010} and polar
	dust emission strength \citep{Honig2013} for the type-1 AGN in NGC
	3783;

    \item fitted the IR SEDs of a large sample of low-$z$ Seyfert-1 nuclei with
	the mid-IR emission dominated by the AGN;

    \item reproduced the UV to mid-IR composite SEDs of the SDSS/{\it Spitzer}
	type-1 AGNs;

    \item matched the composite SED of extremely red luminous quasars in
	\citet{Hamann2017};

    \item explained the warm-excess dust emission seen in some high-$z$ type-1
	AGNs, as first reported by \citet{Xu2015a};

    \item reproduced the IR emission template of hot dust-obscured AGNs proposed by
	\citet{Fan2016}.
\end{enumerate}

Our main conclusions are as follows: 
\\

1. The broad-band IR SEDs of low-$z$ Seyfert-1 nuclei and some high-$z$ type-1
AGNs with peculiar SED features can be reconciled with the quasar intrinsic AGN
templates by adding the IR-reprocessed emission of polar dust. It is possible
that AGNs over a broad range of redshift and luminosity have a similar dust
environment that features a circumnuclear torus plus an extended dust
component. 

2. The polar dust emission could be a natural result of the commonly-seen
UV-optical obscuration among Seyfert-1 nuclei. This statement is supported by
our detailed SED analysis of NGC 3783 and the our successful fitting of the
composite SEDs of the SDSS/{\it Spitzer} Seyfert-1 AGN sample. Surprisingly, we
found the same radial density profile ($n\propto r^{-0.5}$, $r_{\rm out}/r_{\rm
in}\sim500$) and dust grain properties ($a_{\rm max}=10~\mum$, $a_{\rm
min}=0.04~\mum$) of the extended dust component works for most Seyfert-1
nuclei. 

3. The primary factor to determine the behavior of the polar dust emission is
the overall optical depth. The emission SED of the AGN polar dust component
peaks around 26~$\mum$ with a characteristic temperature at $T\sim113~$K,
leaving strong mid- to far-IR warm-excess emission signatures above intrinsic
AGN templates for some objects. The observed size of polar dust emission is a
function of wavelength.

4. Our SED analysis provides an effective method to search for AGNs with polar
dust emission. Among the 64 Seyfert-1 nuclei studied in this work, we found
$\sim$1/3 of them have significant polar dust emission that contributes at
least half of the AGN emission at $\sim10~\mum$, about another 1/3 show no
evidence with IR SEDs matched by the intrinsic templates, and the remaining 1/3
have weaker or absent polar dust emission - they are ambiguous. 

5. The reddened type-1 AGN model trained for NGC~3783 not only fitted the SEDs of most
Seyfert-1 nuclei but also worked reasonably well to reproduce the SEDs for
high-$z$ extremely red quasars and type-1 AGNs with mid-IR warm-excess. These
results suggest that most AGNs could share similar properties of extended dust
environment, which may indicate that AGN-driven outflows dominate the
large-scale structure and grain properties of the nuclear extended dust
environment.

6. The reddened type-1 AGN model can also reproduce the SEDs of hot dust-obscured
galaxies when combined with a star-forming galaxy template. However, the
best-fit parameters suggest a more extended AGN-heated dust distribution with a
steeper density profile compared with objects like NGC~3782. This indicates the
AGN dust environment in hot DOGs might be quite different from other systems.
Objects with similar hot DOG SED features are also found at $z\sim0.1$.

7. The general success of our reddened type-1 AGN model suggests the IR-reprocessed
emission of the extended dust distribution in some objects can complicate the
interpretation of the integrated AGN IR emission, such as the behavior of the
torus radiative transfer models  and constraints on their dust covering
factors.

8. The SED shape of the AGN thermal IR emission might be a reflection of its
dust environment. Depending on the obscuration level along the face-on
direction, we propose three basic scenarios, in which the AGN and galaxy
feedbacks may play different roles, that lead to the diverse IR SED properties
among type-1 AGNs.  
\\

From a technical perspective, contrasting with other alternatives to fit the
AGN-heated dust emission in the literature, the semi-empirical model proposed
here is able to fit a wide range of AGN SEDs using relatively few free
parameters. Combined with well-constrained empirical templates of star-forming
galaxies, we can get reasonably accurate separations of the AGNs and their host
galaxies.  Meanwhile, these observationally calibrated semi-empirical templates
are a valuable tool to characterize the IR color space of AGNs, which could be
useful to look for abnormal objects in the era of {\it JWST}.

\acknowledgments

We thank Dr. Aigen Li and Dr. Robert Nikutta for helpful discussions, Dr. Lei
Hao for providing the spectral measurements of the SDSS DR7 Main Galaxy Sample,
and Dr. Almudena Prieto for the clarification of NGC 3783 SED data. We are also
grateful to the anonymous referee for a very constructive report and helpful
comments.

This work was supported by NASA grants NNX13AD82G and 1255094. 

This work is based in part on observations made with the Spitzer Space
Telescope, which is operated by the Jet Propulsion Laboratory, California
Institute of Technology under a contract with NASA. The Combined Atlas of
Sources with Spitzer IRS Spectra (CASSIS) is a product of the IRS instrument
team, supported by NASA and JPL.

This publication has made use of data products from the {\it Wide-field
Infrared Survey Explorer}, which is a joint project of the University of
California, Los Angeles, and the Jet Propulsion Laboratory/California Institute
of Technology, funded by the National Aeronautics and Space Administration.
This publication also makes use of data products from NEOWISE, which is a
project of the Jet Propulsion Laboratory/California Institute of Technology,
funded by the Planetary Science Division of the National Aeronautics and Space
Administration. This publication makes use of data products from the Two Micron
All Sky Survey, which is a joint project of the University of Massachusetts and
the Infrared Processing and Analysis Center/California Institute of Technology,
funded by the National Aeronautics and Space Administration and the National
Science Foundation. 

Funding for the SDSS and SDSS-II has been provided by the Alfred P. Sloan
Foundation, the Participating Institutions, the National Science Foundation,
the U.S. Department of Energy, the National Aeronautics and Space
Administration, the Japanese Monbukagakusho, the Max Planck Society, and the
Higher Education Funding Council for England. The SDSS Web Site is
http://www.sdss.org/. The SDSS is managed by the Astrophysical Research
Consortium for the Participating Institutions. The Participating Institutions
are the American Museum of Natural History, Astrophysical Institute Potsdam,
University of Basel, University of Cambridge, Case Western Reserve University,
University of Chicago, Drexel University, Fermilab, the Institute for Advanced
Study, the Japan Participation Group, Johns Hopkins University, the Joint
Institute for Nuclear Astrophysics, the Kavli Institute for Particle
Astrophysics and Cosmology, the Korean Scientist Group, the Chinese Academy of
Sciences (LAMOST), Los Alamos National Laboratory, the Max-Planck-Institute for
Astronomy (MPIA), the Max-Planck-Institute for Astrophysics (MPA), New Mexico
State University, Ohio State University, University of Pittsburgh, University
of Portsmouth, Princeton University, the United States Naval Observatory, and
the University of Washington.

We acknowledge the use of the NASA/IPAC Extragalactic Database (NED) which is
operated by the Jet Propulsion Laboratory, California Institute of Technology,
under contract with the National Aeronautics and Space Administration. This
work has also made use of the VizieR catalogue access tool, CDS, Strasbourg,
France.

\software{DUSTY \citep{dusty_new}, SKIRT \citep{Baes2003, Baes2011}}

\appendix

\section{The Effects of Optical Thickness}\label{sec:model-thickness}

We use 3D radiative transfer simulations to demonstrate how the optical
thickness of the dusty medium can affect its SED. Detailed simulation and
analysis of the IR emission from clumpy clouds is described in the literature
(e.g., \citealt{Nenkova2008a, Stalevski2012}).  We illustrate several key
concepts focusing on a single dusty cloud heated by the emission from an
accretion disk at some distance.

\begin{figure}[!htb] 
    \begin{center}
	\includegraphics[width=0.495\hsize]{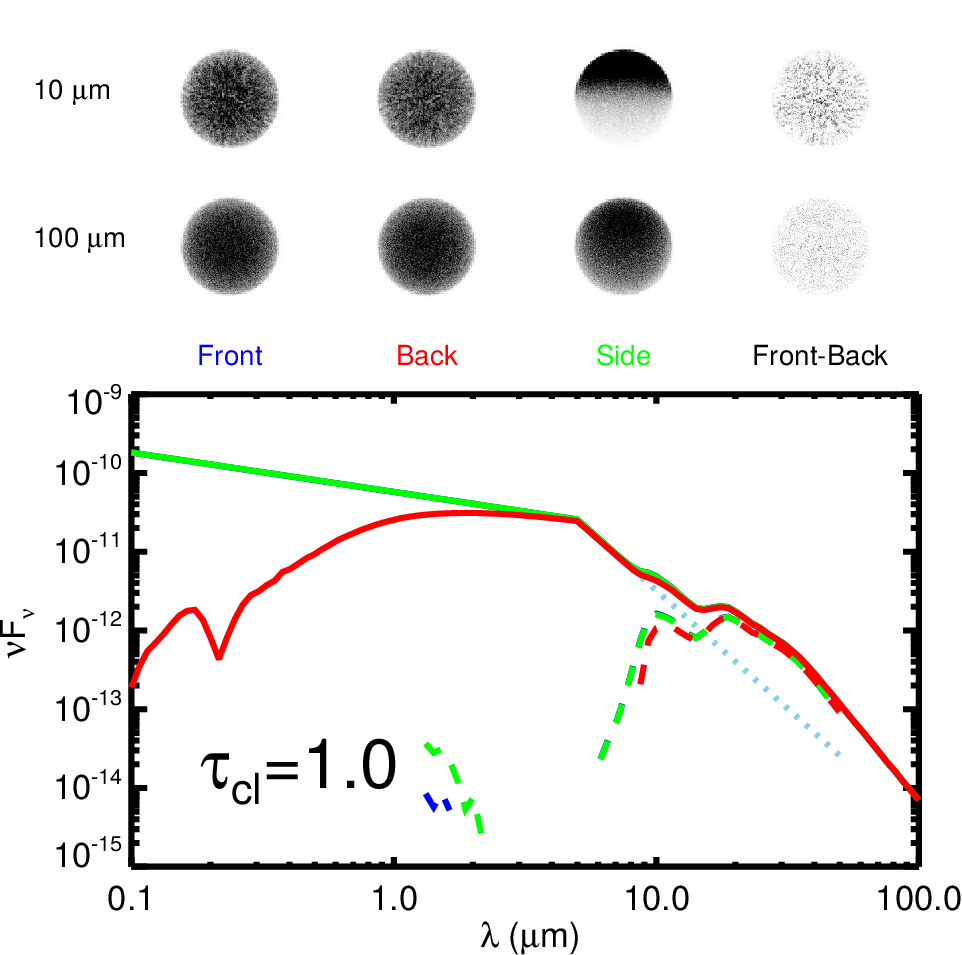} 
	\includegraphics[width=0.495\hsize]{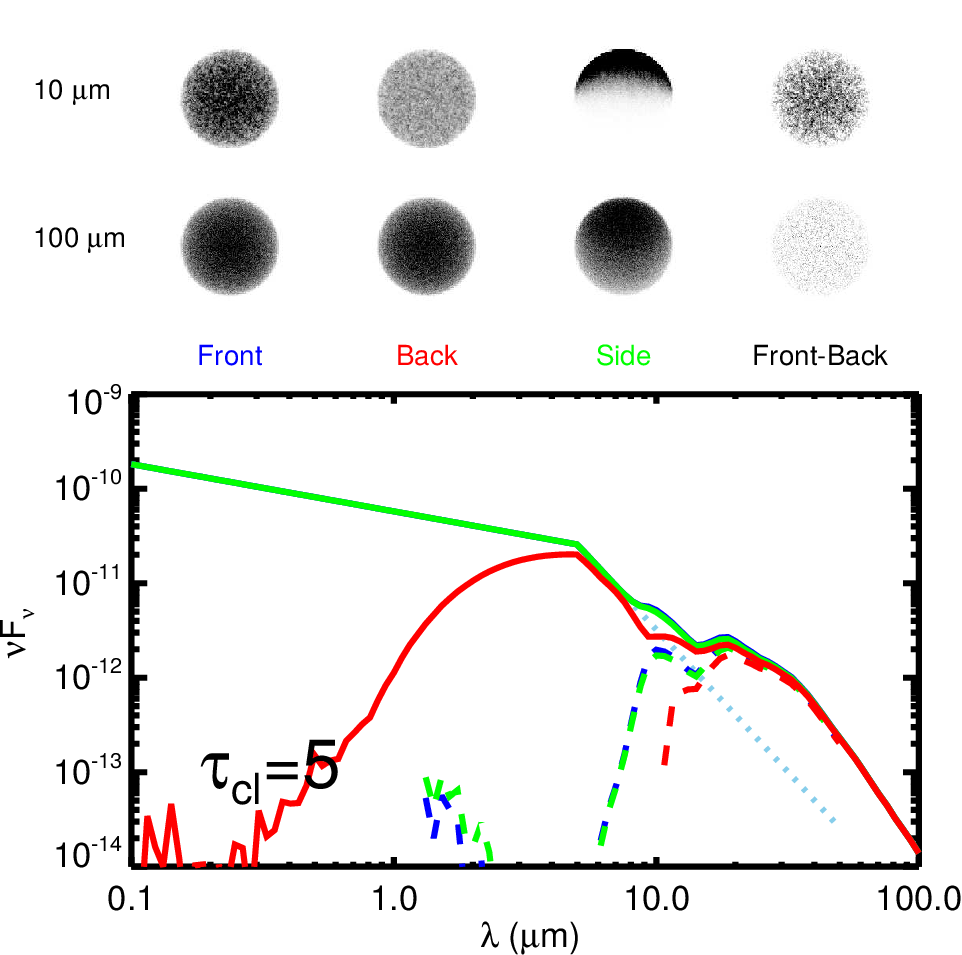} 
	\includegraphics[width=0.495\hsize]{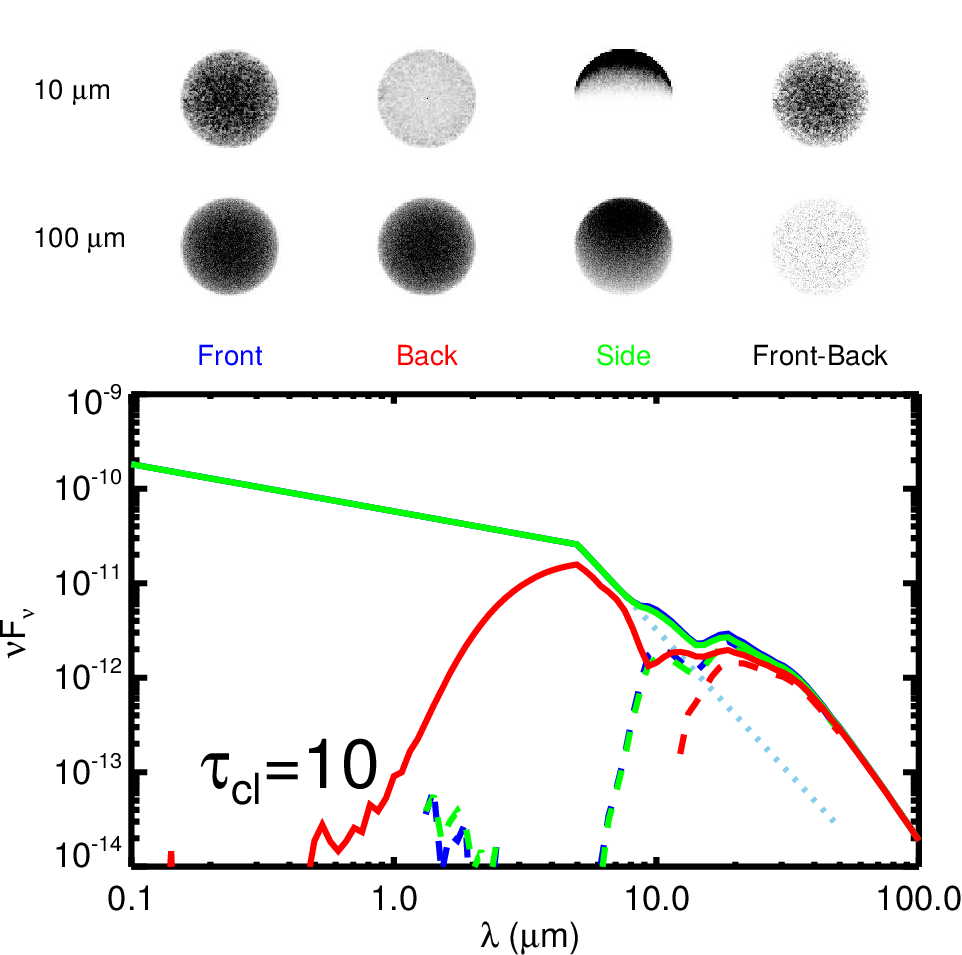} 
	\includegraphics[width=0.495\hsize]{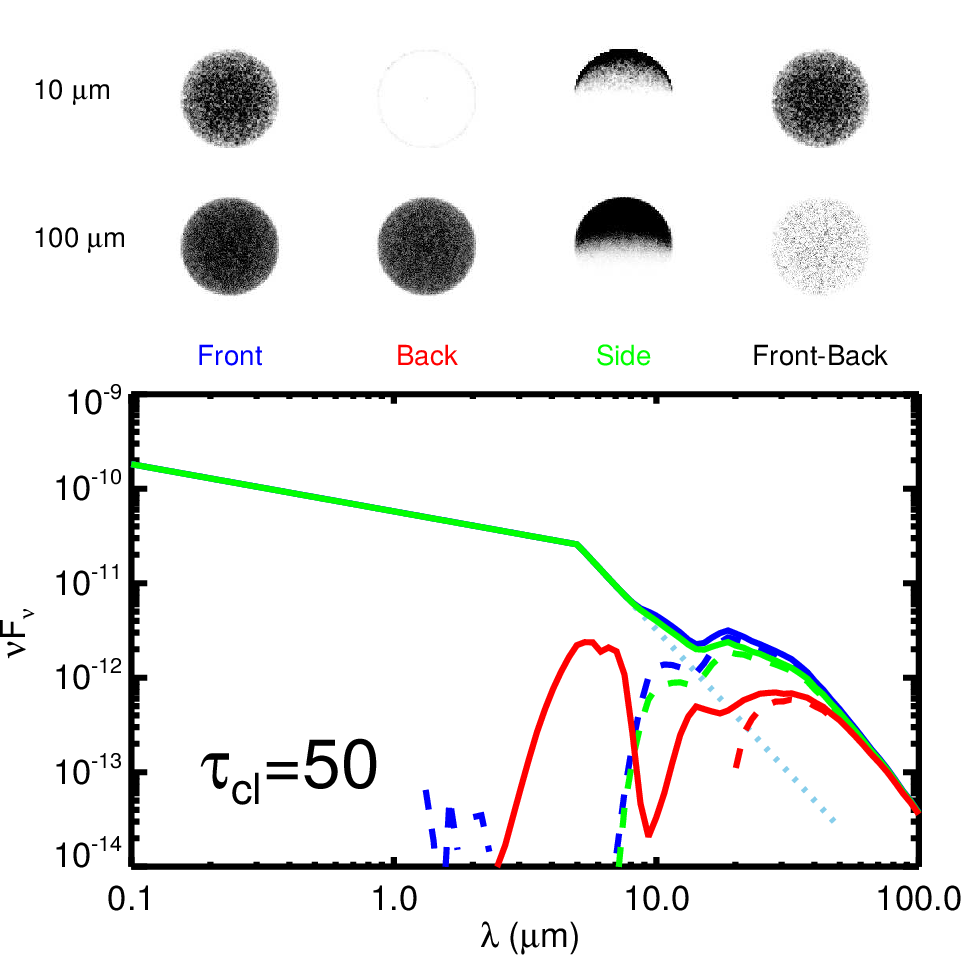} 
	\caption{
	    The output SEDs and images of a single dusty cloud plus the
	    accretion disk emission viewing from different angles. The
	    integrated SEDs are plotted as solid lines. The accretion disk
	    emission is represented as the blue dotted line. We also show the
	    pseudo IR emission SEDs (dashed lines) of the cloud by subtracting
	    the accretion disk SED from the total SED. We show the model images
	    of the dusty cloud at 10~$\mum$ and 100~$\mum$. The brightness is
	    linearly scaled with the darkness of the pixel. To demonstrate the
	    transparency of the dust cloud, a differential image `Front-Back'
	    is also made by subtracting the `Back' emission from the `Front'
	    emission at corresponding pixels.
	    }
	    \label{fig:single_cloud} 
    \end{center} 
\end{figure}

For simplicity, the cloud is assumed to be a sphere with a radius $r=1$~pc and
a homogeneous density distribution of typical ISM dust grains. The accretion
disk is approximated as a point source with a broken power-law SED following
\cite{Stalevski2016}, and located at a distance of 5 pc to the cloud. We use
the radiative transfer code SKIRT \citep{Baes2003, Baes2011} to compute the
output SEDs as well as the images observed from different angles as a function
of the optical thickness of the cloud ($\tau_\text{cl}$).  The results can be
found in Figure~\ref{fig:single_cloud}.  If the cloud is optically thin
($\tau_\text{cl}\lesssim1$), the IR SEDs would be identical for any viewing
angles. In fact, if the V-band optical thickness is not very large
($\tau_\text{cl}\lesssim10$), the cloud would be almost transparent for its own
emission, given the rapidly decreasing extinction at longer wavelengths. These
conclusions can be seen from the differential images between the front view of
the bright side of the cloud and the back view of the dark side of the cloud.
In the case of blocking, the dust emission SED will not change once the
optical depths at the corresponding wavelength are low (e.g., $\lambda
\sim100~\mum$).

%%% Figure 20 %%%

\begin{figure*}[!htb]
    \begin{center}
	\includegraphics[width=1.0\hsize]{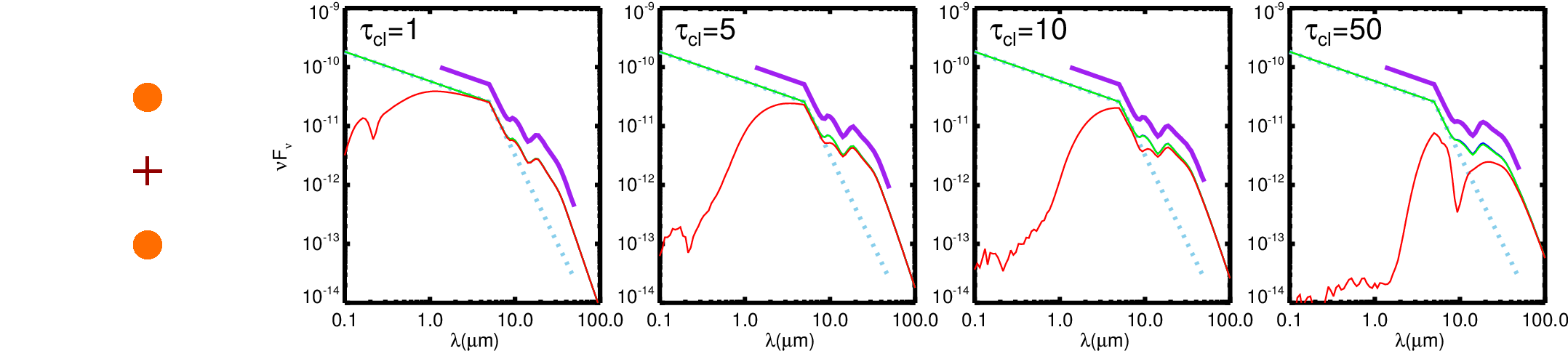}
	\includegraphics[width=1.0\hsize]{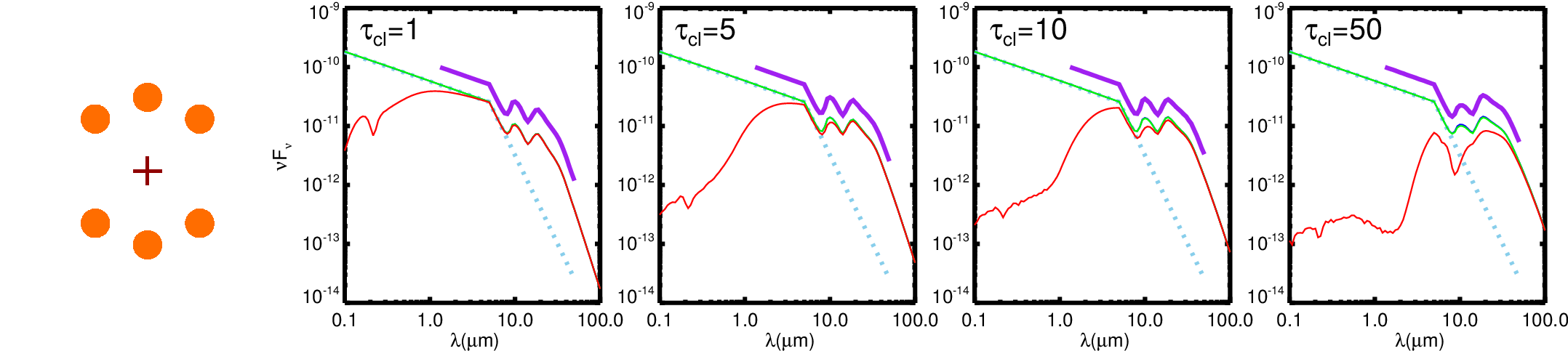}
	\includegraphics[width=1.0\hsize]{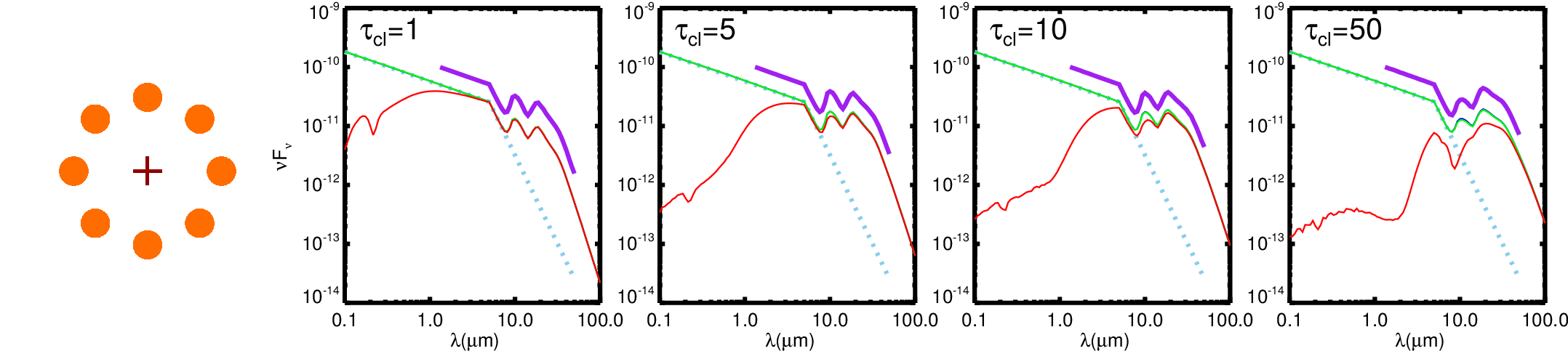}
		\caption{
		    The output SEDs of systems with different numbers of dusty
		    clumps with optical depth $\tau_{\rm V}=1, 5, 10, 50$,
		    heated by the accretion disk emission observed from
		    different lines of sight (blue: face-on, red: edge-on,
		    green: 45-degree inclination angle).  The intrinsic
		    accretion disk emission is represented by a broken power
		    and shown as the dashed blue line. The purple line is the
		    SED of the accretion disk emission linearly combined with
		    the emission from a single cloud but multiplied by N times,
		    where N is the number of the clouds in the corresponding
		    system.  We arbitrarily scaled the purple by 2 times for
		    clarity.
		    }
	\label{fig:eight_cloud}
    \end{center}
\end{figure*}

The optically-thin assumption also makes the effects of the dust-covering
factor relatively easy to be accounted. In Figure~\ref{fig:eight_cloud}, we
symmetrically distribute spherical dusty spheres at the same distance and
calculate the output SEDs. The shape of the dust-processed SED is the same as
the case of a single cloud, and the intensity can be matched by linearly
scaling the emission from the single cloud.

%%% Figure 21 %%%

\begin{figure*}[!htb]
    \begin{center}
	\includegraphics[width=1.0\hsize]{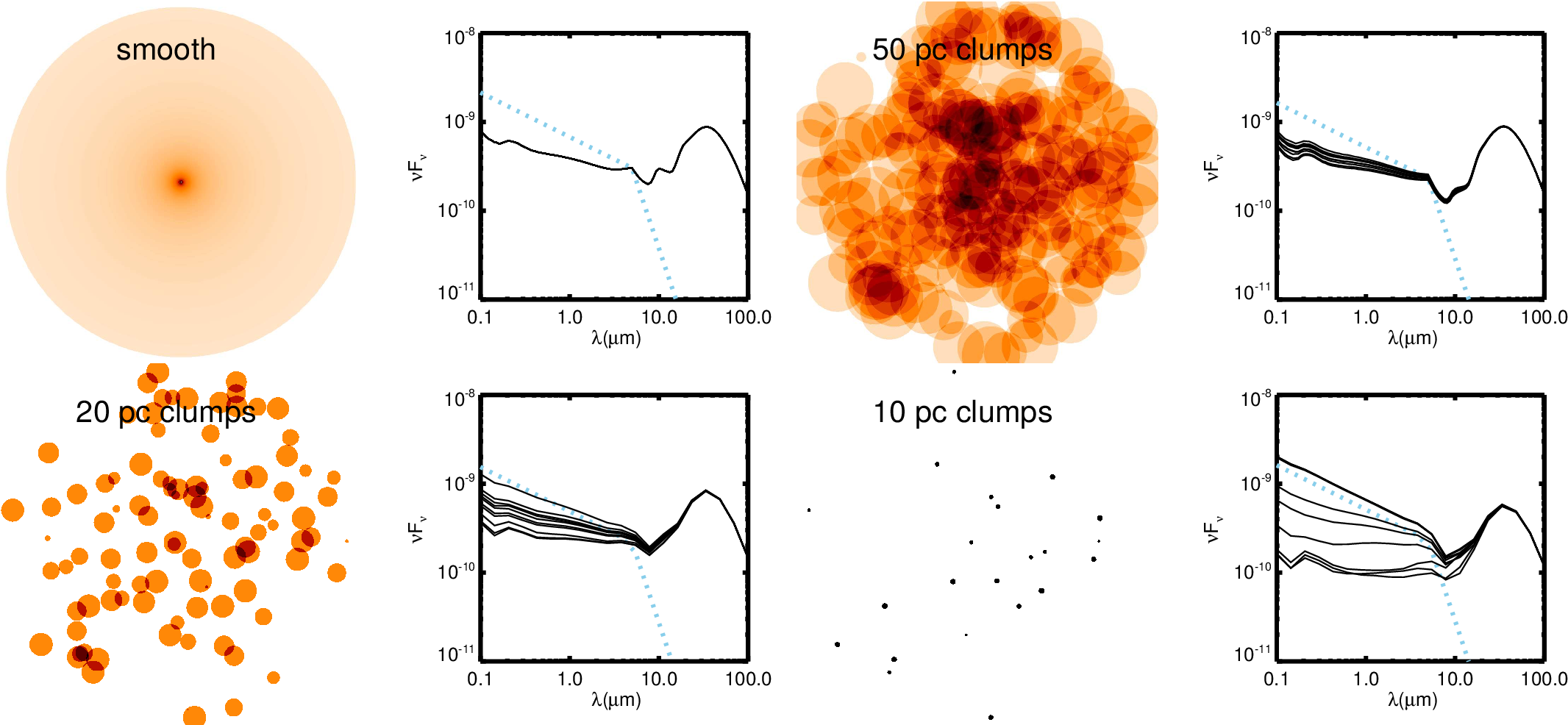}
		\caption{
		    Demonstration on the effect of clumpiness. The
		    top-left panel shows the projected dust density
		    distribution and the SEDs at different line-of-sights
		    (inclination angle 0--90), assuming a $n\propto
		    r^{-0.5}$ dusty sphere with $R_\text{in}=0.6pc$,
		    $R_\text{out}=300pc$ and radial optical depth
		    $\tau_\text{V}=1.5$, heated by a point source in the
		    center with luminosity
		    $L_\text{AGN}=10^{11}~L_\odot$. We adopt a broken
		    power-law SED for the intrinsic accretion disk
		    emission, following e.g, \citet{Stalevski2012} (blue
		    dotted line).  In all these simulations, large dust
		    grains in AGN environment are assumed (see text for
		    details). In the other panels, we redistribute the
		    dust in the sphere into 1000 clumps with different
		    sizes (50 pc, 20 pc and 10 pc), following
		    the same large geometry distribution with the same total
		    dust mass. The simulations are run with a
		    $100\times100\times100$ bin linear mesh of a
		    box size $600\times600\times600$ pc.
		    }
	\label{fig:sphere_test}
    \end{center}
\end{figure*}

For a low-optical-depth dust component, the output SEDs are not sensitive to
the clumpiness of the dusty clouds. We demonstrate the effect of clumpiness in
Figure~\ref{fig:sphere_test}. We first use the SKIRT code to produce the output
SED of a $\tau_{\rm V}=1.5$ dusty sphere with density profile $n\sim r^{-0.5}$,
$r_\text{in}=0.6~pc$ and $r_\text{out}=300~pc$, assuming the large dust grain
distribution $a_\text{max}=10~\mum$, $a_\text{min}=0.04~\mum$. With the same
total dust grain mass, we break the smooth geometry into 1000 randomly
distributed spherical clumps with different size $R_\text{cl}$, but following
the same density profile. Depending on whether the line-of-sight is blocked by
the dust, the UV-optical SEDs present large variations for different viewing
angles.  However, the dust-reprocessed IR SEDs are nearly identical.  We
conclude the clumpiness would not influence the IR SED shapes.

%%% Figure 22 %%%

In summary, we have demonstrated that optical depth up to $\tau_{\rm
V}\lesssim5$ does not strongly modify the infrared output of a cloud composed
of classical ISM dust and directly heated by an AGN. For larger grains
suggested for polar dust, such effects would be further minimized. At the same
radius, the integrated IR SED of $N$ identical clouds, whatever their
placement, can be matched by linearly scaling the emission SED of single cloud
by a factor of $N$. Assuming modest optical depths, the output SED is also
insensitive to the clumpiness of the dusty medium. Thus, the primary
determination of the infrared SED is the placement of the polar dust clouds
along the radial direction, i.e., the radial density profile.

\section{Sample and Data Compilation for Low-$z$ Seyfert-1 nuclei}\label{sec:data}

\subsection{AGNs with HSR SED observations}

We first constructed a sample of AGN with high spatial resolution (HSR)
measurements as follows.  \citet{Asmus2014} presented a comprehensive mid-IR
imaging atlas of 253 nearby AGNs with publicly available data from
sub-arcsec-resolution observations carried out by ground-based 8-m class
telescopes.  We adopted this atlas as the parent sample and selected all type
1, 1.2, 1.5 and 1n objects. To reduce the possibility of mid-IR host galaxy
contamination, we cross-matched this sample with the ALLWISE source catalog
\citep{wise} and compared the 12 $\mu m$ flux from ground-based subarcsec
resolution observation, $F_{g}(12~\mum)$, and that from space-based
6.5\arcsec~resolution WISE band 3 observation, $F_{s}(12~\mum)$.  The W3 flux
is known to have a systematic bias that is color-dependent (Section 2.2 in
\citealt{wise}).  Our objects have $f_\nu\propto\nu^{-\beta}$ with
$\beta\sim$0--3, which corresponds to a flux correction factor of 0.92--1.10.
Considering this and other uncertainties from e.g., flux zeropoints and filter
differences, we selected objects that satisfy
$F(12~\mum)_{g}/F(12~\mum)_{s}>0.9$. In addition, we required that the source
shape is consistent with a point-source in all four WISE bands. Any objects
falling within the extrapolated isophotal footprint of a 2MASS extended source
have been removed to avoid strong stellar contamination in the near-IR.
\footnote{We removed any ALLWISE sources with ext\_flg=1,2,3,5.} Since the SEDs
of Palomar-Green (PG) quasars whose SEDs have been studied in our previous work
\citep{Lyu2017}, we remove them from this study.

Since we are also interested in the AGN polar dust emission, we included all
the type-1 objects with mid-IR interferometry observations in
\citet{Lopez-Gonzaga2016} besides 3C 273 (PG 1226+023). Finally, some
well-known nearby Seyfert-1 nuclei with high-spatial-resolution IR SEDs in the
literature \citep[e.g.,][]{Alonso-Herrero2003, RamosAlmeida2009, Prieto2010,
Fuller2016} are also included in this study. This sample is summarized in
Table~\ref{tab:sample}. 

For the above sample selected from \citet{Asmus2014}, we adopted the 2MASS
profile-fit photometry and the WISE profile-fit photometry to sample the
near-IR to mid-IR SED. The FWHM of a typical 2MASS point-spread-function (PSF)
is about 2.5 arcsec and the FWHM of the WISE band 1--3 is about 6 arcsec. We
replace them with the higher spatial resolution data from, e.g., {\it
Spitzer}/IRAC or HST/NICMOS if available. It is known that the ground-based
mid-IR imaging could have unstable PSF \citep[e.g.,][]{Radomski2008}, thus we
retained the WISE band-3 (12~$\mum$) flux. The FWHM of the WISE W4 band is
about 12 arcsec, we replace that with {\it Spitzer}/MIPS 24 micron data
wherever possible. Most of the other objects have (sub)arcsec resolution SEDs
presented in the literature. We complemented the incomplete SEDs with either
WISE or 2MASS photometry.

To further constrain the SED shape of the dust emission, we collected the
mid-IR spectra for objects observed by {\it Spitzer}/IRS. For the staring mode
observations, we adopt the {\it CASSIS} products \citep{cassis}. For the
mapping mode, we used {\it CUBISM} \citep{cubism} to reprocess the data,
following the standard pipeline given in the software document.

\subsection{The SDSS-{\it Spitzer}/IRS Type-1 AGNs}\label{sec:saga}

We adopted the spectral decomposition results of the Main Galaxy Sample
\citep{Strauss2002} in the SDSS Seventh Data Release (DR7; \citealt{SDSS_DR7})
processed with the optical spectral data reduction pipeline developed by
\citet{Hao2005a}.  After subtracting the host galaxy continuum and the
power-law AGN components, type-1 AGNs were selected as objects with the
H$\alpha$ broad component with full width at half maximum (FWHM) $>$ 1200 km/s
from multi-Gaussian line-profile fit with a rest-frame H$\alpha$ equivalent
width (EW)$>3~\AA$. Due to the requirement of a direction of H$\alpha$
information, the sample was limited to $z<0.33$ by the SDSS spectral coverage. 

After compiling the SDSS Seyfert-1 sample, we cross-matched their SDSS
coordinates with the {\it Combined Atlas of Sources with Spitzer IRS Spectra}
(CASSIS, \citealt{cassis}) within a 3\arcsec~search radius to get the mid-IR
data.  The CASSIS contains all {\it Spitzer}/IRS staring mode observations with
enough signal for a useful spectrum. Since a strong host galaxy contamination
in the mid-IR would make the interpretations ambiguous, only objects without
evidence for strong aromatic features or silicate absorption(s) were selected.
All of the sample have 11.3~$\mum$ aromatic features with equivalent width
(EQW) less than 0.1~$\mum$ from {\it Spitzer}/IRS spectral decomposition with a
modified version of {\it PAHFIT} \citep{Smith2007}, following
\cite{Gallimore2010} to add an optically-thin, warm dust component to reproduce
the AGN silicate emission features (see their Section 4.1). Finally, we ended
up with 33 SDSS-{\it Spitzer}/IRS type-1 AGNs.  We used the IDL routine {\it
DeblendIRS} \citep{Hernan-Caballero2015} to double check the level of star
formation; the average luminosity contribution at 5.5--15~$\mum$ is about 3\%
with the maximum value $<9\%$.

To cover the full SED, we collected other multi-band photometry data for this
sample. Seventeen objects ($\sim$52\%) have X-ray observations from XMM-{\it
Newton} or {\it Chandra}. We searched the {\it GALEX}, {\it SDSS}, {\it 2MASS}
and {\it WISE} archives for the corresponding UV/optical/IR photometric data.
Over 80\% of these objects are resolved in the 2MASS images; for them, we
adopted the Standard Photometry with isophotal apertures based on the K$_s$ 20
mag arcsec$^{-2}$ elliptical isophote from the 2MASS extended source catalog.
For the WISE W1 ($\sim3.4~\mum$)and W2 ($\sim4.6~\mum$) bands, we used the
scaled-2MASS-aperture photometry with the largest aperture as long as no other
sources was included if the object was in the 2MASS extended source catalog and
the WISE aperture photometry is larger than the default profile-fit photometry.
We adopted the profile-fit photometry of the J, H, K, W1 and W2 band in other
cases. 

\subsection{IR Light Curves from WISE/NEOWISE}

To address the infrared variabilities of these AGNs, we derive light curves
based on the W1 and W2 band observations from the {\it WISE} \citep{wise} and
the Near-Earth Object {\it WISE} Reactivation ({\it NEOWISE-R};
\citealt{neowise}) missions, following similar procedures in our previous work
\citep{Lyu2017}. For most objects, the light curves cover seven to nine
different epochs with 10--20 individual exposures for each epoch. We computed
the mean photometry for each epoch and derived the maximum variability
amplitudes. These results are included in Table~\ref{tab:sample}.

\bibliographystyle{apj.bst}

%\bibliography{../../../journal/_bib/hdd.bib}

\end{document}